\documentclass{emulateapj}

\begin{document}
\slugcomment{Submitted to ApJ: 2008---02---01; Accepted: 2008---06---28}
\shorttitle{UV Survey of CO and H$_2$}
\shortauthors{Sheffer et al.}

\title{Ultraviolet Survey of CO and H$_2$ in Diffuse Molecular Clouds:
The Reflection of Two Photochemistry Regimes in Abundance Relationships}
\author{Y. Sheffer\altaffilmark{1}, M. Rogers\altaffilmark{1},
S. R. Federman\altaffilmark{1,2},
N. P. Abel\altaffilmark{3}, R. Gredel\altaffilmark{4},
D. L. Lambert\altaffilmark{5}, and G. Shaw\altaffilmark{6}}

\altaffiltext{1}{Department of Physics and Astronomy, University of Toledo,
Toledo, OH 43606; ysheffe@utnet.utoledo.edu; steven.federman@utoledo.edu}
\altaffiltext{2}{Guest Observer, McDonald Observatory, University of Texas,
Austin, TX 78712}
\altaffiltext{3}{Department of Physics, University of Cincinnati, Cincinnati,
OH 45221; npabel2@gmail.com}
\altaffiltext{4}{Max Planck Institute for Astronomy, Koenigstuhl 17, D69117
Heidelberg, Germany; gredel@mpia.de}
\altaffiltext{5}{W. J. McDonald Observatory, University of Texas, Austin, TX
78712; dll@astro.as.utexas.edu}
\altaffiltext{6}{Department of Astronomy and Astrophysics, Tata Institute of
Fundamental Research, Mumbai 400005, India; gargishaw@gmail.com}

\begin{abstract}
We carried out a comprehensive far-ultraviolet (UV) survey of $^{12}$CO and H$_2$ column
densities along diffuse molecular Galactic sight lines in order to explore in detail
the relationship between CO and H$_2$.
For this survey we measured new CO abundances from absorption bands detected in
{\it Hubble Space Telescope} spectra for 62 sight lines, and new H$_2$ abundances from
absorption bands in {\it Far Ultraviolet Spectroscopy Explorer} data for 58 sight lines. 
In addition, high-resolution optical data were obtained at the McDonald and European Southern
Observatories, yielding new abundances for CH, CH$^+$, and CN along 42 sight lines to aid in
interpreting the CO results.
A plot of log $N$(CO) versus log $N$(H$_2$) shows that two power-law relationships
are needed for a good fit of the entire sample, with a break located at
log $N$(CO, cm$^{-2}$) = 14.1 and log $N$(H$_2$) = 20.4,
corresponding to a change in production route for CO in higher-density gas.
Similar logarithmic plots among all five diatomic molecules allow us to probe their
relationships, revealing additional examples of dual slopes in the cases of CO versus CH
(break at log $N$ = 14.1, 13.0), CH$^+$ versus H$_2$ (13.1, 20.3),
and CH$^+$ versus CO (13.2, 14.1).
These breaks are all in excellent agreement with each other, confirming the
break in the CO versus H$_2$ relationship, as well as the one-to-one
correspondence between CH and H$_2$ abundances.
Our new sight lines were selected according to detectable amounts of CO in their
spectra and they provide information on both lower-density ($\leq$ 100 cm$^{-3}$)
and higher-density diffuse clouds.
The CO versus H$_2$ correlation and its intrinsic width are shown to be empirically related to
the changing total gas density among the sight lines of the sample.
We employ both analytical and numerical chemical schemes in order to derive
details of the molecular environments.
In the denser gas, where C$_2$ and CN molecules also reside, reactions involving C$^+$
and OH are the dominant factor leading to CO formation via equilibrium chemistry.
In the low-density gas, where equilibrium-chemistry studies have failed
to reproduce the abundance of CH$^+$, our numerical analysis shows that nonequilibrium
chemistry must be employed for correctly predicting the abundances of both CH$^+$ and CO.
\end{abstract}

\keywords{astrochemistry --- ISM: abundances --- ISM: molecules --- 
ultraviolet: ISM}

\section{Introduction}

The most abundant molecule in the Cosmos, H$_2$, has no permanent dipole moment
and is thus lacking permitted pure rotation and vibration-rotation transitions.
On the other hand, rotational transitions of CO, such as $J$~= 1$-$0 at 115
GHz, have been routinely and extensively observed in molecular clouds, which
are also too cold for detection of excited levels of H$_2$ (\citealt{Papad02} and
references therein).
Radio telescopes are thus used to map CO emission in the interstellar medium (ISM)
in order to delineate global distributions of molecular clouds in our Galaxy
and in other galaxies.

It is an empirical and theoretical foundation of radio mapping that the velocity-integrated
emission intensity of CO ($W_{\rm CO}$, in units of K km s$^{-1}$) from a
molecular cloud is proportional to the total virialized mass of the cloud, and hence to
its hydrogen content \citep{Lars81, Youn91}.
Radio astronomers thus utilize $W_{\rm CO}$ as a proxy for H$_2$,
by employing the ``$X$-factor'' $X_{\rm CO}$ = $N$(H$_2$)/$W_{\rm CO}(1-0)$,
where $N$ is the observed column density and $X_{\rm CO}$ will be
assumed hereafter to be in units of 10$^{20}$ cm$^{-2}$ (K km s$^{-1}$)$^{-1}$.

An average value is ascribed to giant molecular clouds (GMCs)
in the Milky Way, $X_{\rm CO}$ $\approx$ 4
\citep{Youn82, Dick86}.
\citet{Polk88} found significant millimeter-wave emission from CO that was not
associated with GMCs. This lowers the value of $X_{\rm CO}$, having
a mean value in the solar neighborhood of 1.8 $\pm$ 0.3 \citep{Dame01}.
A dependence of $X_{\rm CO}$ on the metallicity (primarily C/H) of the gas has been found in,
e.g., studies of $\gamma$-ray emission across the Galaxy.
This intensity is the product of interactions between cosmic rays and
``stationary'' gas; thus the $\gamma$-ray intensity is proportional to the
amount of gas along the line of sight.
\citet{Stro04} found that $X_{\rm CO}$ varies between 0.4 and 10.0
from the inner Galaxy to its outer regions,
indicating lower gas metallicity in the outer Galaxy.

The GMCs, as well as the smaller dark clouds, are opaque enough that the CO in their
cores is not dissociated by the interstellar far-ultraviolet (far-UV) radiation field.
Consequently, CO is unobservable in the UV, necessitating its
detection via millimeter-wave emission.
On the other hand, diffuse molecular clouds (as well as envelopes of dark clouds)
have visual extinctions lower than 5 mag, enabling a direct determination
of column densities via line absorption in the UV.
Both CO and H$_2$ are photodissociated by far-UV radiation,
resulting in a variable $X_{\rm CO}$ that depends on the efficiency
of self shielding (as well as mutual shielding) of the two species
\citep{vanD88}, and thus on their column densities.
One of our goals here is to study the behavior of $X_{\rm CO}$
under diffuse-ISM conditions.

The \citet{Fede80} study of diffuse interstellar clouds showed that there is an approximate
quadratic relationship between CO and H$_2$, such that $N$(CO) $\propto$
[$N$(H$_2$)]$^B$, with $B$ $\approx$ 2.
Furthermore, \citet{Fede80} remarked that for a group of lower-$N$
sight lines, a shallower slope with $B$ $\approx$ 1.5 was more appropriate,
thus signaling the possibility of $B$ varying with $N$.
Indeed, for sight lines with log $N$(CO) $\leq$ 16, or values higher than those
that were available to \citet{Fede80}, an even steeper relationship with $B$ $\approx$ 3
was found in the study of \citet{Pan05}.
Both \citet{Burg07} and \citet{Sonn07} confirmed the steepening of the slope near
log $N$(CO) $\approx$ 15, attributing it to self-shielding of CO.
Thus a variable power law in CO versus H$_2$ shows that the abundance of CO
relative to H$_2$, and hence $X_{\rm CO}$, are increasing with $N$(H$_2$)
along diffuse molecular sight lines.
In this paper we explore the CO versus H$_2$ relationship in more detail, namely,
the trend of CO (as well as of CH, CH$^+$, and CN) versus H$_2$ in diffuse
molecular clouds and the dependence of these correlations on $N$(H$_2$)
and on physical parameters, such as the total gas (hydrogen) density,
$n_{\rm H}$.

Our study differs from the recent work of \citet{Burg07} and
\citet{Sonn07} in a number of ways.
First, unlike \citeauthor{Burg07} and like \citeauthor{Sonn07}, our
methodology is based on profile fitting of the data, with detailed decomposition
into cloud component structures \citep{Shef07}.
We do not consider apparent optical depth nor curve of growth treatments,
which are always less preferred to spectrum synthesis by profile fitting
\citep{Sonn07}, nor do we build a grid of models to look
for a solution with a single effective $b$-value \citep{Burg07}.
Second, we follow up on these measurements of $N$ values with two
methods of chemical analysis, analytical and numerical, in order to
derive $n_{\rm H}$ at the sites where CO is detected.
Third, we are able to discern two regimes of CO formation in terms
of $n_{\rm H}$.
CO is associated with the similarly heavy diatomic molecules C$_2$ and CN
inside denser and colder clumps of gas \citep{Fede94, Pan05, Sonn07},
whereas in low-density clouds CO
is related to the formation and chemistry of CH$^+$
\citep{Zsar03}.
Remarkably, this transition in the photochemistry of CO will be shown ($\S$ 3)
to affect also the trends of other correlations among the diatomic
molecules analyzed here.

In $\S$ 2 we detail our sources for data and our methods of reduction
and analysis. Next, in $\S$ 3 observational results are presented in terms
of derived component structures and correlations between molecular column
densities. In $\S$ 4, 5, and 6 we explore some of the physical conditions
of the CO-harboring gas in terms of empirical relationships, analytical
chemical analysis, and detailed numerical modeling with Cloudy, respectively.
A discussion will be given in $\S$ 7, followed by the conclusions in $\S$ 8.

\section{Data and Modeling}

Our primary effort was to detect and measure $N$(CO) for new sight lines from
archival \textit{Hubble Space Telescope} ($HST$) data.
For most of these sight lines the value of $N$(H$_2$) was already known from
previous surveys with the \textit{Copernicus} satellite \citep{Sava77}
or \textit{Far Ultraviolet Spectroscopic Explorer} ($FUSE$)
\citep{Rach02, Andr03, Cart04, Pan05}.
However, for consistency, we determined $N$(H$_2$) also for sight lines with
previously published results.
Only two new sight lines (HD 36841 and HD 43818) lack any $N$(H$_2$) data.
For these, predicted values of $N$(H$_2$) will be provided in $\S$ 4.4 after
exploring the H$_2$ relationships with CO and CH.
We obtained new high-resolution optical spectra of CH$^+$, CH, and CN
by observing 42 sight lines at either McDonald Observatory or at the
European Southern Observatory (ESO).
Results for $^{13}$CO, which is also present in the
\textit{HST} spectra, were published in \citet{Shef07}.

Table 1 provides a list of all sight lines in terms of stars
observed, their spectral types, visual magnitudes, Galactic coordinates, local
standard of rest (LSR)
corrections, $E(\bv)$ reddening values, and heliocentric distances.
Table 2 lists the UV data sets from $HST$ and $FUSE$ for our stellar targets,
and STIS optical setups in terms of gratings and apertures.

\subsection{\textit{HST} Data}

Initially, our sample included 66 sight lines without previous measurements of
$N$(CO). The UV data for 63 of these consist of archival
STIS
observations, from which we extracted spectra of $A-X$ bands of CO between
1229$-$1544 \AA.
The remaining three sight lines have archival GHRS data.
Results on $N$($^{12}$CO) were subsequently published for 23 sight lines:
12 in \citet{Burg07}, three in \citet{Sonn07} (with one sight
line in common with \citeauthor{Burg07}), and 12 in \citet{Shef07} (with two
in common with \citeauthor{Burg07} and one in common with both \citeauthor{Burg07} and
\citeauthor{Sonn07}). Thus, this paper presents new $N$($^{12}$CO)
results for 43 sight lines.
To the entire CO sample of 66 sight lines we added previously published $N$(CO)
values for 48 directions,
yielding a sample of 114 sight lines with UV data.
Figure 1 presents the view of CO absorption along two \textit{HST}/STIS
sight lines that differ by a factor of 700 in $N$(CO).

\subsection{\textit{FUSE} Data}

Our initial sample of 58 sight lines was obtained from archival \textit{FUSE}
observations of H$_2$ absorption at $\lambda <$ 1100 \AA.
Of these, 33 sight lines did not have published $N$(H$_2$) results.
In the meantime, $N$(H$_2$) results were published for five sight lines
in \citet{Burg07} and five more in \citet{Shef07} (with a single
sight line in common with \citeauthor{Burg07}).
This paper, therefore, presents first $N$(H$_2$) results for 24 sight lines.
As described in \citet{Fede05}, our $N$(H$_2$) values are obtained from
spectrum synthesis of the (2$-$0), (3$-$0), and (4$-$0) bands of
the Lyman $B-X$ transitions of H$_2$.
The total column density $N$(H$_2$) listed is based on the absorption from all
rotational levels with $J^\prime$$^\prime$ = 0 through 5.
Roughly 95\% of the total is found in the two $J^\prime$$^\prime$ = 0 and 1
ground states of para- and ortho-H$_2$, respectively.
Figure 2 presents a sample of two \textit{FUSE} sight lines with H$_2$ absorption
profiles that differ by a factor of 13 in $N$(H$_2$).

The spectral coverage of \textit{FUSE} also contains
absorption features from CO \citep{Shef03, Cren04}.
Thus for three sight lines (HD 208905, HD 209481, and HD 209975) with no
{\it HST} spectroscopy we determined
$N$(CO) from the $B-X$ (0$-$0), $C-X$ (0$-$0), and $E-X$  (0$-$0) bands of CO,
as well as confirmed the
CO content along the line of sight toward HD 200775 that was
previously based on $IUE$ data \citep{Knau01}.
The former three stars were included in the high-resolution
optical study of CH, CH$^+$, and CN by \citet{Pan04, Pan05}. 

\subsection{McDonald Data}

High-resolution optical observations of CH$^+$, CH, and CN were obtained with the 2dcoude
cross-dispersed echelle spectrometer \citep{Tull95} for the purpose of
deriving cloud structure templates for CO and H$_2$, without which one cannot
reliably derive the line optical depth for sight lines with very high column
densities.
In addition, the echelle spectra included absorption from \ion{Ca}{2} and CN,
the first provides a high signal-to-noise confirmation of the cloud structure,
while $N$(CN) is used to model the total gas density in the
absorbing cloud, based on the CH and C$_2$ chemical reaction network
described in $\S$ 5.1.
Additional absorption from CH$^+$ provides a check on the component structure for
directions with low molecular concentrations.

Sight lines toward 20 stars were observed at $R \sim$ 170,000 with the 2.7-m
Harlan J. Smith telescope at McDonald Observatory, Texas, during observing
runs in 2004 January, October, and December, and 2005 May and October.
Each echelle exposure included nine orders, which simultaneously recorded two
atomic transitions, \ion{Ca}{1} at 4226 \AA\ and the K line of \ion{Ca}{2} at
3933 \AA, as well as absorption lines from three molecules, CH at 4300 \AA,
CH$^+$ at 4232 \AA, and CN at 3784 \AA.
Two-dimensional reduction tasks in IRAF were used to correct these for bias,
scattered light, pixel-to-pixel variations, and finally to
calibrate the wavelength scale based on accompanying exposures
of a Th-Ar lamp.
The latter step yielded residuals smaller than 0.001 \AA, or $<$ 0.07
km s$^{-1}$.

\subsection{ESO Data}

For the 17 stars in our sample that are located too far south
to be observable from McDonald, data were
obtained at ESO\footnote[7]{Based on observations collected at the European Southern
Observatory, La Silla, Chile, programs nos. 075.C-0025(A) and 077.C-0116(A).} in Chile.
Five more sight lines were added to the ESO observing program to complement
CH results given in \citet{Ande02} with new CH$^+$ and CN acquisitions.
For the 22 sight lines we obtained exposures on CH for 16 sight lines,
on CH$^+$ for 19 sight lines, and on CN for five sight lines.
The observations were carried out at the 3.6-m telescope at
LaSilla in 2005 June and 2006 June using the Coud\'{e} Echelle
Spectrograph (CES) \citep{Enar82}.
The CES is fed with an optical fiber with an aperture of 2\arcsec\ on the sky
and provides $R$ of 220,000.
The data were bias subtracted, flat-fielded, and rebinned to a linear
wavelength scale using the MIDAS long package.
Figures 3, 4, and 5, provide comparisons of sight lines with weak and strong absorption from
CH$^+$, CH, and CN, respectively.
HD 23478 and HD 210121 were observed at McDonald, whereas the acquisition of data
for HD 99872 and HD 116852 occurred at ESO.

\subsection{Ismod.f Spectrum Synthesis}

We used the Y.S. code, Ismod.f, to model Voigt absorption profiles via spectrum
synthesis and automatic rms-minimizations of (data minus fit) residuals.
Besides presenting the data, Figs. 1--5 all include spectrum synthesis fits
that were performed with Ismod.f.
The basic absorption equations were adapted in 1990 from \citet{Blac88}.
Besides fitting radial velocity, excitation temperature ($T_{\rm ex}$), and total
$N$ for any absorption
feature, Ismod.f provided solutions for cloud structures along each sight
line, i.e., the number of cloud components, their relative shifts and
fractions, and their Doppler widths ($b$-values).
This information is critical for proper evaluation of large optical depth
effects and has to be derived \textit{ab initio} whenever not known from
previous investigations.
Table 3 presents all cloud components that we were able to identify via
molecular absorption, while Figs. 3 and 5 present spectra of CH$^+$ and CN
toward HD 23478, with each species clearly showing two cloud components.
Our criteria for detection were a 2-$\sigma$ limit for molecular column
density, as well as simultaneous detection in \ion{Ca}{2} \citep{Pan05}.
All radial velocities have been transformed from the heliocentric scale to the
LSR reference frame.
For CO transition strengths we used the $f$-values of \citet{Chan93},
which in a global sense have been verified to a level of a few percent
by \citet{Eide99}, while wavelengths were obtained from \citet{Mort94}.
Both $f$-values and wavelengths for H$_2$ were obtained from
\citet{Abgr93a, Abgr93b} via files available on the Dr. McCandliss
website\footnote[8]{www.pha.jhu.edu/$\sim$stephan/h2ools2.html}.
As for the species with optical transitions, the corresponding input values for
\ion{Ca}{1}, \ion{Ca}{2}, CH, CH$^+$, and CN were taken from Table 3.4 in
\citet{Pan02}.

A subset sample of $N$($^{12}$CO) and $N$($^{13}$CO) for 25 sight lines
was published by \citet{Shef07},
who showed that a minimal number of absorption bands is needed for
a robust modeling of $N$(CO).
Specifically, a few bands that are optically thin and a few that are optically
thick should be simultaneously synthesized to yield a good measure of $N$
and of those parameters that affect line saturation in the bands,
such as $b$ and $T_{\rm ex}$.
Based on that sample of CO results, we find that our uncertainties in $N$
range from a few to $\sim$20\%. Thus to be on the conservative side,
we shall assume that the 1-$\sigma$ errors are $\pm$20\%, plotting this value
in all of our relevant figures.   

All H$_2$ lines from the $J^\prime$$^\prime$ $\leq$ 5 levels were modeled
simultaneously with Ismod.f.
The major difference in modeling methodology between CO and H$_2$ is that for
the latter we do not attempt to fit any parameters that are associated with the
cloud component structure along the line of sight.
This is a direct result of the relatively low spectral resolution of
\textit{FUSE}
($R$ $\sim$ 20,000) as well as the restricted available range (compared to CO) in
$f$-values for the Lyman and Werner bands of H$_2$.
Since it had been discovered \citep{Fede82} and will be verified in $\S$ 3.3
that $N$(CH) and $N$(H$_2$) have a linear
relationship, our method is to apply any cloud structure already
known from high-resolution CH data to the modeling of H$_2$, while keeping such
structure parameters fixed during the fit.
As in our previous work \citep{Pan05,Fede05} we prefer known cloud structures
to effective $b$-values as the proper solution to H$_2$ line saturation.
This is in contrast to CO, where parameters such as relative strength and width
of components are allowed to vary during the fit.
Further details can be found in our earlier paper \citep{Shef07}.
In a similar fashion to CO, we shall show global $\pm$20\% 1-$\sigma$ error
bars in all our figures that present values for $N$(H$_2$). This uncertainty
is consistent with published results.

Whenever repeated multiple exposures are available, we combined them in
wavelength space for an improved signal-to-noise (S/N) ratio.
In addition, when a feature (band) appears in two adjacent orders, we combined
them after correcting for any small wavelength inconsistencies by subtracting
the wavelength shifts as measured from the absorption line positions.
All our reductions were in IRAF and STSDAS.
A single Gaussian was used to describe the instrumental profile of STIS, but
$R$ was allowed to be a free parameter in the CO fits.
This revealed that $R$ is a decreasing function of slit size for both the E140H
and the E140M gratings of STIS (see Fig. 4 of \citealt{Shef07}).
The range of fitted resolving powers per aperture agrees well with the range of
values given by \citet{Bowe97}, showing that Ismod.f has a good handle on $R$.
Table 4 provides logarithmic values for total line of sight $N$ of all five
diatomic molecules that were modeled with Ismod.f spectrum synthesis in this
study, as well as supplementary $N$ values from the literature.
Throughout the paper, all log $N$ values are expressed in units of cm$^{-2}$.

\section{Observed Relationships}

\subsection{Component Structures}

Our aim of deriving accurate column densities was the main reason for obtaining
high-resolution spectra, both for CO from $HST$ UV exposures and for CH, CH$^+$,
and CN from new optical data.
These spectra reach velocity resolutions between 1.5--2.0 km s$^{-1}$,
enough to resolve many sight lines into multiple cloud components.
The unveiling of such structures is important both for correctly treating
the optical depth along the sight lines (by deriving $b$ values for the line
widths) and for distinguishing characteristics of individual parcels of gas
that would otherwise be lost in integrated line of sight values \citep{Pan05}.

Our earlier analysis of individual cloud components toward Cep OB2
in \citet{Pan05} showed that fits of CO cloud structures consistently resulted in structures
that were similar to CN structures, even though the input for
the synthesis of CO was based on CH cloud structures that have more components than those found for CN.
However, the sight lines sampled toward Cep OB2 were molecule rich,
with a median value of $N$(CO) = 2.5 $\times$ 10$^{15}$ cm$^{-2}$, and
with CN detections along 73\% of the sight lines.
The new CO sample presented here is molecule poor, with a median value
of $N$(CO) = 1.0 $\times$ 10$^{14}$ cm$^{-2}$, i.e., a factor of 25
lower in CO abundance relative to the Cep OB2 sample.
Most of the poorest sight lines here (log $N$(CO) $\lesssim$ 14) are without detected
$N$(CN) but with CO cloud structures that are very similar to those of both
CH and CH$^+$, whereas
for sight lines with CN detections, CN is found in only about
half of the components detected in CO.

The lower molecular abundance along the new sight lines is also reflected in the
generally low number of cloud components for all observed molecular species, even though
these are mostly sight lines with large pathlengths that have a higher chance of
intersecting molecular clouds. Whereas 53\% of the
sample stars are farther than 1 kpc, and 23\% are farther than 3 kpc, the derived cloud
structures have low means of components per sight line: 1.9 $\pm$ 1.0,
1.8 $\pm$ 0.9, and 1.7 $\pm$ 0.8 for CH$^+$, CH, and CO, respectively.
This shows that CO is in excellent agreement with both CH and CH$^+$,
underscoring the prevalence of low-density gas along these sight lines.
The mean of CN components per sight line is smaller, 1.2 $\pm$ 0.4.
CN is detected along sight lines with the higher values of $N$(CO) and
$N$(H$_2$), and then, on average, inside a single component.
Among the 15 new CO sight lines with CN detections, 13 (87\%)
have $N$(CO) $\geq$ 14.0, which is the median value for this sample.
There are only three CO components that have CN with no detected CH$^+$, and
seven that are associated with CH$^+$ but not with CN.
We shall see in $\S$ 4.3 that this dichotomy between $N$(CN) and $N$(CH$^+$)
can be employed as a good qualitative indicator of $n_{\rm H}$. 

In the rest of this section we explore the correlations among logarithms
of observed column densities by deriving power-law parameters
from regression analyses of the form
log $N$(M$_{\rm Y}$) = log $A$ + $B$ $\times$ log $N$(M$_{\rm X}$) \citep{Fede90}, where
M$_{\rm X}$ and M$_{\rm Y}$ are two molecular species.
Unless otherwise indicated, our BCES least-squares fits \citep{Akri96}
are done on detections only, excluding the small number of upper limits.

\subsection{CO versus H$_2$}

Panel (a) of Fig. 6 shows 105 sight lines with CO and H$_2$ detections taken from our
sample and from the samples of \citet{Cren04}, \citet{Pan05}, and
\citet{Shef07}, as well as including results toward bright
stars from \citet{Fede03}. 
A single-slope global correlation
returns a slope of $B$ = 1.89 $\pm$ 0.15, having a correlation
coefficient $r$ = 0.834 and confidence level (CL) $>$ 99.99\%, see Table 5.
The first indication of a global correlation between $N$(CO) and
$N$(H$_2$) was provided by \citet{Fede80},
who found a $B$ of $\approx$2 for 19 $<$ log $N$(H$_2$) $<$ 21.
Our fit of the new sample also confirms the result of \citet{Lisz98},
who plotted $N$ values of CO and H$_2$ from an updated version of the \citet{Fede94}
compilation to derive $B$ = 2.0 $\pm$ 0.3.

From the start, the need for a variable slope description of CO versus H$_2$
was present.
\citet{Fede80} found that lower-$N$ sight lines have a shallower $B$ of $\approx$1.5,
implying a slope break between 20.0 $<$ log $N$(H$_2$) $<$ 20.6
and between 13.6 $<$ log $N$(CO) $<$ 14.8.
\citet{Rach02} were the first to analyze higher-$N$(H$_2$) sight lines from \textit{FUSE},
together with the data from \citet{Fede94}.
They showed qualitatively that the slope of H$_2$ versus CO becomes
shallower above log $N$(H$_2$) $\sim$ 20.5.
According to our inspection of their Fig. 3, the slope of the CO versus H$_2$ relationship
appears to get as high as $\approx$3.5. 
Another exploration of this relationship by \citet{Pan05} was
based on a sample of \textit{FUSE} sight lines toward the Cep OB2 and Cep OB3
associations.
Despite gas density differences, these two associations presented similar
CO/H$_2$ slopes that also indicated a steeper relationship for log $N$(H$_2$) $\ga$ 20,
i.e., $B$ = 3.2 $\pm$ 0.3 and 2.9 $\pm$ 0.6 toward Cep OB2 and OB3, respectively.

\citet{Burg07} examined 19 sight lines and plotted these together
with the results of \citet{Cren04} and \cite{Pan05}, confirming
that CO versus H$_2$ was described by a relationship with
$B$ $\approx$ 2. 
Likewise, \citet{Sonn07} agreed that the overall appearance of
CO versus H$_2$ is as steep as found by \citet{Fede80}, but that it
also appears to have a steeper increase of CO with H$_2$ for $N$(CO) $\ga$
10$^{15}$ cm$^{-2}$, in agreement with the findings of \citet{Rach02},
\citet{Pan05}, and of \citet{Burg07}.

Overall, the indications are that the log $N$(CO) versus log $N$(H$_2$)
relationship is not strictly linear (single-sloped) but that the slope itself
(i.e., the exponent $B$) is also a function of $N$(H$_2$).
Indeed, when we restrict the fit to the lower end of the distribution,
the returned slope is shallower, while the higher end reveals a steeper slope.
In Fig. 6(a) we also show the 10-point means of the CO-H$_2$ sample, revealing a clear
signature of two slopes, or a dual power-law correlation between CO and H$_2$.
Using the two slopes to solve for their intersection point, we find
the break between slopes occurs at log $N$(H$_2$) = 20.4 $\pm$ 0.2 and
log $N$(CO) = 14.1 $\pm$ 0.1.
Finally, employing the break location, we fit 
the two resulting sub-samples to find two highly significant ($>$4 $\sigma$)
correlations (Table 5).
Thus below the break in slope, log $N$(CO) $\propto$ (1.46 $\pm$ 0.23) $\times$ log $N$(H$_2$),
while above it, log $N$(CO) $\propto$ (3.07 $\pm$ 0.73) $\times$ log $N$(H$_2$).
These two slopes are in excellent agreement with previous estimates, confirming
all indications that a steeper slope was needed for sight lines with higher values of $N$. 

The global behavior of $N$(CO) versus $N$(H$_2$) may be understood better when
the UV data
is complemented with $N$ values for 293 dark clouds detected by millimeter-wave
CO emission, and taken from the compilation of \citet{Fede90}.
We note that $N$(H$_2$) values for dark clouds were not directly observed, but
were inferred \citep{Fede90} from the corresponding visual extinction ($A_V$),
which is due solely to H$_2$ in these clouds.
This procedure cannot be applied to diffuse clouds (where $A_V$ $<$ 5 mag) because the
(hydrogen) molecular fraction,
$f$(H$_2$) $\equiv$ 2$N$(H$_2$)/[$N$(\ion{H}{1}) + 2$N$(H$_2$)], is $<$1.
As seen in panel (b) of Fig. 6, beyond the highest end of the diffuse molecular
cloud distribution one encounters the dark clouds, which have higher values of
molecular $N$ and of total gas density.

The previous version of this connection between cloud classes was based on only
20 diffuse cloud data points \citep{Fede90}.
At the time of the \citet{Fede90} study, only five CO values were known above
log $N$ = 15, with none available above log $N$ = 16. 
At this time, however, we have 22 data points with log $N$(CO) $>$ 15.0, of which
five are above log $N$(CO)  = 16, and one (HD 200775) is higher than log $N$ = 17.
Thus during the intervening 18 years the gap between diffuse sight lines
and dark clouds has been filling up with observations.
[Simultaneously, recent radio observations are becoming more sensitive in their ability to
measure smaller CO column densities that approach the diffuse cloud regime, e.g.,
\citet{Gold08}.]
Not only are the two distributions seen to be stretching toward each other, but the
steeper slope of the CO versus H$_2$ distribution, when extended to higher $N$,
is seen to pass near the center of the dark cloud distribution.
However, the rise of $N$(CO) versus $N$(H$_2$) cannot increase without limit
because the supply of atomic carbon for CO formation will be exhausted.
Thus the highest possible CO/H$_2$ ratio is set by C/H$_2$ = 2 $\times$ C/H
or (2.8 $\pm$ 0.4) $\times$ 10$^{-4}$, based on the gas-phase C/H abundance from
\citet{Card96} and shown in Fig. 6(b) as a double dashed line enclosing the
$\pm$1 $\sigma$ range.
Only a single datum out of 398 is seen to be slightly
above the C/H$_2$ limit.

One may need to go beyond fits with linear logarithmic slopes in order to
allow for a better description of the CO-H$_2$ relationship.
Our test of a fourth-order polynomial fit returned a continuously variable
slope that ranged from $B$ = 0.8 to 3.4 for the lowest to highest $N$(CO) values in the
sample of diffuse clouds,
at which point the slope declined while ``connecting'' with the dark
cloud distribution.
Such higher-order fits can only approximate the more realistic slopes predicted by detailed CO
photochemistry models.
Both \citet{Rach02} and \citet{Sonn07} noted the agreement between the observed
trend of CO versus H$_2$ and results from the CO photochemical modeling of \citet{vanD88}
for the CO-rich sight lines with log $N$(CO) $\ga$ 15.
In Fig. 7 we show that the theoretical models of ``translucent clouds'' from \citet{vanD88}
provide a functional variation that closely mimics the observed distribution
of both diffuse and dark clouds, as well as along the transition region (i.e., their
relative locations on the plot).
These curves have steeper slopes than those obtained in our fourth-order polynomial fit,
reaching as high as $B$ = 4.3, 5.4, or 7.3 for the $I_{\rm UV}$ = 0.5, 1.0, or 10 models,
respectively,
where $I_{\rm UV}$ denotes the enhancement factor over the mean interstellar UV radiation field.
All three curves end up at the highest column densities with $B$ between 1.5--1.9, i.e.,
bracketing the dark-cloud slope of $B$ = 1.62 $\pm$ 0.07.
Our modeling with Cloudy of the CO versus H$_2$ relationship that results
from CH$^+$ chemistry will be described in $\S$ 6.

\subsection{CH versus H$_2$ (and CO versus CH)}

Based on 19 data points, \citet{Fede82} demonstrated that $N$(CH) is
proportional to $N$(H$_2$), finding a slope of 1.0 $\pm$ 0.1, while \citet{Dank84}
confirmed this result by finding a slope of 0.85 $\pm$ 0.15 based
on a slightly larger sample with lower-S/N data.
(A combination of the two samples resulted in a slope of 0.90 $\pm$ 0.10.)
As for CO, \citet{Rach02} presented this relationship qualitatively, confirming
its nearly linear appearance and its agreement with the models of \citet{vanD89}
for the highest values of $N$(CH) and $N$(H$_2$).
A slope of 0.95 $\pm$ 0.10 was found by \citet{Pan05}
toward 11 stars in Cep OB2, but the only four data points that were available from
the Cep OB3 sample did not provide a clear case for a CH versus
H$_2$ correlation.
Our log-log plot of CH relative to H$_2$ (Fig. 8) shows a
well-correlated 90-point sample with a single slope of $B$ = 0.97 $\pm$ 0.07.
Thus CH is definitely linearly related to H$_2$, but as is the case with CO,
the width of the correlation is appreciably larger
than individual measurement uncertainties.
These correlations have CL above 99.9\%, and thus our methodology
used above in $\S$ 2.5, of importing CH cloud structures into spectrum syntheses
of H$_2$, is vindicated.

Consequently, the ratio CH/H$_2$ is a quantity that shows no correlation with H$_2$.
Our sample average is CH/H$_2$ = 3.5$^{+2.1}_{-1.4}$ $\times$ 10$^{-8}$, or 
log CH/H$_2$ = $-$7.46 $\pm$ 0.21.
This value can be seen to agree well with the data plotted in Fig. 2 of
\citet{Fede82}.
Thus $N$(H$_2$) can be predicted from optical observations of $N$(CH),
possibly with the exception of certain prominent photon-dominated,
or photo-dissociation, regions (PDRs).
The PDR targets HD 34078 and HD 37903 are found
on the outskirts of the distribution, deviating from the average by about
+3 and $-$2.5 $\sigma$, respectively.

It would be interesting to include CH and H$_2$ values for dark clouds to see
how they relate to the plotted distribution of diffuse sight lines.
\citet{Matt86} found that $N$(CH)/$N$(H$_2$) = 10$^{-7.4}$ for a small sample of dark
clouds, i.e., in excellent agreement with our average above for diffuse
molecular clouds.
\citet{Matt86} compared CH versus H$_2$ for both types of clouds and found
that they are in complete agreement, the dark cloud data being a monotonic
extension of the CH-H$_2$ diffuse cloud relatioship, with a global
slope of $B$ = 1.02 $\pm$ 0.04.
This is confirmed here with the inclusion of the \citet{Matt86} sample of dark clouds
into Fig. 8, yielding $B$ = 1.03 $\pm$ 0.03.
[This relationship breaks down, however, for dense molecular cloud cores
with $N$(H$_2$) $\ga$ 3 $\times$ 10$^{22}$ cm$^{-2}$ \citep{Matt86}.]
In summary, the linear relationship between CH and H$_2$ for diffuse and
dark molecular clouds is bound to be very
useful for determinations of $N$(H$_2$) along sight lines where no CO data
are available, or to corroborate such determinations based on CO data,
as we shall show in $\S$ 4.4.

The tight correspondence between CH and H$_2$ was employed by \cite{Magn95}
in order to derive $N$(H$_2$) from millimeter-wave detections of $N$(CH)
toward diffuse and dark molecular clouds.
Thus they were able to find variation by a factor of 20 in $X_{\rm CO}$
for diffuse clouds ($A_V <$ 4).
Another relationship between CH and $E(\bv)$ was presented by \citet{Magn03},
showing that while $N$(H$_2$) can be predicted from millimeter-wave
CH and reddening measurements, CO cannot be derived from linear
relationships with these parameters. 
Reddening values were already known to correlate with the total proton density
$N$(H) = $N$(\ion{H}{1}) + 2$N$(H$_2$) in diffuse clouds
\citep{Bohl78}, and with $N$(H$_2$) along translucent sight lines
\citep{Rach02}.

The linear relationship between CH and H$_2$ means that a plot of CO versus
CH should be similar to the plot of CO versus H$_2$.
Indeed, our single-slope fit of 92 data points shows that $B$ = 2.05 $\pm$ 0.21,
i.e., $N$(CO) varies as the square of $N$(CH).
Originally, \citet{Fede88} fitted a sample of 19 data points to find
$B$ = 1.97 for CO versus CH.
Employing a sample twice as large, this quadratic relationship was confirmed by
\citet{Fede94}, who
commented that above about $N$(CH) $\sim$ 3 $\times$ 10$^{13}$ and $N$(CO)
$\sim$ 10$^{15}$ cm$^{-2}$, $N$(CO) is increasing more rapidly (higher $B$).
In our sample, which is more than double in size yet again, one can see (Fig. 9)
that the slope appears to be increasing above log $N$(CH) $\sim$ 13.3 and log
$N$(CO) $\sim$ 15 toward the locus
of HD 200775 in very good agreement with \citet{Fede94}, who also noted
the outlying position of the PDR sight line toward AE Aur (HD 34078).

\citet{Sonn07} found a single slope of $B$ = 4.0 $\pm$ 0.3 for
a CO versus CH sample that tended to have higher column densities, i.e.,
mostly with log $N$(CO) $>$ 14 and log $N$(CH) $>$ 13.
In fact, as shown in Fig. 9, the 10-point means of our sample reveal the presence of
two slopes with CL $>$ 99.99\%, $B$ = 1.50 $\pm$ 0.30
and 2.80 $\pm$ 0.85, below and above a break at log $N$(CH, CO) = (13.0, 14.1), respectively.
There is excellent agreement with the power-law break found for CO versus H$_2$,
since according to the CH/H$_2$ ratio found above, log $N$(CH) = 13.0 corresponds to
log $N$(H$_2$) = 20.5.
As is the case for CO versus H$_2$, the steeper power law can be extended
to reach near the center of the dark cloud distribution, which was plotted after
converting its H$_2$ coordinate into a CH location. 
These characteristics of the CO versus CH plot confirm that CH can be used as a
dependable proxy for H$_2$, and that a distinct change in CO photochemistry
occurs at log $N$(CO) = 14.1 $\pm$ 0.1.

\subsection{CH$^+$ versus H$_2$ (and CH$^+$ versus CO)}

\citet{Fede82} also compiled and plotted 25 sight lines with detected
CH$^+$ and log $N$(H$_2$) $>$ 19, finding an insignificant correlation ($r$ = 0.3 and
CL $<$90\%) with an unspecified $B$.
\citet{Rach02} also found a linear relationship between the two species,
but with much increased scatter above log $N$(H$_2$) $\sim$ 20.
They commented that \citet{Gred97} found $N$(CH$^+$) and $N$(CH) to be
correlated, which in the light of the tight correlation between CH and H$_2$
presented in the previous section is consistent with a correlation
between CH$^+$ and H$_2$.
Indeed, our sample of 86 points returns a single-slope $B$ = 0.42 $\pm$ 0.10
with CL $>$ 99.99\%, confirming that a global correlation
exists between CH$^+$ and H$_2$.

However, in agreement with \citet{Rach02}, the current sample also shows
a marked increase in data scatter or a loss of correlation above
$N$(H$_2$) $\approx$ 2. $\times$ 10$^{20}$ cm$^{-2}$.
Panel (a) of Fig. 10 shows that employing 10-point means reveals yet another
broken-slope relationship, with $B$ = 0.78 $\pm$ 0.22 for log $N$(H$_2$) $<$ 20.3,
and $B$ = 0.15 $\pm$ 0.21 above that break. 
The steeper power law indicates (at CL = 99.95\% or 3.5 $\sigma$) that CH$^+$
varies nearly linearly with H$_2$ at lower column densities.
Above the break in slope, which is in excellent agreement with the H$_2$
break from Fig. 6, both $B$ and its CL indicate that CH$^+$ and H$_2$ are
no longer correlated. 

\citet{Lamb86} presented correlations between log $N$(CH$^+$) and
log $N$(H$_2$*), i.e., column densities of excited states of H$_2$
involving the $J$ = 3 and 5 levels.
Here we are unable to confirm these findings, because for our sight lines we do not find any
CH$^+$ and H$_2$* correlations for all $J$ = 1 to 4 levels.
However, we note that the \citet{Lamb86} study included sight lines with a range
of $N$ values much larger than ours, including sight lines that are H$_2$-poorer by
at least 3 orders of magnitude than those studied here.
We shall comment further about excited H$_2$ in $\S$ 6.2 when discussing
the formation of CH$^+$.

When CH$^+$ is plotted against CO, the behavior is similar to that found
just above for CH$^+$ versus H$_2$, namely, presenting a slope break that is flanked by
two different power-law fits (Fig. 10b).
Below log $N$(CO) = 14.1 we find $B$ = 0.46 $\pm$ 0.10 (CL $>$ 99.99\%)
while above the break $B$ = $-$0.14 $\pm$ 0.07 (CL = 93\%).
Again, showing excellent agreement with the CO break in Fig. 6, CH$^+$
is definitely correlated with CO for lower column densities, but is
insignificantly (1.8 $\sigma$) anti-correlated with CO above the break.
We believe that this behavior reflects the importance of CH$^+$ reactions when
$N$(CO) is low, so that a different production route operates at higher $N$(CO).
It is interesting to note further that with log $N$(CH$^+$) $\propto$ $B_1$
$\times$ log $N$(H$_2$)
and with log $N$(CH$^+$) $\propto$ $B_2$ $\times$ log $N$(CO), one expects
log $N$(CO) to be $\propto$ $B_1$/$B_2$ $\times$ log $N$(H$_2$).
Thus 0.78/0.46 = 1.7 $\pm$ 0.6 is in excellent agreement with $B$ = 1.46 $\pm$ 0.23
found in $\S$ 3.2 for the low-$N$ sight lines. 
Finally, both CH$^+$ breaks relative to H$_2$ (Fig. 10a) and to CO (Fig. 10b)
are in complete mutual agreement that the abundance of CH$^+$ presents a power-law change at
log $N$ = 13.2 $\pm$ 0.1.
A more detailed and chemically-motivated treatment of CH$^+$ and its relationships
with H$_2$ and CO will be
provided in $\S$ 6 based on numerical modeling with Cloudy and the
incorporation of a nonequilibrium term in the chemical formation of CH$^+$.

\subsection{CN versus H$_2$ (and CO versus CN)}

CN detections here encompass a smaller
sample that includes less than half of the sight lines that are included
in the H$_2$, CO, CH, and CH$^+$ samples.
\citet{Fede84} analyzed a smaller sample still, and concluded that
the CN abundance was proportional to the third power of H$_2$.
\citet{Rach02} showed that a strong correlation exists here as well,
with an estimated $B$ $\sim$ 2.5 according to our inspection. 
Such steep slopes are not confirmed here, because we find $B$ = 1.5 $\pm$ 0.4
from our regression fit, with $r$ = 0.67 (CL $>$ 99.99\%).
However, owing to the relatively large number of upper limits on CN, we
decided to employ the Buckley-James method of linear regression with censored data,
available from the ASURV statistical package \citep{Isob86}.
This fit returned a steeper slope of $B$ = 1.8 $\pm$ 0.4 (Fig. 11), which
was also confirmed by the EM algorithm and by Schmitt's method,
but is also appreciably shallower than $B$ $\approx$ 3.
One possible explanation for the disagreement among fitted slopes may involve the
smaller number of sight lines with detected CN in previous studies.
Our sample includes 40 sight lines with detected CN, and as a result there
is a significant ``re-''population of the plot with log $N$(H$_2$) $<$ 20.5.
However, the sample is still too small and restricted in range to reveal
information about any power-law break.
Thus arbitrarily breaking the sample at mid-point results in similar slopes,
both of which have CL below 90\% (Table 5).

Figure 12 shows the run of log $N$(CO)
versus log $N$(CN) with $B$ = 1.4 $\pm$ 0.2, fitting the CN detections only.
This 42-point sample has $r$ = 0.84 and thus CL $>$ 99.99\%.
Again, owing to the presence of 17 CN upper limits, we employed the Buckley-James
censored-data fit and derived $B$ = 1.9 $\pm$ 0.2, a slope that seems to
treat the upper limits as quasi-detections. On the other hand, using the
Schmitt method returned a fit with the significantly lower $B$ = 0.8 $\pm$ 0.2, which
seems to exclude all upper limits in the independent variable (i.e., in CN).
Thus our result based on detections only is bracketed
by the two slopes based on censored-data methods.
A smaller sample of (uncensored) 31 points was presented by \citet{Sonn07},
yielding (in an unweighted fit) $B$ = 1.5 $\pm$ 0.1, and $r$ = 0.90
(CL $>$ 99.9\%), thus agreeing more with our detections-only fit than with our Buckley-James fit.
Again, splitting the sample into two equal sub-samples returns two $\la$2 $\sigma$
fits with identical slopes, i.e., with no evidence for a break.
Since the CN sample is derived for the most part from high-$N$ sight lines
that are above the CO versus H$_2$ break, consistency is still preserved.

\section{Derived Physical Conditions}

\subsection{UV Shielding of CO}

Earlier ($\S$ 3.2) we described the finding of a power-law break in the correlation
analysis of CO versus H$_2$ at log $N$ = 14.1 versus 20.4.
This break in slope was also confirmed above through analysis of CO versus CH,
CH$^+$ versus H$_2$, and CH$^+$ versus CO.
This power-law break corresponds to a change in CO photochemistry.
Our value is similar to log $N$(CO) $\ga$ 14 found by \citet{Frer82}
for two CO isotopologues, C$^{18}$O and $^{13}$CO, through their correlations
with $A_V$ in molecular clouds.
In addition,
a value of log $N$(CO) $\ga$ 14 was supported by the self-shielding
computations of \citet{Ball82}, who showed the importance of line photodissociation
in steepening the increase of $N$(CO) with depth into a molecular cloud.

\citet{vanD88} presented the most detailed modeling of CO photochemistry in the
regime of translucent (1 $<$ $A_V$ $<$ 5 mag) sight lines.
In $\S$ 3.2 (and Fig. 7) we also compared the observed distribution
of diffuse and dark clouds with three families of \citet{vanD88} models
for translucent sight lines differing in $I_{\rm UV}$.
This showed the good global agreement of abundance trends for CO versus H$_2$.
In their modeling, \citet{vanD88} incorporated a detailed description of numerous CO absorption
bands in the far-UV, owing to their importance in diminishing CO
photodissociation rates through self shielding.
This detailed band structure was also needed for precise accounting of
the shielding of CO bands by H$_2$ absorption lines, since
the total UV shielding of CO is controlled by both $N$(CO) and $N$(H$_2$).
This two-parameter shielding function of CO, $\Theta$, was tabulated in
Table 5 of \citet{vanD88}.
The values of $\Theta$ are smaller than 1 since they provide the reduction in the
photodissociation rate of CO.
When $\Theta$ $\la$ 0.1, or log $N$(CO) $\ga$ 15, total UV shielding of CO
results in rapid steepening of the CO versus H$_2$ relationship. 

In order to determine the effect that $\Theta$ has on the observed distribution
of CO versus H$_2$, we present again in Fig. 13 the global view of the diffuse and dark cloud
distribution, together with our dual power-law fits, all overlaid by
contours of theoretical $\Theta$ values based on an interpolation of Table 5 of \citet{vanD88}.
It can be seen that for decreasing values of $\Theta$ the values of $N$(CO) are
increasing, as the photodestruction of the molecule is being diminished.
This provides a demonstration of the applicability of the
\citet{vanD88} shielding function to diffuse and dark sight lines.
Quantitatively, the location of the slope break at log $N$(H$_2$, CO) = (20.4, 14.1)
is seen to be near $\Theta$ $\approx$ 0.4, i.e., where the reduction in
the photodissociation of CO equals 1 mag.
Here, we interpret the slope break as the locus of a transition between
lower-density and higher-density regimes, but the comparison with $\Theta$
suggests some contribution from the UV shielding of CO to the steepening
of the slope near the break.
Future calculations of $\Theta$, which should include
updated $f$-values of predissociating CO bands \citep{Fede01,Shef03,Eide04,Eide06},
may clarify the association between $\Theta$ and the observed break in the
slope of CO versus H$_2$.

\subsection{Excitation Temperatures}

Each molecule in the ISM is influenced by both collisions with other molecules
and atoms (matter) and by interactions with photons (radiation).
The former processes will tend to thermalize the internal level populations of
the molecules so that their $T_{\rm ex}$ will reflect
the kinetic energy of the colliding gas particles.
Such a case is reflected in the $J$ = 0 and 1 populations of H$_2$ owing to
the lack of permitted dipole transitions that lead to cooling of the molecule.
As can be seen in Table 6, H$_2$ along all sight lines has relatively high values of
$T_{\rm ex}$($J$ = 0,1) (hereafter $T_{01}$) since they reflect the prevailing kinetic
temperature of the gas.
The average of 56 sight lines with newly-derived H$_2$ parameters is
$T_{01}$(H$_2$) = 76 $\pm$ 14 K, which is in excellent agreement with the
\citet{Sava77} result of 77 $\pm$ 17 K.
As we pointed out in \citet{Shef07}, sight lines with lower $T_{01}$
values for H$_2$ are associated with detected amounts of $^{13}$CO.
The sub-sample of 25 sight lines with $^{13}$CO was shown to have an average $T_{01}$(H$_2$) that
is 20 K below the average $T_{01}$ for sight lines without detected $^{13}$CO.
The entire sample here is not constrained by the presence of $^{13}$CO,
just like the original sample of \citet{Sava77}.

The YS Ismod.f code also returned fitted $T_{\rm ex}$ values for the higher-$J$
levels of H$_2$, as listed in Table 6.
A cursory inspection shows that for H$_2$, as detected in diffuse
molecular gas, $T_{01}$ $<$ $T_{02}$ $<$ $T_{03}$ $<$ $T_{04}$.
This is confirmed by their means from all sight lines:
76 $\pm$ 15, 101 $\pm$ 15, 140 $\pm$ 23, and 213 $\pm$ 31 K, respectively.
As can be seen in Fig. 14, logarithmic correlations exist between each $T_{0J}$
(for $J$ $>$ 1) and $T_{01}$, all with CL $>$ 99.99\%.
The respective slopes of the regressions are mutually identical within their uncertainties:
$B$ = 0.48 $\pm$ 0.07, 0.52 $\pm$ 0.09, and 0.46 $\pm$ 0.08, in order of increasing $J$.
The positive slopes may be indicating a connection between increasing
$T_{0J}$ and decreasing gas density at cloud edges, where there
is more efficient pumping by FUV photons.

CO presents the opposite case of sub-thermal excitation,
with $T_{\rm ex}$ values rarely rising above $\approx$5 K.
The only interesting case of warmer CO is found along the
$\rho$ Oph D (HD 147888) sight line: $T_{01}$(CO) = 13.6 K.
In this case the CO being probed is near the $\rho$ Oph molecular cloud,
and CO emission from the latter is able to raise $T_{\rm ex}$(CO) along
the diffuse part of the cloud \citep{Wann97}.
The average $T_{01}$(CO) from 61 (62) sight lines without (with) $\rho$ Oph D
is 3.5 $\pm$ 0.7 (3.6 $\pm$ 1.5) K.
Slightly higher values are found for $T_{02}$(CO) = 4.2 $\pm$ 0.8 (4.4 $\pm$ 1.4) K and
$T_{03}$(CO) = 5.3 $\pm$ 1.3 (6.0 $\pm$ 2.0) K, without (with) $\rho$ Oph D,
but more likely showing that $T_{ex}$(CO) is constant within the uncertainties.
We also see no dependence of $T_{01}$(CO) on the density indicator CN/CH$^+$.
Furthermore, the same mean value for $T_{01}$(CO) is found
for sight lines with or without detected CN, despite a density
difference of a factor of 10.
These differences from $T_{ex}$(H$_2$) arise because in diffuse molecular clouds
densities remain below the critical density of $\sim$2000 cm$^{-3}$ for CO.

\subsection{Total Gas Density: Empirical Indicators}

The total gas (proton) density, $n_{\rm H}$ $\equiv$ $n$(\ion{H}{1}) + 2$n$(H$_2$), controls the
chemical reaction networks via the density dependence of molecule-production terms,
and thus affects the resultant molecular abundances.
We may explore such density effects by deriving $n_{\rm H}$ for sight lines
in a variety of ways.
The simplest and most empirical involves the ratio of two observables,
$N$(CH) and $N$(CN).
\citet{Card91} showed that the CN/CH ratio is correlated with $n_{\rm H}^2$
since CN is formed inside denser and colder clumps of gas out of
pre-existing CH.
Thus plotting other quantities versus CN/CH is tantamount to showing the
relationship of those quantities with $n_{\rm H}$ \citep{Sonn07}.

As will be reinforced below, the CH$^+$ molecule is typically formed in lower
density regimes, leading to a dependence opposite to that of CN.
This was shown empirically by \citet{Card90} who found
that CN/CH was anti-correlated with CH$^+$/CH.
Thus, instead of using CH, which is less dependent on density thanks to its
connection to gas containing CN or CH$^+$ \citep{Lamb90, Pan05},
CN/CH$^+$ should be a more effective indicator of the density
than the CN/CH ratio.
In Fig. 15 we show that both empirical density indicators
are well correlated with each other ($r$ = 0.77, CL $>$ 99.99\%).
Note, however, that CN/CH extends over less than two orders of magnitude, whereas CN/CH$^+$
spans three orders, suggesting that CN/CH$^+$ responds better to changes in $n_{\rm H}$,
a picture consistent with the presence of CH in both high- and low-$n_{\rm H}$
gas.

Panel (a) of Fig. 16 shows the abundance of CO relative to H$_2$ versus
the empirical density indicator CN/CH.
With $r$ = 0.60 the plot shows a very good correlation (CL $>$ 99.99\%) between the two quantities,
having $B$ = 1.16 $\pm$ 0.25.
For comparison, Fig. 16(b) shows CO/H$_2$ versus CN/CH$^+$, and the
correlation is found to be even tighter, having a larger $r$ of 0.78 and $B$ = 0.85 $\pm$ 0.11.
Similar plots of species other than CO also exhibit larger $r$ values
and visibly tighter
relationships, confirming the better role of CN/CH$^+$ in sorting diffuse sight
lines according to $n_{\rm H}$.
Furthermore, the tighter correlation seen in Fig. 16(b) is an indication
that the CO/H$_2$ ratio is controlled significantly by the local gas density.

Given the better association between $n_{\rm H}$ and the observed CN/CH$^+$
ratio, one may imitate a 3-dimensional (3-D) plot by employing proportionately-sized symbols
to represent values of the latter quantity on
the 2-D surface of, e.g., the $N$(CO) versus $N$(H$_2$) plot that was shown
in Fig. 6.
From the resulting Fig. 17 one can
discern two general trends involving variations in gas density (as given by CN/CH$^+$).
First, density is clearly the lowest toward the lower left corner of the
plot (where many of the values are upper limits) and vice versa, showing that
both $N$(CO) and $N$(H$_2$) are correlated with $n_{\rm H}$.
Second, the density clearly varies in an orthogonal direction to its first
gradient, i.e., it is increasing from the lower envelope to the upper envelope
of the distribution.

\citet{Fede80} were the first to find that the dispersion in the relationship of CO versus H$_2$
is significantly larger
than the measurement uncertainties associated with individual data points,
as confirmed in Fig. 6.
These authors were also able to show that applying high-density and low-density
chemical models to this relationship indicated that its width was
affected by gas density, such that $n_{\rm H}$ is higher at the
upper envelope, in agreement with our findings
using CN/CH$^+$ as the density indicator. 

The two Cepheus samples in \citet{Pan05} appeared to occupy
nonoverlapping positions on the CO versus H$_2$ plot.
However, when compared with the current, much larger sample that includes
the sight lines from \citet{Pan05}, all Cep OB2 (having higher density gas) and OB3 (lower density)
data points are part of the global distribution of points, 
although they seem to belong to the upper and lower envelopes of the
distribution, respectively, thus providing more support to the overall picture.

Recently, \citet{Lisz07} suggested that the CO versus H$_2$ relationship
in diffuse clouds directly reflects the
formation of CO from a HCO$^+$ precursor.
However, while the second (cross-wise) variation agrees qualitatively with the models of \citet{Lisz07},
see his Fig. 1, those models do not reproduce the first variation,
i.e., the rise in density in tandem with increasing $N$ values.
In fact, the \citet{Lisz07} models have \textit{constant} density values between
the lower left and upper right corners of the CO versus H$_2$ plot.
Perhaps this difference is an indication that \citeauthor{Lisz07}'s assumption that CO
production is controlled by recombination of HCO$^+$ with a constant
abundance of 2 $\times$ 10$^{-9}$ relative to H$_2$ is inadequate. 

\subsection{Predicted Column Densities for H$_2$}

After fitting a correlation plot between two observed column densities ($N_{\rm o}$),
fit parameters may be used to predict one of the column densities ($N_{\rm p}$) in the absence
of the other.
The value of $N_{\rm p}$(H$_2$) = 7.4--10.0 $\times$ 10$^{20}$ cm$^{-2}$ toward HD 208266
was given in \citet{Pan05},
based on their fits of $N_{\rm o}$(CO) and $N_{\rm o}$(CH) versus $N_{\rm o}$(H$_2$) for the
small sample of sight lines toward Cep OB2.
Here we use the dual-slope relationship between CO and H$_2$ in Fig. 6(a), as well
as the single-slope relationship between CH and H$_2$ in Fig. 8, to predict
(the unobserved) $N$(H$_2$)
toward the two stars without H$_2$ data but with CO and CH data, HD 36841 and
HD 43818/11 Gem, which are near the CO versus H$_2$ break in slope, as well as toward HD 208266.
These predictions employ the global $\pm$20\% 1-$\sigma$ uncertainties in $N_{\rm o}$ values.

The CO-based log $N_{\rm p}$(H$_2$) values
for HD 36841 and HD 43818 are 20.46 $\pm$ 0.06 and 20.32 $\pm$ 0.06, respectively.
The same exercise for HD 208266 yields log $N_{\rm p}$(H$_2$) = 21.12 $\pm$ 0.03,
which is 2 $\sigma$ away from the predicted range (20.87--21.00) given in \citet{Pan05}.
The CH-based log $N_{\rm p}$(H$_2$) values for HD 36841 and HD 43818
are 20.41 $\pm$ 0.08 and 20.45 $\pm$ 0.08, respectively, which are 0.05 lower and
0.13 higher than the CO-based log $N_{\rm p}$(H$_2$).
Conceivably, when the difference is larger than 20\% (or $>$0.08 in the log) it is reflecting the
additional uncertainties introduced by the intrinsic widths of the correlations.
The same exercise for HD 208266 yields log $N_{\rm p}$(H$_2$) = 20.95 $\pm$ 0.08,
which is 0.17 lower than the CO-based prediction but in excellent agreement
with the predicted mid-range value given in \citet{Pan05}.
Combining results from CO and CH, both HD 36841 and HD 43818 are predicted here to have
log $N_{\rm p}$(H$_2$) = 20.4 $\pm$ 0.1, while for HD 208266 the prediction is 21.0 $\pm$ 0.1.
The corresponding 3-$\sigma$ uncertainties are provided by
the full width ($\pm$0.3) of the horizontal spread of $N_{\rm o}$(H$_2$) in both Fig. 6 and
Fig. 8.  

\section{Analytical Chemistry}

We examined the results presented here from two chemical
perspectives.  First, two sets of analytical expressions from previous 
work are used in this section to extract $n_{\rm H}$ associated with the 
material containing CO, one set involves the equilibrium chemistry leading 
to CN and another set describes the (equilibrium) synthesis of CH from CH$^+$.  
In $\S$ 6 we provide a more general chemical analysis based on the use 
of the Cloudy code.

\subsection{CN Chemistry}

Analytical expressions for the chemistry connecting CH, C$_2$, 
and CN in diffuse interstellar clouds \citep{Fede94}, with 
updated rate coefficients \citep{Knau01, Pan01}, are 
used to extract estimates for gas density, $n_{\rm H}$(CN). To summarize, the production
of CN is primarily given by the reactions C$_2$(N,C)CN, CH(N,H)CN, and
the chain C$^+$(NH,H)CN$^+$(H$_2$,H)HCN$^+$(e,H)CN, which has a parallel path where
HCN$^+$ reacts with H$_2$ to produce H$_2$CN$^+$ and then CN via
electron recombination.
$Observed$ $N$(CH) and $N$(C$_2$) (when available) are adopted for the 
comparison between observed and predicted CN column densities.  
A steady-state rate equation involving terms for chemical production and photo-destruction
of CN is employed in the determination of $n_{\rm H}$ \citep{Fede94}.

As in our recent papers (e.g., \citealt{Gred02, Pan05, Welt06}), 
results are presented for individual velocity components, whenever 
possible.  Like \citet{Gred02}, we determine upper limits on 
$n_{\rm H}$(CN) for components without detectable amounts of CN 
absorption.  This is especially important for the results 
presented here because many new detections of CO are found in 
directions that are not very rich in molecules.
We do not repeat the chemical 
analysis for sight lines in Ophiuchus and in Cep OB2 and OB3 
described in \citet{Pan05}, nor the analyses of the 
photodissociation regions illuminated by HD~37903 and HD~200775 
found in \citet{Knau01}.  Similarly, many of the directions 
contained in a reanalysis of spectra acquired with the \textit{Copernicus} 
satellite \citep{Cren04} are discussed in \citet{Zsar03}.
Updates are given for some of the sight lines 
examined by \citet{Fede94} and \citet{Wann99} in 
order to provide a self-consistent analysis and comparisons are 
presented below.

A key ingredient in this analysis is the value for the amount 
of extinction at UV wavelengths caused by interstellar grains, 
$\tau_{\rm UV}$, for each sight line.  This was determined by 
examining various measures for grain properties: the ratio of 
total to selective extinction \citep{CC91, Lars00, Barb01, Patr01,
Whit01, Duca03, Patr03, Vale04, Fitz05, Lars05, Sofi05},
the shape of the UV extinction curve 
\citep{Mass83, Witt84, Fitz90, Papaj91, Welt92, Lars96, Patr99},
and a comparison of the ratio $E(15-V)$/$E(\bv)$
\citep{Krel83, Sava85, Papaj92}.
For most sight lines, typical grain properties apply and we adopted 
$\tau_{\rm UV}$ = 2 $\times$ 3.1$E(\bv)$, where 2 is a prefactor that
depends on characteristics of the extinction curve \citep{Fede94} and the amount of 
reddening came from the work cited above or from \citet{Seab84}, 
\citet{Carn86}, and \citet{Aiel88}.  When grain 
properties suggested enhanced UV extinction, we used prefactors 
of 3 (for the directions toward HD~12323, HD~15137, HD~36841, 
HD~163758, HD~185418, HD~198781, and HD~210121) or 2.5 (for HD~14434) 
instead of 2, depending on the severity of the difference from 
typical values.  Several directions (HD~96675, HD~99872, HD~102065, 
and HD~124314) indicated below typical UV extinction; here the prefactor 
was set to 1.7.  For HD~93840, an intermediate value
seems to be appropriate, and a prefactor of 1.85 was adopted.  
For sight lines without information on grain 
properties (HD~24190, HD~30122, HD~137595, HD~190918, HD~192035, 
and HD~192369), the typical relation was employed.  One further 
constraint was considered: components having separations less 
than 20 km s$^{-1}$ were assumed to arise from nearby complexes 
where shadowing would be present and each component would 
experience the full amount of extinction as a result.  The lone 
exception was the direction toward HD~13745, where the 
components are 26 km s$^{-1}$ apart.  The results for HD~30122 
given in Table 7 show the effects that uncertainties in 
$\tau_{\rm UV}$ have on $n_{\rm H}$(CN).  In general, 
photodissociation is the dominant destruction pathway for the 
clouds in our study, and therefore uncertainties in $\tau_{\rm UV}$
lead to inferred uncertainties of $\sim$30\% in $n_{\rm H}$(CN).

For many of the directions, $N$ data for
CH, CN, as well as CH$^+$ used in the next section, were obtained 
as part of the present study.  Much of the remaining data on $N$(CH)
and $N$(CN) come from the compilation of \citet{Fede94},
but there are a number of updates 
now available.  For $\xi$ Per, we included the results 
of \citet{Cran95} for CH and of \citet{Lamb95} for C$_2$. The results of 
\citet{Lamb95} for $\zeta$ Oph were also used here.  We 
adopted the results of \citet{Kacz00} for the C$_2$ column 
toward X Per.  The CH results of \citet{Ande02}
for HD~99872, HD~115455, and HD137595 are included, as are
the CH and CH$^+$ results of \citet{Gred97} for HD 114886.
For the sight line toward HD~154368 we incorporated $N$(CH) 
from Welty (2005, private communication) and $N$(CN) from
\citet{Roth93} and \citet{Roth95}. For the gas toward 
HD~185418 and HD~192639, we used the results from \citet{Sonn02, Sonn03},
supplemented by those of \citet{Thor03}.
Since \citet{Pan05} did not consider directions 
without detectable amounts of CN in their analyses, we do so 
here for HD~208440, HD~208905, HD~209339, 19 Cep, and 
HD~217035A.  For stars in Per OB2 (40 Per, HD~23478, 
and HD~24190) we used our unpublished results.  We also note that 
for gas toward $o$ Per, X Per, and 62 Tau, component structure 
is available for CN, but not C$_2$.  Since these species appear 
to coexist (e.g., \citealt{Fede94}), we scaled the C$_2$ 
results so that CN/C$_2$ was the same for each component.  
Finally, as indicated in the Table, line-of-sight results are 
given for directions where component information is missing for 
C$_2$ as well as CN.

The results of this analysis appear in Table 7.  For each cloud
in a specific direction, we list the observed values
$N_{\rm o}$(CH), $N_{\rm o}$(C$_2$), and $N_{\rm o}$(CN),
and the predicted values $N_{\rm p}$(C$_2$) and $N_{\rm p}$(CN) that best match the
observations, the kinetic temperature 
($T$), $I_{\rm UV}$, $\tau_{\rm UV}$, and $n_{\rm H}$(CN).
$N$ values
are given in units of 10$^{12}$ cm$^{-2}$.  Most calculations are
based on $I_{\rm UV}$ equaling 1 and on $T$ $=$ 65 K.  The latter
value is not critical because the results for $n_{\rm H}$ are not very
sensitive to $T$.  For especially molecule-rich clouds and for 
some clouds studied by us in the past, lower values for $T$ are 
adopted.  The $N_{\rm p}$ values are generally in very 
good agreement, and are always within a factor of two, 
of the $N_{\rm o}$ values.

\subsection{CH$^+$ Chemistry}

For many of the directions listed in Table 7, only upper limits 
on CN are available.  For nearly all of these, CO production via 
reactions involving CH$^+$ appears likely (see below).  We 
therefore considered estimating the gas density from the chemical 
scheme linking CH and CH$^+$ \citep{Welt06, Ritc06}
as follows: CH$^+$(H$_2$,H)CH$_2^+$(H$_2$,H)CH$_3^+$ and the dissociative
recombination CH$_3^+$(e,H$_2$)CH.
In particular, we used the analytical expression in \citeauthor{Ritc06},

\[ n_{\rm H} = \frac{N({\rm CH})}{N({\rm CH^+)}}
\frac{2 I_{\rm UV} \Gamma({\rm CH})}{0.67 k f({\rm H_2})}, \]

\noindent where $\Gamma$(CH) is the CH photodissociation rate 
[1.3 $\times$ 10$^{-9}$ exp($-\tau_{\rm UV}$) s$^{-1}$], $k$ is the 
rate coefficient for the reaction CH$^+$(H$_2$,H)CH$_2^+$
(1.2 $\times$ 10$^{-9}$ cm$^3$ s$^{-1}$), and $f$(H$_2$) 
is the molecular fraction.
In addition to the present study, $N$(H$_2$) values come 
from \cite{Sava77}, \citet{Rach02}, and \citet{Pan05}.
The column densities of atomic hydrogen, $N$(\ion{H}{1}), 
are from \citet{Sava77} for the bright stars, and from \citet{Rach02},
\citet{Andr03}, \citet{Cart04}, and 
\citet{Jens07} for sight lines studied with \textit{FUSE}.
Data on atomic hydrogen do not exist 
for the sight lines toward HD~114886 and HD~137595.  For 
these stars, we estimated $N$(\ion{H}{1}) from 
$E(\bv)$ using the relationship between reddening and total 
proton column density of \citet{Bohl78} and accounting for 
the amount of H$_2$.  Values of $N$ for the carbon-bearing 
molecules are taken from the sources given in the previous 
section for the most part or from those compiled by \citet{Cren04}
for the bright stars.  The results appear in Table 8.

\subsection{Comparison of Results}

Many of the sight lines listed in Table 7 were analyzed in our 
previous work.  For $o$ Per, X Per, and 62 Tau, we now 
incorporated results for individual velocity components by 
scaling the values for $N$(C$_2$) to those measured for $N$(CN).  
The updated chemistry does not significantly affect the conclusions 
of \citet{Fede94} for the three sight lines, nor for the 
gas toward $\zeta$ Per, $\xi$ Per, $\chi$ Oph, and $\zeta$ Oph.  
There is also reasonable correspondence between the present chemical 
results for 40 Per and those of \citet{Wann99}, which are 
based on C$_2$ (and C~{\small I}) excitation.  Finally, our inferred 
density for the main component toward HD~154368 is about a factor of 
2 larger than our previous estimate \citep{Fede88}, which 
was based on an earlier, higher measure for $N$(CH).  This refined 
value for $n_{\rm H}$(CN) is consistent with the results of \citet{vanD84}
from C$_2$ excitation (300--1000 cm$^{-3}$) and \citet{Blac91}
from CN excitation (750--2000 cm$^{-3}$).

Comparisons for several other sight lines with other work are 
also available.  The most well-studied of these directions is 
toward HD~210121.  Our chemical analysis indicates that $n_{\rm H}$ =
1425 cm$^{-3}$, one of the highest values in Table 7.  The large 
value agrees with results from analyses of molecular excitation.  
For C$_2$ and CN, \citet{Gred92} found densities of 500 to 
1000 cm$^{-3}$ and 1500 to 2500 cm$^{-3}$, respectively, while 
\citet{Roue02} obtained densities of about 2000 cm$^{-3}$ from 
the distribution of C$_3$ levels.  Our upper limits on density for 
the gas toward HD~185418 and HD~192639 ($\le$ 30 to 40 cm$^{-3}$) 
are consistent with densities inferred from C~{\small I} excitation 
\citep{Sonn02, Sonn03}.

For most directions with CN upper limits, results from CN chemistry 
are not very constraining (e.g., \citealt{Fede97b}).  
Instead, comparisons with C~{\small I} excitation seem to be more 
appropriate, as in the cases of HD~185418 and HD~192639.  For such 
directions, CH$^+$ chemistry usually dominates, and so the results 
from Table 8 would be more meaningful.  For the nearby bright stars, 
$\alpha$ Cam, $\nu$ Sco, and $\mu$ Nor, a comparison with the 
results from \citet{Jenk83} is possible.  For 
the three directions, densities are derived from the quoted pressures 
assuming that $T_{01}$ is the kinetic temperature.  Consistency 
between results occurs for $\alpha$ Cam and $\mu$ Nor, where upper 
limits from \ion{C}{1} excitation are 10 and 370 cm$^{-3}$, 
respectively, versus our values of about 1 for no enhancement in 
the strength of the UV radiation field.  However, the $lower$ limit 
toward $\nu$ Sco of 70 cm$^{-3}$ contrasts with our density of 
10 cm$^{-3}$. If the cloud were near the star, such that $I_{\rm UV}$ 
was greater than 1, the two measures could be brought into agreement.  
This points out a deficiency in the current analysis of CH$^+$-like 
CH \citep{Lamb90} for many of the sight lines in Table 8: Most stars in the Table 
lie at least a kpc away.  Thus, a significant amount of atomic 
hydrogen and a corresponding amount of UV extinction are not likely 
associated with the gas containing CH and CH$^+$.  The most important 
factor is the exponential change in extinction, thereby increasing 
the estimate for $n_{\rm H}$.  The typical increase is about a factor of 
10, but can be significantly more.  For the clouds toward HD~185418 
and HD~192639, densities approaching 10 are possible, comparable to 
the results from C~{\small I} \citep{Sonn02, Sonn03}.  
Overall, the analytical results presented in Tables 7 and 8 are 
consistent with the detailed models described next.  In particular, 
the gas rich in CN and CO tends to have larger densities.

\subsection{The Role of C$_2$ Data}

Solving for $n_{\rm H}$(CN) via the analytical chemistry scheme depends on
input of $N$(CH) and $N$(CN), but not necessarily on the availability of $N$(C$_2$).
When $N$(C$_2$) is not available, it is predicted by the
model via CH(C$^+$,H)C$_2^+$, which is followed by
C$_2^+$(H$_2$,H)C$_2$H$^+$ and C$_2$H$^+$(H$_2$,H)C$_2$H$_2^+$,
both leading to the C$_2$ molecule by dissociative recombination.
The predicted value of $N$(C$_2$) is then determined from a steady-state rate
equation involving terms for C$_2$ chemical production and photodestruction.
However, when $N$(C$_2$) is known, it introduces
a constraint on the chemical formation of C$_2$, which in turn is affecting
the production (and predicted abundance) of CN.
It would be interesting to investigate the differences, if any, between sight
lines with or without observed values of $N$(C$_2$).

Figure 18(a) shows a logarithmic plot of CN-derived gas densities, $n_{\rm H}$(CN),
versus observed $N$(CN) for sight lines with detected CN components, fitted with
$B$ = 0.31 $\pm$ 0.10 (solid line).
(Data were gathered from the chemical analyses of this paper, of \citealt{Knau01}
and of \citealt{Pan05}.)
Whereas sight lines without observed values of $N$(C$_2$) (empty circles) show
a loose correlation with $r$ = 0.32 between $n_{\rm H}$(CN) and $N$(CN), a much tighter
correlation with $r$ = 0.61 is seen in the distribution of the filled circles, which denote
sight lines with C$_2$ measurements.
Thus it is apparent that more robust derivations of $n_{\rm H}$(CN) require
not only observed $N$(CH) and $N$(CN) values, but also observed $N$(C$_2$) data.
Similarly, Fig. 18(b) shows the run of $n_{\rm H}$(CN) versus the ratio
CN/CH$^+$,
which is the better empirical proxy for gas density ($\S$ 4.3).
Again it is obvious that sight lines with observed $N$(C$_2$) show a tighter
correlation than those without C$_2$ data.

Finally, as a confirmation of the close affinity between resultant gas densities
and input C$_2$ observations, we plot in Fig. 19(a) these two quantities.
It shows that $n_{\rm H}$(CN) predictions correlate well with $N$(C$_2$)
for sight lines with C$_2$ detections ($r$ = 0.63, CL $>$ 99.5\%).
This relationship, with $B$ = 0.48 $\pm$ 0.15, is in line with the
expectation that CN and C$_2$ molecules
are formed in higher density (and colder) clumps of gas because of the
correlation found above between $n_{\rm H}$(CN) and $N$(CN).
Figure 19(b) indeed shows that the correspondence between the observables $N$(CN)
and $N$(C$_2$) has a slope of 0.97 $\pm$ 0.28 and $r$ = 0.64, or
CL $\geq$ 99.7\%.
This result agrees at the 2-$\sigma$ level with the slope of 1.6 $\pm$ 0.2
found by \citet{Fede94} from a larger 33-point sample that showed
a tighter correlation ($r$ = 0.85 and CL $>$ 99.99\%).
\citet{Gred04} showed that a small sample of sight lines toward Cep OB4 is probing
a high-$n_{\rm H}$ molecular cloud and providing tight correlations between all three
molecules that are involved in equilibrium chemistry: CH, C$_2$, and CN.
It is clear that CN chemistry is dependent on $N$(C$_2$) and that robust
$n_{\rm H}$(CN) values can be derived from these two observables.

\section{Numerical Models with Cloudy}

\subsection{Computational Details}

We performed a series of model calculations designed to cover a range of 
physical parameters characteristic of diffuse and molecular clouds.
Our calculations used version C07.02 of the spectral synthesis code Cloudy,
last described by \citet{Ferl98}.
\citet{vanH04}, \citet{Abel05}, \citet{Shaw05}, and
\citet{Roll07} discuss in detail the Cloudy treatment of various physical
processes important in modeling atomic and molecular phases of the ISM.
\citet{Roll07} compare the predictions made by various
PDR codes and find excellent agreement between
Cloudy and the codes used in \citet{Kauf99}, \citet{Boge05},
and the Meudon PDR code \citep{LePe06}.

The geometry of our model is a plane-parallel slab illuminated from both sides
by far-UV radiation.
This geometry is appropriate for diffuse environments bathed on all
sides by the far-UV radiation field, and is identical to that used in both
\citet{vanD88} and \citet{LePe06}.
\citet{LePe06} showed that $N_{\rm p}$ values for species
like CO and CH in a single versus double sided calculation vary by up to
a factor of two for an $A_V$ between 0.2--5 mag.

Our choice of explored ranges in physical parameters such as density, radiation
field intensity, cosmic ray ionization rate, and stopping criterion
(the physical thickness of our slab model) are
determined by the need to compare our results with observations and with
results from previous studies, and by typical diffuse cloud conditions.
Diffuse clouds generally have $n_{\rm H}$ ranging from 10 to 5000 cm$^{-3}$
\citep{Snow06},
and therefore we vary $n_{\rm H}$ from 10 to 1000 cm$^{-3}$, in
increments of 1 dex.
For simplicity, all calculations are performed at constant (depth-independent)
density.
We use the \citet{Drai78} radiation field in our calculations, which is also
used in \citet{vanD88} and \citet{LePe06}.
We vary the far-UV intensity from $I_{\rm UV}$ = 0.1 to 10 times the average value of the
interstellar radiation field, which equals $1.6 \times 10^{-3}$ erg
cm$^{-2}$ s$^{-1}$ \citep{Habi68}, also in increments of 1 dex.
We also include the effect of cosmic rays,
for which we use a cosmic ray ionization rate $\zeta = 3 \times 10^{-17}$ cm$^3$
s$^{-1}$.
Higher values of $\zeta$ were found by \citet{McCa03} and \citet{Shaw06}
studying H$_3^+$
toward $\zeta$ Per and HD185418, respectively; by \citet{Lisz03}, who inferred a higher $\zeta$
based on analysis of HD and H$_3^+$ along a sample of sight lines; and by
\citet{Fede96b} based on cosmic ray-induced chemistry of OH toward $o$ Per.
Nevertheless, since our goal is to model global trends, we decided to use a value of
$\zeta$ consistent with the average value of $\zeta =2.5 \times 10^{-17}$
cm$^3$ s$^{-1}$ determined by \citet{Will98}.
We stop all calculations once $N$(H$_2$) = 2 $\times 10^{21}$ cm$^{-2}$, a value
high enough to include all the diffuse cloud observational data in our sample.
This stopping criterion corresponds to $A_V$ of 1--5 mag.
We integrate molecular $N$ for all $N$(H$_2)$ values up to the
stopping criterion, thus including all the phases of ISM clouds as classified by
\citet{Snow06}.

The thermal and ionization balance are both computed self-consistently.
The temperature is computed from energy conservation consisting of a host of
microphysical processes \citep{Ferl98, Abel05, Roll07}.
All atomic photo-processes are calculated by integrating the product of the
incident radiation field intensity over the cross section for the
photo-interaction rate.
We also integrate the cross section for photodissociation of H$_2$, using a
detailed H$_2$ model incorporated into Cloudy \citep{Shaw05}.
For CO, we use the shielding function described in \citet{vanD88}
and \citet{Holl91}.
\citet{LePe06} show the predicted $N$(CO) is about a factor of two
smaller when using a shielding function versus an exact treatment of CO
photodissociation in diffuse clouds.
This is small, however, when compared to the increases by a factor of 100 that
nonequilibrium CH$^+$ formation can contribute to the formation of CO in diffuse environments
(\citealt{Zsar03} and this work). 

Our assumed gas and dust abundances are consistent with average ISM values. 
For the gas phase, we include the 30 lightest elements.
The abundance relative to hydrogen for each species is an average of the
abundance taken from \citet{Cowi86} and $\zeta$ Oph \citep{Sava96}. 
The only exceptions are C/H and O/H.
For C/H we use the value determined by \citet{Sava96} without
averaging, while for O/H we use the value determined by \citet{Meye98}.
Some of the more important abundances by number are: He/H = 0.098,
C/H = 1.3 $\times$ 10$^{-4}$, O/H = 3.2 $\times$ 10$^{-4}$,
N/H = 8 $\times$ 10$^{-5}$, Ne/H = 1.2 $\times$ 10$^{-4}$,
Si/H = 3.2 $\times$ 10$^{-5}$, S/H = 3.2 $\times$ 10$^{-6}$, and
Cl/H = 1 $\times$ 10$^{-7}$.

Our network includes all known important chemical channels leading 
to CO and CH$^+$ formation (Fig. 20).
The chemical reaction network consists of approximately 1200 reactions
involving 89 molecules made up of H, He, C, N, O, Si, S, and Cl.
A complete list of molecules and reactions, along with rates, can be found on
the Cloudy website\footnote[9]{www.nublado.org}.
Most of the rate coefficients come from the UMIST database
\citep{LeTe00, Wood07}, although there are a few exceptions.
For the important C$^+$(OH,H)CO$^+$ reaction, we use a temperature-dependent
rate based on the data of \citet{Dube92}, with an equation derived in
\citet{Abel05}.
H$_2$ is known to form primarily through catalysis on grain surfaces, and we
compute the rate of H$_2$ formation using the temperature- and
material-dependent rates given in \citet{Caza02}.

The most important aspect to our calculations is the modeling of
nonequilibrium chemistry in order to simulate the formation of CH$^+$ and its
trickle-down effects on CO.
To this end, we use the method given in \citet{Fede96a} that
incorporates a coupling between the ions and neutrals.
The physical model for the nonequilibrium chemistry involves Alfv\'{e}n waves
that, upon entering the cloud, dissipate over some physical scale, as described
in $\S$ 6.2.
We model this effect by reducing the coupling by one-third
for $N$(H$_2) \ge 4 \times 10^{20}$ cm$^{-2}$. 
This roughly corresponds to the transition from the Diffuse Atomic to Diffuse
Molecular phase \citep{Snow06}, and effectively ``turns off''
nonequilibrium effects for $N(H_2) \ge 4 \times 10^{20}$ cm$^{-2}$.
The coupling in terms of
$T_{\rm eff}$ depends most critically on $\Delta v_{\rm turb}$, which is
the turbulent velocity of the gas.
Therefore, we study the effects of $\Delta v_{\rm turb}$ on model predictions
by using three different values for it, 2.0, 3.3, and 4.0
km s$^{-1}$.
These values for $\Delta v_{\rm turb}$ are physically motivated from a
number of considerations.
A typical CH$^+$ linewidth is $\sim $2.5 km s$^{-1}$
\citep{Craw94, Cran95, Craw95, Pan04}.
The character of $ \Delta v_{\rm turb} $ in the nonequilibrium chemistry is
3-D; therefore if the CH$^+$ linewidth is completely described
by a 1-D turbulence driven by Alfv\'{e}n waves, $ \Delta v_{\rm turb} =
\sqrt{3} \times {\rm linewidth} \sim $ 4.3 km s$^{-1}$ \citep{Heil05}.
These authors found that a typical 1-D value for $ \Delta v_{\rm turb} 
\sim $ 1.2--1.3 km s$^{-1}$ in the CNM, corresponding to 2.1--2.3 km s$^{-1}$
for 3-D turbulence.
The average of the CH$^+$ and CNM turbulence is $\sim $3.3 km s$^{-1}$.
For all combinations of $n_{\rm H}$ and $I_{\rm UV}$ considered, we used this average
value for $ \Delta v_{\rm turb} $ (3.3 km s$^{-1})$, and only compute models
with $ \Delta v_{\rm turb} $ = 2.0 or 4.0 km s$^{-1}$ for $n_{\rm H}$ = 100
cm$^{-3}$ and $I_{\rm UV}$ = 1. 
More details about the nonequilibrium CH$^+$ chemistry follow.

\subsection{Forming and Modeling CH$^+$}

Despite being one of the earliest molecules detected in the ISM, the
formation of CH$^+$ in the diffuse ISM remains one of the biggest challenges in 
astrochemistry.
The fundamental issue is that equilibrium chemical models
under-predict $N$(CH$^+)$ by 3--4 orders of magnitude.
The main problem in the formation of CH$^+$ is the primary formation channel
leading to CH$^+$ in the reaction
\begin{equation}
{\rm C}^+({\rm H}_2,{\rm H}){\rm CH}^+.
\end{equation}
This reaction is highly endothermic, with a rate of $1 \times 10^{-10} \exp 
(-4640/T_{\rm kin})$ cm$^3$ s$^{-1}$ \citep{Fede96a}.
One way around the endothermicity of (1) is to have C$^+$ react with excited
H$_2$ (H$_2$*), which reduces or eliminates the exponential temperature dependence
in the rate constant.
As was mentioned in $\S$ 3.4, observations by \citet{Lamb86} provided correlations between
$N$(CH$^+$) and $N$(H$_2$*) for $J$ = 3 and 5, but our sample of sight lines does not
show any similar correlations, a fact that could be the result of our much narrower range
of examined $N$(H$_2$*) values.
In any case, models of CH$^+$ chemistry, which include the formation process via H$_2$*,
still do not reproduce the observed CH$^+$ in diffuse clouds \citep{Garr03}.
CH$^+$ can also form through the radiative association of C$^+$ and H.
The rate for this reaction is 1.7 $\times$ 10$^{-17}$ cm$^3$ s$^{-1}$,
which exceeds the rate for reaction (1) for temperatures lower than 300 K.
CH$^+$ is easily destroyed, however, through reactions with H and H$_2$.
Therefore, the only way to efficiently produce CH$^+$ to observed levels
is to increase the temperature above the value used in models of equilibrium chemistry.

It is generally agreed that nonequilibrium chemistry is the key to solving the
CH$^+$ abundance problem in diffuse clouds.
However, the exact physical mechanism producing CH$^+$ is still unclear.
Hydrodynamic or Magnetohydrodynamic shock models \citep{Elit78, Drai86}
generate large amounts of CH$^+$ by heating the gas to
where CH$^+$ efficiently forms by equation (1).
However, the lack of velocity differences between CH and CH$^+$ \citep{Gred93, Fede96a, Fede97b}
argues against shocks, while excitation
analysis of interstellar C$_2$ \citep{Gred99} suggests CH$^+$ production occurs
in regions where the gas temperature is 50--100 K.
Recently, \citet{Lesa07} modeled the effects of turbulent diffusion on
diffuse cloud chemistry, determining that this mechanism can increase the
CH$^+$ abundance by up to an order of magnitude, which is still $\sim$2 orders
of magnitude lower than observed.

It has been suggested that CH$^+$-formation is driven by non-Maxwellian velocity
distributions of H$_2$ and/or C$^+$ \citep{Gred93}.
One possible solution to the problem of forming large quantities of CH$^+$ in
cold ($T_{\rm kin} <$ 100 K) regions is discussed in \citet{Fede96a}.
In this work, the authors propose Alfv\'{e}n waves entering a diffuse cloud
from the intercloud medium are coupled to the cold gas through the Lorentz
force (for ions) and collisions with the ions (for neutral atoms/molecules).
The coupling results in significant nonthermal motion of the gas along the
physical extent over which the MHD waves do not dissipate, consisting of a
boundary layer on the cloud-intercloud surface.
As a result, an effective temperature can be defined that characterizes the
reaction between two species undergoing nonthermal motions \citep{Flow85, Fede96a}
\begin{equation}
T_{\rm eff} = T_{\rm kin} + \frac{\mu}{3k} (\Delta v_{\rm turb})^2.
\end{equation}
In this equation, $k$ is the Boltzmann constant, $\mu$ is the reduced mass of
the system, and $\Delta v_{\rm turb}$ is assumed to equal the Alfv\'{e}n speed.
For turbulent velocities consistent with the observed linewidths of CH$^+$,
$T_{\rm eff}$ is large enough to significantly increase the reaction rate of
equation (1), increasing $N$(CH$^+)$ to values consistent with observation. 

While this physical process is not the only possible explanation for the 
observed CH$^+$ abundance, this method does have several important 
characteristics.
One is that it allows for the formation of CH$^+$ without 
heating the gas to temperatures inconsistent with the observed level of 
molecular excitation.
This mechanism also explains the lack of OH toward $\xi$ Per as a result of
ion-neutral decoupling, where the magnetic field was coupled to the ions but
not the neutrals.
Thus $T_{\rm eff}$ increased the reaction rate of ion-neutral reactions
such as equation (1), but not neutral-neutral reactions such as O(H$_2$,H)OH,
an important pathway to OH production at high temperatures.
Finally, Alfv\'{e}n wave propagation and dissipation in a cold diffuse cloud
is a relatively simple way to model nonequilibrium effects in a calculation
designed to model equilibrium chemistry.
All one needs to do is compute $T_{\rm eff}$ for each reaction using equation
(2) and replace $T$ with $T_{\rm eff}$ when calculating the rate coefficient. 

\subsection{Effects of CH$^+$ on Other Molecules }

The regions where CH$^+$ forms also contain significant quantities of other 
molecules.
This conclusion is independent of the actual physical processes controlling
CH$^+$ formation.
\citet{Fede97b} and \citet{Zsar03} estimated the
contribution to the formation of CH and CO due to equilibrium processes alone
(i.e. due to regions that do not form CH$^+$) using a simple chemical model of
a diffuse cloud.
These studies found that most CH and CO (over 90\% in many cases) could not be
explained through equilibrium processes.
The conclusion is that CH and CO in low-density ($n_{\rm H} \le $ 100 cm$^{-3}$)
sight lines form in regions where
nonequilibrium processes dominate the chemistry.
So the same physical process that controls CH$^+$ formation is also likely to
contribute to the formation of these molecules.

Almost all explanations to account for CH$^+$ involve increasing the
temperature in order to activate the formation channel given by equation (1).
However, increasing the temperature also increases the rates of other reactions,
leading to increased formation of certain molecules.
One example, OH, has already been mentioned.
Forming CH$^+$ via equation (1) also leads to increased formation of CH through
the chain involving CH$_2^+$ and CH$_3^+$, as was given in $\S$ 5.2.
Forming CH$^+$ also leads to the formation of CO$^+$ via the reaction
CH$^+$(O,H)CO$^+$, which is then followed by these two CO-forming channels:
CO$^+$(H,H$^+$)CO and CO$^+$(H$_2$,H)HCO$^+$ together with HCO$^+$(e,H)CO.
Finally, CO and CH are coupled through the neutral-neutral reaction CH(O,H)CO.
This last reaction is an efficient formation route of CO and destruction route
for CH at the high effective temperatures required for CH$^+$ formation. 
At higher density ($n_{\rm H} \ge $ 100 cm$^{-3}$), C$^+$(OH,H)CO$^+$, followed by either
CO$^+$-to-CO channel above, becomes the primary route for CO formation.
Regardless of $n_{\rm H}$, photodissociation is the primary destruction process for
CO. 

Calculations made by \citet{vanD88}, \citet{Wari96}, and
\citet{LePe06} did not consider nonequilibrium effects.
The models of \citet{vanD88} and \citet{LePe06} were used
by \citet{Sonn07} to show that, if $n_{\rm H}$ is sufficiently
high, the correlation between CO and H$_2$ observed through UV absorption
from \textit{Copernicus, IUE, FUSE}, and \textit{HST} can be reproduced (see also Fig. 7).
However, since neither model considered CH$^+$ formation nor nonequilibrium
chemistry (although the Meudon PDR group did in the past take CH$^+$ into
account through a shock model, see \citealt{LePe04}), the CO relationship
with H$_2$ is likely to be much different in a model that also reproduces
trends in CH$^+$ versus H$_2$ and CH$^+$ versus CO.
Such modeling and comparisons with observed trends in H$_2$, CO, and CH$^+$
abundances are the goals of our analysis.

\subsection{Comparing Model Results to Observation}

The results of our calculations are shown in Figs. 21, 22, and 23.
Each one shows plots of $N$(CH$^+)$ versus $N$(H$_2)$, $N$(CH$^+)$ versus $N$(CO),
and $N$(CO) versus $N$(H$_2)$.
Figure 21 shows the results for $I_{\rm UV}$ = 1, log $n_{\rm H}$ = 1, 2, and 3
($ \Delta v_{\rm turb} $ = 3.3 km s$^{-1})$; Fig. 22 the results for
log $n_{\rm H}$ = 2, $I_{\rm UV}$ = 0.1, 1, and 10 ($ \Delta v_{\rm turb} $ = 3.3 km
s$^{-1})$; and Fig. 23 the results for log $n_{\rm H}$ = 2, $I_{\rm UV}$ = 1,
and $ \Delta v_{\rm turb} $ = 2, 3.3, and 4 km s$^{-1}$.

\subsubsection{CH$^+$ versus H$_2$}

The general observed trend in this plot is that $N$(CH$^+)$ appears to saturate
around 2 $\times 10^{13}$ cm$^{-2}$, after which increasing $N$(H$_2)$ does
not result in increased $N$(CH$^+)$ (recall Fig. 10).
This trend is likely due to CH$^+$ formation by a mechanism that is acting over
only a portion of the cloud.
Our model mimics this effect by reducing $T_{\rm eff}$ for $N$(H$_2) \ge 4
\times 10^{20}$ cm$^{-2}$.
Once $N$(H$_2)$ becomes greater than this limit, the combination of smaller
$T_{\rm eff}$ (due to ``turning off'' the nonequilibrium chemistry) and
increased destruction of CH$^+$ through reactions with H$_2$ leads to a
decreased CH$^+$ density and hence a saturated $N$(CH$^+)$.
Figures. 21 and 22 show that for any value of $N$(H$_2)$ the value of
$N$(CH$^+)$ is inversely related to $n_{\rm H}$/$I_{\rm UV}$.
As $n_{\rm H}$/$I_{\rm UV}$ increases, the depth at which hydrogen becomes
predominately molecular [$f$(H$_2$) $>$ 0.8] decreases (see Fig. 15
of \citealt{LePe06}), while an increased H$_2$ abundance increases the
destruction rate of CH$^+$ through the formation of CH$_2^+$ and, eventually,
CH ($\S$ 6.3).
This effect is greatest for $n_{\rm H}$/$I_{\rm UV}$ = 1000 cm$^{-3}$ (corresponding
to log $n_{\rm H}$ = 3, $I_{\rm UV}$ = 1 in Fig. 21 or log $n_{\rm H}$ = 2,
$I_{\rm UV}$ = 0.1 in Fig. 22) because $f$(H$_2$) is then nearly 1,
whereas for smaller $n_{\rm H}$/$I_{\rm UV}$ values $f$(H$_2$) never exceeds 
0.8 \citep{LePe06}.
Our models predict (in Fig. 23) that $N$(CH$^+)$ increases as
$ \Delta v_{\rm turb} $ increases.
This trend is easy to understand as a temperature effect.
Increasing $ \Delta v_{\rm turb} $ increases $T_{\rm eff}$, which increases the
rate of reaction (1). 

We find that reasonable assumptions of model parameters can explain the
observed distribution in the $N$(CH$^+)$ versus $N$(H$_2)$ plot.
If we limit ourselves to only the case where $I_{\rm UV}$ = 1 (Fig. 21), then
over half of the data points lie in the region between the log $n_{\rm H}$ = 1
and 3 lines.
About 75\% of the rest of the data fall above the log $n_{\rm H}$ = 1 line,
i.e., where $n_{\rm H}$ $<$ 10 cm$^{-3}$.
When the effects of turbulence (Fig. 23) are considered, then essentially
all the observations are consistent with a suitable combination of
$n_{\rm H}$ and $\Delta v_{\rm turb}$.

\subsubsection{CH$^+$ versus CO}

A plot of log $N$(CH$^+$) versus log $N$(CO) shows an important observational trend
that can be understood through our calculations.
Initially, $N$(CH$^+)$ increases with $N$(CO), as was found in Fig. 10.
Once $N$(CO) reaches 10$^{14}$ cm$^{-2}$,
$N$(CH$^+)$ no longer increases,
but levels off at $N$(CH$^+) \sim 2 \times 10^{13}$cm$^{-2}$.
For even larger $N$(CO), $N$(CH$^+)$ appears to decrease.
This trend can be understood as reflecting variations in $n_{\rm H}$ (or
$n_{\rm H}$/$I_{\rm UV}$, if $I_{\rm UV}$ differs significantly from unity).
As $n_{\rm H}$ increases, the amount of CH$^+$ decreases, while the amount of
CO increases (see Fig. 21).
Therefore, the observations are well characterized by variations in $n_{\rm H}$,
since for log $N$(CO) $\la$ 14, where CH$^+$ and CO are coupled through the CH$^+$ + O reaction,
$n_{\rm H}$ = 10--100 cm$^{-3}$, while regions of higher CO and lower
CH$^+$ have $n_{\rm H} >$ 100 cm$^{-3}$.
This conclusion does depend somewhat on $ \Delta v_{\rm turb} $, since for
$ \Delta v_{\rm turb} $ = 2 km s$^{-1}$, the amount of $N$(CH$^+)$ per $N$(CO)
falls off significantly.
However, for regions with values of $I_{\rm UV}$ consistent with the average
interstellar far-UV radiation field, the ratio of $N$(CH$^+)$/$N$(CO)
combined with our models is a good diagnostic of the density and hence the
importance of nonequilibrium effects. 

\subsubsection{CO versus H$_2$}

Our calculations for the variation in $N$(CO) versus $N$(H$_2)$ show several 
important results.
Comparing the results of the equilibrium models of \citet{LePe06} with
the log $n_{\rm H}$ = 2; $I_{\rm UV}$ = 1 calculations, we find that including CH$^+$
chemistry can increase $N$(CO) by a factor of 50--100 at low column densities.
Such a dramatic increase in CO cannot be attributed to a more
rigorous treatment of the CO dissociation rate or geometry effects, both of
which would enter at the factor-of-2 level \citep{LePe06}. 
Instead, this points to the chemistry of CH$^+$ as essential for understanding the
abundance of CO in diffuse clouds, especially for $N$(CO) $<$
10$^{14}$--10$^{15}$ cm$^{-2}$.
Equilibrium models need high $n_{\rm H}$ values between 100--1000 cm$^{-3}$ to 
match the observed $N$(CO) versus $N$(H$_2$) \citep{vanD88, Sonn07}.
Our calculations show that the value of $n_{\rm H}$ can be an order of
magnitude lower.
This supports the conclusion of \citet{Zsar03} that nonequilibrium
chemistry is important to the formation of CO in diffuse environments.
For larger densities (or large $n_{\rm H}$/$I_{\rm UV}$), CH$^+$ chemistry is less
important due to destruction of CH$^+$ through the formation of CH$_2^+$ ($\S$ 6.3).
For regions where $N$(H$_2$) exceeds the cutoff for nonequilibrium effects,
modeled $N$(CO) values remain constant as a result of the decline in
$T_{\rm eff}$ and thus in CO formation rates.
However, observed $N$(CO) can readily reach values higher than 10$^{15}$
cm$^{-2}$ for higher $N$(H$_2)$.
This is because, for large $n_{\rm H}$/$I_{\rm UV}$, the CO photodissociation rate is
less effective in destroying CO while the reaction C$^+$(OH,H)CO$^+$ still
forms CO efficiently.
The effect of $T_{\rm eff}$ on $N$(CO) is easily seen in Fig. 23, where 
increasing $ \Delta v_{\rm turb} $ by a factor of 2 leads to $N$(CO) values
higher by $\sim $4 dex.
Overall, about 35\% of the observations require densities ranging from 100 to
1000 cm$^{-3}$, with the rest requiring lower densities and hence the
effect of nonequilibrium CH$^+$ chemistry on CO production. 

From Fig. 21 we also see that modeled $n_{\rm H}$ values do increase both along
and across the (variable) slope of the CO versus H$_2$ distribution.
This confirms the results in $\S$ 4.3, where these trends were derived
qualitatively based on observed CN/CH$^+$ ratios.
The range of $n_{\rm H}$ between 10 and 1000 cm$^{-3}$ from Cloudy is in
good agreement with the analytic results from the CN chemistry in $\S$ 5.
For sight lines without detected CN, analytic
CH$^+$ chemistry yielded very low values for $n_{\rm H}$, of which some
30\% were between 3 and 30 cm$^{-3}$ and thus in agreement with the lowest
numerical values in Fig. 21.
It is evident from the Cloudy results that CH$^+$ resides in regions of lower
gas density, whereas higher values of $n_{\rm H}$ correlate well with increased
$N$(CO). This is in excellent agreement with our
pointing out in $\S$$\S$ 4 and 5 that higher-density gas is associated both with
CN and CO, as well as with our empirical finding that $n_{\rm H}$
is revealed by the observed ratio $N$(CN)/$N$(CH$^+$).

\section{Discussion}

\subsection{The Ratio of CO to H$_2$}

In this study we showed (Fig. 6) that the power-law relationship, $N(CO) \propto [N(H_2)]^B$, 
is observed to behave differently in two density regimes that control
the production route for CO.
For sight lines below the break at log $N$(H$_2$, CO) = (20.4, 14.1), 
the relationship has $B$ = 1.5 $\pm$ 0.2, while above the break for
higher-$N$, higher-$n_{\rm H}$ sight lines, it becomes
steeper with $B$ = 3.1 $\pm$ 0.7.
The higher value for $B$ is consistent with CO photochemical predictions
\citep{vanD88} of the transition region 
between the diffuse and dark cloud regimes (panel b of Fig. 6 and Fig. 7),
where UV shielding plays an important role ($\Theta$ $\approx$ 0.1) for $N$(CO) $\approx$ 10$^{15}$
and $N$(H$_2$) $\approx$ 10$^{21}$ cm$^{-2}$.
Throughout the plot the vertical dispersion in log $N$(CO)
has a full width of $\pm$1.0, a range that is much
larger than what is expected from observational uncertainties alone.
This intrinsic dispersion is influenced by the value of $I_{\rm UV}$/$n_{\rm H}$,
as can be seen in Figs. 7, 21, and 22.
For $I_{\rm UV}$ that is not far from 1, the width of the dispersion is reflecting
the variability of $n_{\rm H}$, since we showed
in $\S$ 4.3 (and Fig. 17) the changes in CN/CH$^+$ both along and across the
relationship of CO versus H$_2$ (for a constant $I_{\rm UV}$ = 1).
A quantitative numerical confirmation from Cloudy was provided in Fig. 21,
where for log $N$(H$_2$)
$\lesssim$ 20.5 the dispersion in CO is seen to correspond to a range in
$n_{\rm H}$ between 10 and 100 cm$^{-3}$.

The CO versus H$_2$ relationship can be recast into
$$ \frac{N(CO)}{N(H_2)} \propto [N(H_2)]^{(B-1)}, $$
which means that in the regime of diffuse clouds, the abundance of CO relative
to H$_2$ is {\it not} a constant factor but varies between dependence on the square root of
$N$(H$_2$) ($B - 1 \approx$ 0.5) for the lower values of $N$, and dependence
on the second power ($B - 1 \approx$ 2.1) of $N$(H$_2$)
for clouds with log $N$ $\ge$ 20.4.
This steeper dependence, though, gets shallower again once the transition
into the dark cloud regime has occurred, just before CO
uses up all the C atoms in the gas.
In fact, for the assumed full conversion of C atoms into CO, the constant
CO/H$_2$ ratio means that the relative abundance of the two molecules is
independent of $N$(H$_2$) for the highest $N$ values. 
Such global variations have a bearing on $X_{\rm CO}$ whenever measurements
include diffuse and translucent sight lines, since low values of CO/H$_2$
for low $N$ values directly translate into higher values for $X_{\rm CO}$.
There is no doubt that this $X$-factor is dependent on physical conditions that
affect the abundance of CO in diffuse molecular clouds.
However, since our sample involves essentially local clouds, it is not
relevant to the issue of variations in $X_{\rm CO}$ over Galactic scales,
where metallicity can play a role \citep{Stro04}.

Our range of log CO/H$_2$ ratios shows values between $-$7.58
and $-$4.68, with the single exception of HD 200775
having a value of $-$3.88.
The latter value is 47\% of the value obtained from a full conversion of
all carbon atoms into CO molecules, log (2 $\times$ C/H) = $-$3.55.
As remarked earlier, full conversion is expected inside dark clouds, a regime
associated with the PDR illuminated by HD 200775.
\citet{Fede80} presented log CO/H$_2$ values between $-$7.37
and $-$5.30, i.e., overall lower values that were the result of their
sample of sight lines with lower $N$(CO).
For the small samples toward Cep OB2 and OB3 \citet{Pan05} obtained
values from $-$6.31 to $-$4.85 and from $-$6.42 to $-$5.95, respectively,
with the former range clearly including higher-$N$(CO) sight lines.
\citet{Burg07} and \citet{Sonn07} presented restricted
samples that ranged from $-$7.00 to $-$4.74
and from $-$6.56 to $-$4.56, respectively.
The results from the smaller samples show good agreement with ours, albeit
their ranges are narrower, as expected.

\subsection{Connections to Molecular Clouds}

We showed the correspondence between significant molecular absorption 
and the presence of molecular clouds seen in emission for a number 
of directions in the past.  \citet{Gred92} mapped the high 
latitude cloud responsible for the absorption seen toward HD~210121, while 
\citet{Gred94} mapped the CO emission around HD~154368. 
\citet{Fede94} indicated the sight lines probing molecular 
clouds associated with stars in Taurus, Ophiuchus, and Cep OB3, while 
\citet{Wann99} did the same for the dark cloud B5 and stars 
in Per OB2.  Most recently, \citet{Pan05} examined the correspondence 
between CO cloudlets seen in emission and stars in Cep OB2.

We can do the same for additional sight lines from the current survey.  
The most extensive sets of measurements probe molecular clouds in 
Chamaeleon (HD~93237, HD~94454, HD~96675, HD~99872, and HD~102065), 
the Southern Coalsack (HD~106943, HD~108002, HD~108639, HD~110434, 
HD~114886, HD~115071, and HD~115455), and Lupus (HD~137595, HD~140037, 
HD~144965, HD~147683).  From the emission maps compiled by \citet{Ande02},
an interesting trend is discerned.  Only directions 
with significant $N$(CO) values (greater than 10$^{15}$ 
cm$^{-2}$) and CN ($\sim$ 10$^{12}$ cm$^{-2}$, when available) lie 
within the CO contours.  These are HD~96675, which probes the 
Cham I cloud, and HD~144965 and HD~147683, which pass through a 
cloud in Lupus.  A particularly interesting sight line for future 
study is toward HD~147683, where $N$(CO) is about 10$^{16}$ cm$^{-2}$.  
Unfortunately, no CN data exist at the present time.

With the aid of the SIMBAD site at the Centre de Donn\'{e}es 
astronomiques de Stasbourg, we found other likely associations 
based on the similarity in $v_{\rm LSR}$.  The direction toward 
HD~30122 appears to be probing the envelope of L1538 seen in 
CO emission \citep{Unge87}.  The gas toward 
HD~36841 in the Ori OB1 association may be related to that seen 
in emission from the reflection nebula IC~423 \citep{Madd86}
in the dark cloud L913 \citep{Clem88}.  For all 
other sight lines, no clear correspondence could be found.

\subsection{Further Chemical Considerations}

Having applied the analytical expressions to extract gas densities 
from chemical schemes involving CN and CH$^+$ to numerous 
sight lines, we now have a 
clearer understanding of their limitations.  While higher densities 
are found for CN-rich directions, as also found from our more 
comprehensive models, the correspondence between density and points 
on plots of $N$(CO) versus $N$(H$_2$), etc. is rather weak (Fig. 17).  The 
relationships involving $N$(CN)/$N$(CH) are stronger (see left panel of Fig. 16
and the discussion in \citealt{Pan05}).  The best correspondence is 
seen when $N$(CO)/$N$(H$_2$) is plotted against $N$(CN)/$N$(CH$^+$) in the right panel
of Fig. 16. This arises because CN only probes denser diffuse gas
(e.g., \citealt{Card91, Pan05}),
in which CH$^+$ is more likely to be destroyed by H$_2$.

The limitations involve a number of factors.  For a given velocity 
component, the amount of CH$^+$-like CH \citep{Lamb90} is not easily obtained; the 
dispersion in the relationship between $N$(CH) and $N$(CH$^+$) for 
directions without detectable amounts of CN is too large \citep{Pan05}.
Most sight lines, however, reveal absorption from all three molecules at 
a given velocity (e.g., Table 3).
 As for the CH$^+$ chemistry, the amount of material 
along a line of sight and the strength of the local interstellar 
radiation field are not well known, especially for stars greater than 
a kpc away.  More comprehensive models, combining the synthesis of 
CH$^+$ and CN, are needed for the next level of understanding.  Our 
goal is to apply models based on Cloudy to this problem.

In this work, we presented a series of Cloudy-based calculations of diffuse cloud conditions
that simultaneously reproduce the observed H$_2$, CH$^+$, and CO abundances in
these environments.
Diffuse sight lines with $N$(CO) $<$ 10$^{14}$--10$^{15}$
cm$^{-2}$ are well characterized by regions with $n_{\rm H}$/$I_{\rm UV} <$ 100
cm$^{-3}$, \textit{but only if the effects of CH$^+$ are taken into account}.
Without the effects of nonequilibrium CH$^+$ chemistry, equilibrium calculations predict too
little CO per H$_2$.
This result appears robust to uncertainties in the C/H abundance or the CO
photodissociation rate. Furthermore,
$N$(CH$^+$) increases with increasing $N$(CO), until $N$(CO) reaches
10$^{14}$--10$^{15}$ cm$^{-2}$.
For $N$(CO) $>$ 10$^{15}$ cm$^{-2}$, $N$(CH$^+$) appears to decrease for
increasing $N$(CO).
Our models show this is likely a density effect, with $N$(CO) increasing,
and $N$(CH$^+$) decreasing, with increasing $n_{\rm H}$. 
Last, the observed trend of $N$(CH$^+$) flattening out at a few times 10$^{13}$
cm$^{-2}$ can be explained if the nonequilibrium chemistry acts only over a
certain physical size, such as the Alfv\'{e}n wave propagation formalism in
\citet{Fede96a}.
The observed scatter in $N$(H$_2$) with $N$(CH$^+$) is best explained through
a combination of density effects and the importance of nonequilibrium
processes, parametrized in this work by $ \Delta v_{\rm turb} $.

\subsection{The Synoptic View}

The sight lines from our study have properties comparable to those inferred
from both (a) \ion{H}{1} self-absorption
(HISA) clouds with weak or no CO emission and (b) CO-poor H$_2$ gas revealed via
$\gamma$-rays and far-infrared (FIR) emission.
The former category has been investigated in recent
21-cm radio $absorption$ surveys of the Galactic plane \citep{Gibs05},
showing that cold atomic hydrogen gas is not necessarily associated
with detections of CO emission.
Since these clouds can be small ($<$ 0.6 pc) with $n$ $\geq$ 100 cm$^{-3}$ and
$T_{\rm spin}$ $<$ 50 K, the physical conditions in them are very similar
to the clouds studied here, or in other words, they correspond
to the intermediate
category of diffuse molecular clouds \citep{Snow06}.
One may, therefore, assume that despite CO nondetections via radio emission, CO is
likely present in these clouds, albeit with low CO/H$_2$ values
determined by small values of $n_{\rm H}/I_{\rm UV}$.
Such clouds with low $N$(CO) should, in principle, be detected via UV
absorption.
For example, \citet{Klaa05} provide $N$(CO) $<$ 6 $\times$ 10$^{15}$
cm$^{-2}$ for a small HISA feature, an upper limit that excludes only the
top 6\% of our diffuse sight lines.

In fact, an analysis of a Galactic plane survey by \citet{Kava05} has
determined that 60\% of HISA features are associated with CO emission,
with $n$ $\sim$ few $\times$ 100 cm$^{-3}$ and 6 K $<$ $T_{\rm spin}$ $<$ 41 K.
Although \citet{Kava05} suggest that these are ``missing link'' clouds
between the atomic and dense molecular varieties, we point out that this region
in parameter space is occupied by diffuse molecular clouds.
With H/H$_2$ $\leq$ 0.01 and CO/H$_2$ $\leq$ 10$^{-5}$ \citep{Klaa05}
these are probably clouds that include the types of carbon-bearing molecular
photochemistries that were explored here.
The inferred $T_{\rm spin}$ values are lower than the $T_{01}$(H$_2$)
kinetic temperatures along diffuse sight lines, resembling more $T_{02}$(C$_2$)
values that are associated with denser diffuse gas and the presence of
$^{13}$CO \citep{Shef07}.
It will be interesting to see if these colder clouds are related to sight lines
with very low $^{12}$CO/$^{13}$CO as observed by \citet{Lisz98}
using millimeter-wave $absorption$ observations.

A second category of CO-poor H$_2$ gas has been revealed in FIR studies
\citep{Reac94, Meye96, Doug07}.
The survey of \citet{Reac94} found infrared excess emission
from cirrus clouds attributed to cold H$_2$ gas and dust, whereas only half
of the clouds showed detectable levels of CO emission.
When detected, CO was found to be subthermally excited with inferred $n_{\rm H}$
$\sim$ 200 cm$^{-3}$ and $^{12}$CO/$^{13}$CO ratios between $\sim$10 and $>$90,
all indicating overlap with the parameter space of diffuse molecular clouds.
CO-poor H$_2$ gas has been inferred also from Galactic surveys of $\gamma$ rays
\citep{Gren05}, which trace the gas content in the ISM and show
an ``excess'' of $\gamma$ emission not associated with CO emission.
Indeed, \citet{Gren05} indicate that the CO-less gas is found around
dense molecular clouds (that are detected via CO emission) and along bridges
between cloud cores and atomic gas, precisely the sites where one would find
gas known as diffuse molecular clouds.
As is the case with the HISA clouds, we believe that
CO is still there, albeit at low levels of abundance relative to H$_2$
that are potentially observable via UV absorption but are not seen via current
methods that detect CO in emission.

\citet{FW82} showed that a connection exists between diffuse
molecular gas and dark clouds, namely, that there is good agreement in CH
abundance and radial velocity between radio emission
for dark clouds and optical absorption along nearby sight lines, the latter
probing the outer envelopes of dark clouds.
This CH connection has been exploited by Magnani and colleagues
\citep{Magn95, Magn03} in deriving CO/H$_2$ ratios for translucent
sight lines, based on the tight correlation between CH and H$_2$
(\citealt{Fede82}; our $\S$ 3.3).
\citet{Magn05} observed radio emission from CH
along the Galactic plane, finding similarities to CO emission
line profiles and inferring that the molecular gas has $n_{\rm H}$ $<$ 1000 cm$^{-3}$.
The material resembles the denser gas in our sample of diffuse molecular
clouds, consistent with the presence of CO emission (see $\S$ 7.2).
This connection is also related to OH emission that has been detected from
intermediate regions
around denser molecular CO-emitting clouds \citep{Wann93}
and the lack of a correlation between $N$($^{13}$CO) and $N$(OH)
in another sample of molecular clouds \citep{Gold05}.

\section{Conclusions}

Our study of diffuse molecular clouds employed a new and extensive sample of sight
lines with UV observations of CO and H$_2$ to explore in detail the power-law relationship
between the two species.
The slope of log $N$(CO) $\propto$ $B$ $\times$ log $N$(H$_2$) was shown to require
two components, one with $B$ = 1.5 $\pm$ 0.2 for log $N$(H$_2$) $\le$ 20.4,
and another $B$ = 3.1 $\pm$ 0.7 for VUV sight lines with higher $N$(H$_2$).
The break in slope arises from a change in CO production, with CH$^+$ + O important
at low $N$(CO) and C$^+$ + OH at $N$(CO) $>$ 10$^{14}$ cm$^{-2}$.
The ratio CO/H$_2$ has a dependence on $N$(H$_2$)
that results in an increase by $\sim$3.5 orders of magnitude over the range
of log $N$(H$_2$) $\approx$ 19.5--22.0.
Causes for variation in $X_{\rm CO}$ include (a) the $n_{\rm H}/I_{\rm UV}$
ratio, which affects production and destruction, including self shielding,
and (b) the metallicity of gas.

Together with the CO and H$_2$ we also analyzed new data for the carbon-bearing diatomic
molecules CH, CH$^+$, and CN (as well as C$_2$) that are accessible through ground-based
spectroscopy.
The linear relationship between $N$(CH) and $N$(H$_2$) was confirmed again, both directly
using these two molecules, as well as indirectly by showing that the CO versus CH
relationship follows that for CO versus H$_2$.
After determining fitted relationships of both CO and CH versus H$_2$ we were able to employ
fit parameters in the prediction of $N_{\rm p}$(H$_2$) for three sight lines without H$_2$ data.
Analyzing $N$(CH$^+$) versus H$_2$ and CO resulted in two more confirmations of the power-law break
displayed by CO versus either H$_2$ or CH, showing that this break separates the
regime of low-density photochemistry from that involving high density.
As for $N$(CN), all our regression fits returned slopes with $B$ $\leq$ 1.8,
somewhat shallower than earlier reports.
Since essentially all CN detections are along high-density sight lines, the absence
of a detected break in slope for the (smaller) CN sample is not surprising.

Many of the sight lines here are helping us to explore
molecular environments that are associated with low $n_{\rm H}$.
As such, these lines of sight probe regions where nonequilibrium CH$^+$ chemistry is dominating
the production of CO, as confirmed by modeling with Cloudy.
For those sight lines with higher $n_{\rm H}$ it was possible to include (equilibrium)
chemistry of CH, C$_2$, and CN to predict molecular abundances and gas density.
Such predictions were found to have tighter correlations when $N$(C$_2$) is part
of the input into the chemical model.
For the entire range of densities we showed that the empirical ratio $N$(CN)/$N$(CH$^+$)
is better suited than $N$(CN)/$N$(CH) as an indicator of the average
$n_{\rm H}$ along diffuse sight lines.

We also considered rotational (excitation) temperatures
in our modeling of CO and H$_2$, showing that $T_{0J}$(CO) does not vary for $J$ = 1--3.
On the other hand, $T_{0J}$(H$_2$) increases with $J$,
with indistinguishable slopes between log $T_{0J}$ and log $T_{01}$ for $J$ = 2--4.
Further analysis of the excitation of both molecules should help constrain the conditions
in diffuse molecular clouds.

As related in $\S$ 7.4, it is our understanding that the regime of low-$N$(CO), low-$n_{\rm H}$
diffuse molecular clouds is also sampled by a variety of non-UV observational methods,
which nonetheless result in a significant number of CO nondetections.
Thus the true nature of diffuse molecular clouds is best revealed by
synoptic knowledge extracted from
studies spanning the electromagnetic spectrum from radio and FIR,
through visible and UV, to gamma ray observations.

\acknowledgments

We thank NASA for grant NNG04GD31G and STScI for grant HST-AR-09921.01-A.
Data files for this paper were accessed through the Multiwavelength Archive at STScI.
Y.S. acknowledges ongoing demand for versions of Ismod.f by members of the S. Federman group
at Toledo, and thanks the Dept. of Physics and Astronomy at the University of Toledo
for having been granted the title of Research Assistant Professor.
M. R. acknowledges support by the National Science Foundation under Grant No.
0353899 for the Research Experience for Undergraduates in the Dept. of Physics
and Astronomy at the University of Toledo.
N. P. A. would like to acknowledge financial support through the National Science Foundation
under Grant No. 0094050, 0607497 to The University of Cincinnati.
We are grateful to the ESO staff who made available one night in service mode
after the technical problems with the dome wheels had been fixed.
Last, but not least, we thank an anonymous referee, whose thorough comments helped
us to improve our analysis and its presentation.

\clearpage
\begin{deluxetable}{llllrrcccc}
\tabletypesize{\scriptsize}
\tablewidth{0pc}
\tablecaption{Stellar Data for Sight Lines with New Detections of CO\tablenotemark{a}}
\tablehead{
\colhead{Star} 
&\colhead{Name} 
&\colhead{Sp.} 
&\colhead{$V$} 
&\colhead{$l$} 
&\colhead{$b$} 
&\colhead{$v_{\rm LSR}$\tablenotemark{b}} 
&\colhead{$E(\bv)$}
&\colhead{$D_{\rm helio}$\tablenotemark{c}} 
&\colhead{Refs\tablenotemark{d}}\\
\colhead{} 
&\colhead{} 
&\colhead{} 
&\colhead{(mag)} 
&\colhead{(deg)} 
&\colhead{(deg)} 
&\colhead{(km s$^{-1}$)} 
&\colhead{(mag)} 
&\colhead{(pc)} 
&\colhead{}}
\startdata
BD+48 3437 &           &B1 Iab  &8.73 & 93.56 &$-$2.06  &14.4    &0.35 &6500 &1,10\\
BD+53 2820 &           &B0 IV:n &9.96 &101.24 &$-$1.69  &12.8    &0.29 &4100 &2,2 \\
CPD$-$69 1743&         &B2 Vn   &9.46 &303.71 &$-$7.35  &$-$8.0  &0.30 &4700 &3,3 \\
CPD$-$59 2603&V572 Car &O7 V    &8.75 &287.59 &$-$0.69  &$-$11.6 &0.46 &2600 &3,3 \\
HD  12323  &           &O9      &8.92 &132.91 &$-$5.87  &3.5     &0.23 &3600 &2,2 \\
HD  13268  &           &O8 Vnn  &8.18 &133.96 &$-$4.99  &3.3     &0.36 &2400 &2,2 \\
HD  13745  &V354 Per   &O9.7 II &7.90 &134.58 &$-$4.96  &3.1     &0.46 &1600 &1,10\\
HD  14434  &           &O6.5    &8.59 &135.08 &$-$3.82  &3.1     &0.48 &4100 &2,2 \\
HD  15137  &           &O9.5 V  &7.86 &137.46 &$-$7.58  &1.8     &0.35 &2700 &1,10\\
HD  23180  &$o$ Per    &B1 III  &3.86 &160.36 &$-$17.74 &$-$6.7  &0.30 & 430 &2,2 \\
HD  23478  &           &B3 IV   &6.69 &160.76 &$-$17.42 &$-$6.7  &0.28 & 240 &1,11\\
HD  24190  &           &B2 V    &7.45 &160.39 &$-$15.18 &$-$6.4  &0.30 & 550 &1,10\\
HD  24398 &$\zeta$ Per &B1 Iab  &2.88 &162.29 &$-$16.69 &$-$7.1  &0.34 & 300 &4,11\\
HD  30122  &HR 1512    &B5 III  &6.34 &176.62 &$-$14.03 &$-$10.9 &0.40 & 220 &5,10\\
HD  34078  &AE Aur     &O9.5 Ve &6.00 &172.08 &$-$2.26  &$-$8.4  &0.53 & 450:&1,11\\
HD  36841  &           &O8      &8.58 &204.26 &$-$17.22 &$-$17.1 &0.35 &1200 &1,10\\
HD  37367  &HR 1924    &B2 IV-V &5.99 &179.04 &$-$1.03  &$-$10.1 &0.42 & 240 &2,2 \\
HD  37903  &           &B1.5 V  &7.84 &206.85 &$-$16.54 &$-$17.6 &0.32 & 790 &4,10\\
HD  43818  &11/LU Gem  &B0 II   &6.92 &188.49 &+3.87    &$-$11.9 &0.52 &1600 &2,2 \\
HD  58510  &           &B1 Iab  &6.80 &235.52 &$-$2.47  &$-$18.8 &0.32 &4500 &1,10\\
HD  63005  &           &O7      &9.13 &242.47 &$-$0.93  &$-$18.5 &0.32 &5200 &1,10\\
HD  91983  &        &O9.5/B0 Ib:&8.58 &285.88 &+0.05    &$-$11.9 &0.29 &7000 &2,2 \\
HD  93205  &V560 Car   &O3 V    &7.76 &287.57 &$-$0.71  &$-$11.6 &0.38 &3200 &2,2 \\
HD  93222  &           &O8      &8.11 &287.74 &$-$1.02  &$-$11.6 &0.36 &1700 &2,2 \\
HD  93237  &DR Cha     &B4 IVe  &5.97 &297.18 &$-$18.39 &$-$10.9 &0.09 & 310 &6,11\\
HD  93840  &           &B1.5 Iab&7.79 &282.14 &+11.10   &$-$11.1 &0.16 &5700 &1,10\\
HD  94454  &           &B8 III  &6.70 &295.69 &$-$14.73 &$-$11.0 &0.18 & 330 &6,11\\
HD  96675  &           &B6 IV/V &7.6  &296.62 &$-$14.57 &$-$10.7 &0.30 & 160 &2,11\\
HD  99872  &HR 4425    &B3 V    &6.11 &296.69 &$-$10.62 &$-$10.4 &0.36 & 230 &2,11\\
HD 102065  &           &B2 V    &6.61 &300.03 &$-$18.00 &$-$10.1 &0.17 & 170 &7,10\\
HD 106943  &           &B7 IV   &7.51 &298.96 &+1.14    &$-$8.3  &0.15 & 500 &1,10\\
HD 108002  &           &B2 Ia/ab&6.95 &300.16 &$-$2.48  &$-$8.4  &0.32 &3400 &1,10\\
HD 108610  &           &B3 IV/V &6.92 &300.28 &+0.88    &$-$7.9  &0.15 & 380 &1,11\\
HD 108639  &           &B1 III  &7.81 &300.22 &+1.95    &$-$7.8  &0.37 & 110 &1,10\\
HD 110434  &           &B8/9 III&7.55 &302.07 &$-$3.60  &$-$8.0  &0.05 & 370:&6,11\\
HD 112999  &V946 Cen   &B6 IIIn &7.38 &304.17 &+2.18    &$-$6.6  &0.23 & 340 &1,11\\
HD 114886  &           &O9 V    &6.89 &305.52 &$-$0.83  &$-$6.6  &0.40 &1000 &8,10\\
HD 115071  &V961 Cen   &O9.5 V  &7.97 &305.76 &+0.15    &$-$6.4  &0.53 &1200 &1,10\\
HD 115455  &           &O7.5 III&7.97 &306.06 &+0.22    &$-$6.3  &0.49 &2000 &6,10\\
HD 116852  &           &O9 III  &8.49 &304.88 &$-$16.13 &$-$8.6  &0.21 &4800 &2,2 \\
HD 122879  &HR 5281    &B0 Ia   &6.43 &312.26 &+1.79    &$-$4.2  &0.36 &2300 &2,2 \\
HD 124314  &           &O7      &6.64 &312.67 &$-$0.42  &$-$4.4  &0.53 &1100 &3,3 \\
HD 137595  &           &B3 Vn   &7.50 &336.72 &+18.86   &5.7     &0.25 & 400 &1,10\\
HD 140037  &           &B5 III  &7.48 &340.15 &+18.04   &6.6     &0.09 & 270:&6,11\\
HD 144965  &           &B3 Vne  &7.12 &339.04 &+08.42   &5.2     &0.35 & 290 &6,10\\
HD 147683  &V760 Sco   &B4 V    &7.05 &344.86 &+10.09   &7.2     &0.39 & 280 &1,10\\
HD 147888&$\rho$ Oph D &B3/B 4V &6.78 &353.65 &+17.71   &10.5    &0.51 & 140 &2,11\\
HD 152590  &           &O7.5 V  &8.48 &344.84 &+1.83    &6.2     &0.48 &1800 &1,10\\
HD 152723  &           &O7/O8   &7.31 &344.81 &+1.61    &6.1     &0.42 &1600 &2,2 \\
HD 157857  &           &O7 e    &7.81 & 12.97 &+13.31   &14.9    &0.43 &1900 &2,2 \\
HD 163758  &           &O6.5    &7.32 &355.36 &$-$6.10  &8.1     &0.33 &2600 &4,10\\
HD 185418  &           &B0.5 V  &7.52 & 53.60 &$-$2.17  &18.1    &0.50 & 910 &2,2 \\
HD 190918  &V1676 Cyg  &WN      &6.81 & 72.65 &+2.07    &18.0    &0.45 &2300 &1,10\\
HD 192035  &RX Cyg     &B0 IIIn &8.22 & 83.33 &+7.76    &17.3    &0.37 &2800 &1,10\\
HD 192639  &           &O8 e    &7.11 & 74.90 &+1.48    &17.7    &0.62 &1600 &4,10\\
HD 195965  &           &B0 V    &6.98 & 85.71 &+5.00    &16.7    &0.25 & 790 &1,10\\
HD 198781  &HR 7993    &B0.5 V  &6.46 & 99.94 &+12.61   &14.7    &0.35 & 730 &2,2 \\
HD 200775  &V380 Cep   &B2 Ve   &7.42 &104.06 &+14.19   &13.9    &0.57 & 430:&9,11\\
HD 203532  &HR 8176    &B3 IV   &6.36 &309.46 &$-$31.74 &$-$8.6  &0.28 & 250 &2,11\\
HD 208905  &           &B1 Vp   &7.01 &103.53 &+5.17    &13.1    &0.37 & 790 &1,10\\
HD 209481  &14/LZ Cep  &O9 V    &5.55 &102.01 &+2.18    &13.1    &0.37 & 690 &1,10\\
HD 209975  &19 Cep     &O9 Ib   &5.11 &104.87 &+5.39    &12.8    &0.34 &1300 &2,2 \\
HD 210121  &           &B9      &7.69 & 56.88 &$-$44.46 &7.9     &0.31 & 210 &2,11\\
HD 210809  &           &O9 Ib   &7.56 & 99.85 &$-$3.13  &13.0    &0.31 &4000 &2,2 \\
HD 220057  &NSV 14513  &B2 IV   &6.95 &112.13 &+0.21    &10.4    &0.23 & 560 &2,2 \\
HD 303308  &           &O3 V    &8.21 &287.59 &$-$0.61  &$-$11.6 &0.45 &3600 &2,2 \\
HD 308813  &           &O9.5 V  &9.32 &294.79 &$-$1.61  &$-$10.0 &0.31 &2400 &1,10\\
\enddata
\tablenotetext{a}{Information from the SIMBAD database is included.}
\tablenotetext{b}{Correction from heliocentric velocity to the LSR frame.}
\tablenotetext{c}{
Distance derived from either a spectroscopic parallax using $M_V$ from Table 3 of
reference 9, unless taken from the $E(B-V)$ reference, or from
a $\geq$4 $\sigma$ Hipparcos parallax from reference 10 as listed by Simbad, unless a
$\geq$3 $\sigma$ parallax was used and flagged with ``:''.}
\tablenotetext{d}{First reference is for $E(B-V)$, the second is for $D_{\rm helio}$.}
\tablerefs{
(1) \citealt{Neck80};
(2) \citealt{Vale04};
(3) \citealt{Dipl94};
(4) \citealt{Wegn03};
(5) \citealt{Carn86};
(6) \citealt{Ande02};
(7) \citealt{Rach02};
(8) \citealt{Sava85};
(9) \citealt{LeCo99};
(10) \citealt{Shul85};
(11) \citealt{Perr97}.
}
\end{deluxetable}

\clearpage
\begin{deluxetable}{lccccclcccc}
\tabletypesize{\scriptsize}
\tablewidth{0pc}
\tablecaption{UV Data Sets for New CO Sight Lines}
\tablehead{\colhead{Star} &\multicolumn{3}{c}{\textit{HST}/STIS} &\textit{FUSE}
&&\colhead{Star} &\multicolumn{3}{c}{\textit{HST}/STIS} &\textit{FUSE} \\
\cline{2-4} \cline{8-10}\\
 &Data set &Grating &Slit &Data set && &Data set &Grating &Slit &Data set \\}
\startdata
BD+48 3437   &o6359s &E140M  &0.2X0.2  &P10184 &&HD 110434    &o6lj0b &E140H  &0.1X0.03 &A12019 \\
BD+53 2820   &o6359q &E140M  &0.2X0.2  &P12232 &&HD 112999    &o6lj0c &E140H  &0.1X0.03 &A12020 \\
CPD$-$69 1743&o63566 &E140M  &0.2X0.2  &P10137 &&HD 114886    &o6lj0d &E140H  &0.1X0.03 &A12018 \\
CPD$-$59 2603&o40p01 &E140H  &0.2X0.09 &P12215 &&HD 115071    &o6lj0e &E140H  &0.2X0.09 &G93215 \\
             &o4qx03 &E140H  &0.2X0.09 &       &&HD 115455    &o6lj0f &E140H  &0.1X0.03 &A12007 \\
HD 12323     &o63505 &E140M  &0.2X0.2  &P10202 &&HD 116852    &o63571 &E140H  &0.2X0.2  &P10138 \\
HD 13268     &o63506 &E140M  &0.2X0.2  &P10203 &&HD 122879    &o6lz57 &E140H  &0.2X0.2  &B07105 \\
HD 13745     &o6lz05 &E140M  &0.2X0.2  &P10204 &&HD 124314    &o54307 &E140H  &0.1X0.03 &P10262 \\
HD 14434     &o63508 &E140M  &0.2X0.2  &P10205 &&             &o6lz58 &E140H  &0.2X0.2  &       \\
HD 15137     &o6lz06 &E140H  &0.2X0.2  &P10206 &&HD 137595    &o6lj03 &E140H  &0.2X0.09 &A12012 \\
HD 23180  &o64801--4 &E140H  &0.2X0.05 &\nodata&&HD 140037    &o6lj04 &E140H  &0.1X0.03 &A12015 \\
HD 23478     &o6lj01 &E140H  &0.1X0.03 &A12002 &&HD 144965    &o6lj05 &E140H  &0.1X0.03 &A12016 \\
HD 24190     &o6lj02 &E140H  &0.1X0.03 &A12001 &&HD 147683    &o6lj06 &E140H  &0.2X0.09 &A12009 \\
HD 24398 &o64810--11 &E140H  &0.2X0.05 &\nodata&&HD 147888    &o59s05 &E140H  &0.2X0.09 &P11615 \\
HD 30122     &o5c065 &E140H  &0.2X0.2  &Q20103 &&HD 152590    &o6lz67 &E140M  &0.2X0.2  &B07106 \\
HD 36841     &o63516 &E140M  &0.2X0.2  &\nodata&&HD 152723    &o63586 &E140H  &0.2X0.2  &P10271 \\
HD 37367     &o5c013 &E140H  &0.2X0.2  &B07102 &&HD 157857    &o5c04d &E140H  &0.2X0.2  &P10275 \\
HD 37903     &o59s04 &E140H  &0.2X0.09 &P11606 &&HD 163758    &o63595 &E140H  &0.2X0.2  &P10159 \\
HD 43818     &o5c07i &E140H  &0.2X0.2  &\nodata&&HD 185418    &o5c01q &E140H  &0.2X0.2  & \\
HD 58510     &o63530 &E140H  &0.2X0.2  &P10219 &&HD 190918    &o6359j &E140M  &0.2X0.2  &P10285 \\
HD 63005     &o63531 &E140M  &0.2X0.2  &P10221 &&HD 192035    &o6359k &E140M  &0.2X0.2  &P10286 \\
HD 91983     &o5c08n &E140H  &0.2X0.2  &B07104 &&HD 192639    &o5c08t &H140H  &0.2X0.2  & \\
HD 93205     &o4qx01 &E140H  &0.2X0.09 &P10236 &&HD 195965    &o6bg01 &E140H  &0.1X0.03 &P10288 \\
HD 93222     &o4qx02 &E140H  &0.2X0.09 &P10237 &&HD 198781    &o5c049 &E140H  &0.2X0.2  &P23102 \\
HD 93237     &o6lj0g &E140H  &0.1X0.03 &A12010 &&HD 200775    &\nodata&\nodata&\nodata  &A05101 \\
HD 93840     &o63549 &E140H  &0.2X0.2  &P10127 &&HD 203532    &o5co1s &E140H  &0.2X0.2  &B07108 \\
HD 94454     &o6lj0h &E140H  &0.1X0.03 &A12005 &&HD 208905    &\nodata&\nodata&\nodata  &D01401 \\
HD 96675\tablenotemark{a}&z19w01\tablenotemark{a} &G160M\tablenotemark{a} &0.25\tablenotemark{a} &Q10102 &&HD 209481    &\nodata&\nodata&\nodata  &D01402 \\
HD 99872     &o6lj0i &E140H  &0.1X0.03 &A12006 &&HD 209975    &\nodata&\nodata&\nodata  &D01403 \\
HD 102065    &o4o001 &E140H  &0.2X0.09 &Q10101 &&HD 210121    &\nodata&\nodata&\nodata  &P24901 \\
HD 106943    &o6lj07 &E140H  &0.1X0.03 &A12011 &&HD 210809    &o6359t &E140M  &0.2X0.2  &P12231 \\
HD 108002    &o6lj08 &E140H  &0.1X0.03 &A12017 &&HD 220057    &o5c01x &E140H  &0.2X0.2  &Z90178 \\
HD 108610    &o6lj09 &E140H  &0.1X0.03 &A12014 &&HD 303308    &o4qx04 &E140H  &0.2X0.09 &P12216 \\
HD 108639    &o6lj0a &E140H  &0.2X0.09 &A12013 &&HD 308813    &o63559 &E140M  &0.2X0.2  &P12219 \\
\enddata
\tablenotetext{a}{The \textit{HST} data set for HD 96675 is from the GHRS, not the STIS.}
\end{deluxetable}

\clearpage
\begin{deluxetable}{lrccccccccccccc}
\tabletypesize{\scriptsize}
\tablewidth{0pc}
\tablecaption{Newly Detected Cloud Components of Molecular Species}
\tablehead{
&&\multicolumn{2}{c}{CO}
&&\multicolumn{2}{c}{CN}
&&\multicolumn{2}{c}{CH}
&&\multicolumn{2}{c}{CH$^+$}\\
\cline{3-4} \cline{6-7} \cline{9-10} \cline{12-13}
\colhead{Star}
&\colhead{$v_{\rm LSR}$\tablenotemark{a}}
&\colhead{$N$\tablenotemark{b}}
&\colhead{$b$\tablenotemark{a}}
&&\colhead{$N$\tablenotemark{b}}
&\colhead{$b$\tablenotemark{a}}
&&\colhead{$N$\tablenotemark{b}}
&\colhead{$b$\tablenotemark{a}}
&&\colhead{$N$\tablenotemark{b}}
&\colhead{$b$\tablenotemark{a}}\\
&&\colhead{(10$^{14}$)}
&&&\colhead{(10$^{12}$)}
&&&\colhead{(10$^{12}$)}
&&&\colhead{(10$^{12}$)}&}
\startdata
BD+483437 &$-$16.4&$\ldots$ &&&$\ldots$ &&&$\ldots$ &&& 4.35&1.8 \\
          &$-$11.4&$\ldots$ &&&$\ldots$ &&&$\ldots$ &&& 0.98&1.0 \\
          &$-$1.0 & 0.23&1.5 &&$\ldots$ &&&$\ldots$ &&& 2.15&1.7 \\
          &   5.2 &$\ldots$ &&& 0.41&0.5 && 6.86&3.2 && 7.25&1.6 \\
BD+532820 &$-$5.4 &$\ldots$ &&&$\ldots$ &&&$\ldots$ &&& 2.15&1.0 \\
          &   0.9 & 0.50&0.5 &&$\ldots$ &&& 1.59&0.5 &&$\ldots$ & \\
          &   6.9 & 0.27&0.4 &&$\ldots$ &&& 2.49&1.1 && 4.26&2.2 \\
CPD$-$691743& 0.8 & 0.11&1.4 &&$\ldots$ &&&$\ldots$ &&&15.42&1.4 \\
CPD$-$592603&$-$5.7& 0.08&1.5 &&$\ldots$ &&& 3.80&1.3 &&$\ldots$ & \\
          &$-$2.8 & 1.13&0.7 &&$\ldots$ &&& 7.91&0.5 &&16.36&2.9 \\
HD 12323  &$-$13.5& 2.53&0.7 && 0.63&0.5 && 1.97&1.0 &&$\ldots$ & \\
          &$-$9.7 &$\ldots$ &&& 0.73&0.5 && 2.37&1.0 &&$\ldots$ & \\
          &$-$5.9 & 0.89&0.8 &&$\ldots$ &&& 1.98&2.3 && 5.52&2.5 \\
          &   0.3 &$\ldots$ &&&$\ldots$ &&&$\ldots$ &&& 2.42&1.5 \\
HD 13268  &$-$36.2&$\ldots$ &&&$\ldots$ &&& 0.94&1.0 &&$\ldots$ & \\
          &$-$19.5&$\ldots$ &&&$\ldots$ &&&$\ldots$ &&& 3.1 &3.0 \\
          &$-$16.4& 0.16&1.3 &&$\ldots$ &&& 1.13&1.0 &&$\ldots$ & \\
          &$-$10.4&$\ldots$ &&& 0.43&0.5 && 2.06&1.9 && 3.5 &2.3 \\
          &$-$7.4 & 1.26&1.0 && 0.94&1.3 && 5.29&2.3 &&$\ldots$ & \\
          &$-$5.1 &$\ldots$ &&&$\ldots$ &&&$\ldots$ &&& 4.8 &1.6 \\
          &$-$1.0 & 0.14&1.0 &&$\ldots$ &&& 2.15&2.5 && 3.5 &1.5 \\
HD 13745  &$-$43.9& 0.33&0.7 && 0.52&1.1 && 5.85&2.3 && 8.59&2.0 \\
          &$-$18.1& 0.36&0.5 &&$\ldots$ &&& 3.97&1.9 &&10.98&2.8 \\
          &$-$10.3& 0.11&1.3 &&$\ldots$ &&&$\ldots$ &&& 7.22&3.0 \\
          &$-$4.3 & 0.07&0.3 &&$\ldots$ &&&$\ldots$ &&&$\ldots$ & \\
          &   0.0 &$\ldots$ &&&$\ldots$ &&&$\ldots$ &&& 5.89&3.0 \\
HD 14434  &$-$6.1 & 0.43&0.6 &&$\ldots$ &&&$\ldots$ &&& 8.52&3.0 \\
          &$-$1.0 & 1.83&0.7 && 0.39&0.8 && 9.24&2.3 &&14.94&2.5 \\
HD 15137  &$-$13.4& 0.11&0.7 &&$\ldots$ &&& 2.17&2.5 && 4.13&2.9 \\
          &$-$7.4 &$\ldots$ &&&$\ldots$ &&& 1.49&1.0 && 7.70&3.0 \\
          &$-$4.0 & 0.03&0.6 &&$\ldots$ &&&$\ldots$ &&&$\ldots$ & \\
          &$-$0.2 & 0.19&1.3 &&$\ldots$ &&& 2.50&2.1 && 2.26&1.7 \\
HD 23180  &   4.6 & 0.81&0.8 &&$\ldots$ &&& 7.01&2.1 && 1.58&2.0 \\
          &   7.3 & 5.97&1.0 && 1.33&1.7 &&11.96&1.5 && 5.72&2.0 & \\
HD 23478  &   4.1 & 6.28&1.0 && 0.67&1.1 &&13.53&1.8 && 1.42&1.2 \\
          &   7.7 & 1.77&0.9 && 0.43&0.5 && 4.72&0.9 && 1.40&1.0 \\
HD 24190  &   6.5 & 0.84&0.5 && 0.50&1.8 &&$\ldots$ &&&$\ldots$ & \\
          &   9.7 & 0.06&1.0 &&$\ldots$ &&&$\ldots$ &&&$\ldots$ & \\
HD 24398  &   6.8 &17.95&0.8 && 3.20&1.0 &&21.41&1.6 && 3.13&2.3 \\
HD 30122  &   4.2 &$\ldots$ &&&$\ldots$ &&&$\ldots$ &&& 2.98&2.9 \\
          &   6.9 & 7.04&0.8 && 0.87&1.3 &&15.66&2.0 &&$\ldots$ & \\
HD 36841  &$-$4.6 & 0.46&0.5 &&$\ldots$ &&&$\ldots$ &&&$\ldots$ & \\
          &   5.7 & 0.35&0.4 &&$\ldots$ &&&$\ldots$ &&&$\ldots$ & \\
          &   10.6&$\ldots$ &&& 0.78&0.5 && 9.98&1.5 && 5.74&2.2 \\
HD 37367  &   3.8 &$\ldots$ &&&$\ldots$ &&& 2.21&2.0 &&$\ldots$ & \\
          &   6.2 & 0.70&1.5 &&$\ldots$ &&& 9.80&2.0 &&32.41&2.2 \\
HD 43818  &$-$7.0 &$\ldots$ &&&$\ldots$ &&& 3.22&2.5 &&$\ldots$ & \\
          &$-$3.9 & 0.74&1.5 &&$\ldots$ &&& 3.23&1.6 && 3.61&1.8 \\
          &   1.2 &$\ldots$ &&&$\ldots$ &&& 2.77&1.6 && 7.36&3.0 \\
          &   5.2 &$\ldots$ &&&$\ldots$ &&& 2.03&1.9 && 4.14&3.0 \\
HD 58510  &   23.6&$\ldots$ &&&$\ldots$ &&&$\ldots$ &&& 1.77&1.5 \\
          &   26.7&$\ldots$ &&&$\ldots$ &&& 2.52&2.0 &&10.74&2.1 \\
          &   29.6& 0.18&1.5 &&$\ldots$ &&& 2.53&1.6 &&$\ldots$ & \\
HD 63005  &   9.9 &$\ldots$ &&&$\ldots$ &&&$\ldots$ &&& 2.67&1.8 \\
          &   14.3& 0.63&0.9 && 0.99&1.1 && 4.51&2.2 && 5.67&2.5 \\
          &   21.0& 0.39&1.5 &&$\ldots$ &&& 4.46&2.0 && 5.69&1.7 \\
HD 91983  &$-$14.0& 0.62&0.5 &&$\ldots$ &&&$\ldots$ &&& 3.27&1.2 \\
HD 93205  &$-$6.5 & 0.06&1.0 &&$\ldots$ &&& 1.33&1.0 &&$\ldots$ & \\
          &$-$2.8 & 0.08&0.3 &&$\ldots$ &&& 1.31&0.5 &&$\ldots$ & \\
HD 93222  &$-$6.3 & 0.18&1.5 &&$\ldots$ &&& 2.28&2.1 &&$\ldots$ & \\
HD 93237\tablenotemark{c}  &   3.5 & 0.25&0.7 &&$\ldots$ &&& 1.2 &0.7 &&$\ldots$ & \\
HD 93840  &$-$7.0 & 0.18&0.8 &&$\ldots$ &&& 1.79&0.5 && 3.28&2.7 \\
HD 94454\tablenotemark{c}  &   3.6 & 2.02&1.3 &&$\ldots$ &&& 9.4 &1.6 &&$\ldots$ & \\
HD 96675  &   4.1 &20.18&0.9 && 4.96&0.5 &&22.76&1.1 && 4.90&2.0 \\
HD 99872\tablenotemark{c}  &   3.2 & 4.54&0.7 &&$\ldots$ &&&12.6 &1.7 &&23.10&1.9 \\
HD 102065 &   1.0 &$\ldots$ &&&$\ldots$ &&& 1.14&0.5 && 4.97&1.9 \\
          &   3.8 & 0.49&1.4 &&$\ldots$ &&& 6.03&1.6 && 5.78&1.2 \\
HD 106943 &$-$0.8 & 0.06&0.8 &&$\ldots$ &&&$\ldots$ &&&$\ldots$ & \\
HD 108002\tablenotemark{c} &$-$10.1& 0.06&0.4 &&$\ldots$ &&&$\ldots$ &&&$\ldots$ & \\
          &   1.6 & 0.35&0.8 &&$\ldots$ &&& 3.2 &1.5 &&$\ldots$ & \\
HD 108639 &$-$1.2 & 0.15&1.0 &&$\ldots$ &&&$\ldots$ &&&$\ldots$ & \\
\enddata
\end{deluxetable}

\clearpage
\begin{deluxetable}{lrccccccccccccc}
\setcounter{table}{3}
\tabletypesize{\scriptsize}
\tablewidth{0pc}
\tablecaption{Newly Detected Cloud Components of Molecular Species (Cont.)}
\tablehead{
&&\multicolumn{2}{c}{CO}
&&\multicolumn{2}{c}{CN}
&&\multicolumn{2}{c}{CH}
&&\multicolumn{2}{c}{CH$^+$}\\
\cline{3-4} \cline{6-7} \cline{9-10} \cline{12-13}
\colhead{Star}
&\colhead{$v_{\rm LSR}$\tablenotemark{a}}
&\colhead{$N$\tablenotemark{b}}
&\colhead{$b$\tablenotemark{a}}
&&\colhead{$N$\tablenotemark{b}}
&\colhead{$b$\tablenotemark{a}}
&&\colhead{$N$\tablenotemark{b}}
&\colhead{$b$\tablenotemark{a}}
&&\colhead{$N$\tablenotemark{b}}
&\colhead{$b$\tablenotemark{a}}\\
&&\colhead{(10$^{14}$)}
&&&\colhead{(10$^{12}$)}
&&&\colhead{(10$^{12}$)}
&&&\colhead{(10$^{12}$)}&}
\startdata
HD 110434 &$-$5.1 & 0.04&0.6 &&$\ldots$ &&&$\ldots$ &&&$\ldots$ & \\
          &$-$1.7 & 0.04&0.3 &&$\ldots$ &&&$\ldots$ &&&$\ldots$ & \\
HD 114886\tablenotemark{d} &$-$27.5& 0.09&0.6 &&$\ldots$ &&& 4. &&& 6. & \\
          & $-$4.3& 0.19&1.5 &&$\ldots$ &&&$\ldots$ &&&13. & \\
          & $-$1.9& 0.13&1.5 &&$\ldots$ &&& 5. &&&$\ldots$ & \\
HD 115071\tablenotemark{c} &$-$3.3 & 3.40&1.3 &&$\ldots$ &&& 8.2 &2.7 &&$\ldots$ & \\
HD 115455\tablenotemark{c} &$-$3.3 & 1.18&1.0 &&$\ldots$ &&&17.  &3.6 && 9.06&1.4 \\
          &$-$0.1 &$\ldots$ &&&$\ldots$ &&&$\ldots$ &&& 3.30&1.0 \\
HD 116852 &   0.6 & 0.20&0.5 &&$\ldots$ &&& 1.73&0.5 && 1.93&1.0 \\
HD 122879 &$-$26.1&$\ldots$ &&&$\ldots$ &&&$\ldots$ &&& 3.94&2.2 \\
          &$-$2.5 & 0.13&1.5 &&$\ldots$ &&& 2.45&1.8 && 7.24&2.3 \\
HD 124314 &$-$23.6& 0.24&1.5 &&$\ldots$ &&& 1.75&0.9 && 8.59&2.4 \\
          &$-$2.1 & 1.33&0.8 &&$\ldots$ &&& 7.07&1.7 && 6.23&2.6 \\
HD 137595\tablenotemark{c} &   5.4 & 0.77&1.0 &&$\ldots$ &&&12.2 &3.5 && 9.63&2.5 \\
HD 140037 &   1.1 & 0.41&1.4 &&$\ldots$ &&&$\ldots$ &&&$\ldots$ & \\
HD 144965\tablenotemark{c} &   4.5 &19.35&0.6 &&$\ldots$ &&&14.4 &1.6 && 7.58&1.9 \\
HD 147683\tablenotemark{c} &   5.0 & 0.54&0.3 &&$\ldots$ &&& 2.0 &$<$0.3 &&11.08&2.5 \\
          &   6.1 &88.55&0.6 &&$\ldots$ &&&15.2 &1.9 && 5.59&2.8 \\
          &  11.8 & 0.18&1.0 &&$\ldots$ &&& 5.0 &2.5 && 2.53&1.6 \\
HD 148937 &$-$13.5& 1.07&0.4 &&$\ldots$ &&&$\ldots$ &&&$\ldots$ & \\ 
          &$-$10.4& 2.51&0.7 &&$\ldots$ &&&$\ldots$ &&&$\ldots$ & \\
          &   3.2 & 0.23&0.8 &&$\ldots$ &&&$\ldots$ &&&$\ldots$ & \\
HD 152590 &$-$0.5 & 0.44&0.5 &&$\ldots$ &&& 7.59&1.9 && 8.45&2.3 \\
          &   5.3 & 0.15&1.5 &&$\ldots$ &&& 3.91&3.0 &&10.69&2.1 \\
HD 152723 &   2.1 & 0.10&1.5 &&$\ldots$ &&& 4.06&5.2 && 2.65&1.5 \\
          &   7.7 & 0.05&0.5 &&$\ldots$ &&& 4.20&0.8 && 3.63&1.9 \\
          &  10.2 & 0.43&1.5 &&$\ldots$ &&& 3.70&1.4 && 4.78&2.2 \\
HD 157857 &$-$5.6 &$\ldots$ &&&$\ldots$ &&&$\ldots$ &&& 3.37&2.3 \\
          &   0.0 & 1.22&0.6 &&$\ldots$ &&& 5.02&1.0 && 7.88&2.8 \\
          &   4.2 &$\ldots$ &&&$\ldots$ &&& 2.83&2.6 && 8.27&3.0 \\
HD 163758 &$-$3.8 & 0.24&0.9 &&$\ldots$ &&& 2.25&1.1 &&$\ldots$ & \\
          &   2.1 &$\ldots$ &&&$\ldots$ &&&$\ldots$ &&& 1.42&1.0 \\
HD 177989 &   6.7 & 0.04&1.0 &&$\ldots$ &&&$\ldots$ &&&$\ldots$ & \\
          &  10.1 & 0.45&0.8 &&$\ldots$ &&&$\ldots$ &&&$\ldots$ & \\
          &  12.1 & 3.96&0.5 &&$\ldots$ &&&$\ldots$ &&&$\ldots$ & \\
HD 185418\tablenotemark{e} &   6.8 & 4.53&0.7 &&$\ldots$ &&& 8.6 &1.1 && 3.9 &1.3 \\
          &  11.0 & 0.80&0.3 &&$\ldots$ &&& 3.2 &0.5 && 8.6 &2.1 \\
HD 190918 &   2.1 &$\ldots$ &&&$\ldots$ &&& 1.74&1.5 && 3.77&1.4 \\
          &   5.8 & 0.05&0.3 &&$\ldots$ &&&$\ldots$ &&& 5.51&2.0 \\
          &  18.3 & 0.09&1.5 &&$\ldots$ &&& 1.16&0.6 && 4.74&2.5 \\
HD 192035 &   1.4 &$\ldots$ &&&$\ldots$ &&& 2.84&1.0 && 3.61&2.0 \\
          &   5.6 &10.76&1.0 && 4.05&0.8 &&10.12&1.0 && 4.11&2.0 \\
          &   9.4 & 0.81&1.5 &&$\ldots$ &&& 3.53&1.0 &&$\ldots$ & \\
HD 192639\tablenotemark{f} &   7.3 & 0.60&1.0 &&$\ldots$ &&& 6.8 &1.2 &&$\ldots$ & \\
HD 195965 &   6.4 & 0.50&0.5 &&$\ldots$ &&&$\ldots$ &&&$\ldots$ & \\
          &  10.2 & 0.70&1.0 &&$\ldots$ &&&$\ldots$ &&&$\ldots$ & \\
HD 198781 &$-$0.1 & 0.66&0.3 &&$\ldots$ &&&$\ldots$ &&&$\ldots$ & \\
          &   5.4 &15.87&0.4 && 2.13&0.5 &&13.17&1.5 && 3.34&1.7 \\
HD 203532 &   5.3 &45.62&0.6 &&$\ldots$ &&& 9.76&3.1 && 2.98&2.0 \\
          &   7.5 &$\ldots$ &&&$\ldots$ &&&14.91&1.0 &&$\ldots$ & \\
HD 210121 &$-$6.2 &$\ldots$ &&&13.34&0.9 &&28.61&1.6 && 5.05&2.8 \\
          &$-$1.7 &$\ldots$ &&&$\ldots$ &&&$\ldots$ &&& 6.55&1.7 \\
HD 210809 &$-$1.2 & 0.23&1.5 &&$\ldots$ &&& 1.61&2.0 &&&$\ldots$ & \\
          &   2.8 &$\ldots$ &&&$\ldots$ &&& 3.86&1.5 && 7.51&1.7 \\
HD 220057 &$-$1.8 & 4.34&1.1 && 0.96&0.6 && 8.22&1.3 && 5.27&3.0 \\
          &   1.9 &$\ldots$ &&&$\ldots$ &&& 4.87&2.5 &&$\ldots$ & \\
HD 303308 &$-$7.7 & 0.10&1.5 &&$\ldots$ &&&$\ldots$ &&&$\ldots$ & \\
          &$-$4.5 & 0.35&1.4 &&$\ldots$ &&& 5.40&2.9 &&$\ldots$ & \\
          &$-$1.4 & 0.07&0.3 &&$\ldots$ &&&$\ldots$ &&&$\ldots$ & \\
HD 308813 &$-$2.8 & 0.69&0.7 &&$\ldots$ &&& 5.09&0.9 &&$\ldots$ & \\
\enddata
\tablenotetext{a}{Units are km s$^{-1}$.}
\tablenotetext{b}{Units are cm$^{-2}$.}
\tablenotetext{c}{CH results are from \citealt{Ande02}.} 
\tablenotetext{d}{CH and CH$^+$ results are from \citealt{Gred97}.}
\tablenotetext{e}{CH and CH$^+$ results are from \citealt{Sonn03}.}
\tablenotetext{f}{CH results are from \citealt{Sonn07}.}
\end{deluxetable}

\clearpage
\begin{deluxetable}{lcccccccccc}
\tabletypesize{\scriptsize}
\tablewidth{0pc}
\tablecaption{New Total Molecular Column Densities\tablenotemark{a}}
\tablehead{
&\multicolumn{10}{c}{log $N$ (cm$^{-2}$)}\\
\cline{2-11}
\colhead{Star} 
&\colhead{H$_2$} 
&\colhead{Ref}
&\colhead{CO} 
&\colhead{Ref}
&\colhead{CH$^+$} 
&\colhead{Ref}
&\colhead{CH}
&\colhead{Ref}
&\colhead{CN}
&\colhead{Ref}
}
\startdata
BD+48 3437   &20.42&  &13.36&  &13.18&  &12.79&  &11.76& \\
BD+53 2820   &20.15&  &13.89&  &12.81&  &12.61&  &$<$11.97& \\
CPD$-$69 1743&19.99&  &13.08&  &13.18&  &\nodata&&\nodata&\\
CPD$-$59 2603&20.15&  &14.08&  &13.20&  &13.08&  &\nodata&\\
HD  12323    &20.32&  &14.53&  &12.90&  &12.63&  &12.30& \\
HD  13268    &20.51&  &14.20&  &13.18&  &13.04&  &12.30& \\
HD  13745    &20.67&  &13.94&  &13.52&  &12.99&  &11.86& \\
HD  14434    &20.43&  &14.36&  &13.38&  &12.96&  &11.74& \\
HD  15137    &20.32&  &13.52&  &13.15&  &12.79&  &$<$11.62& \\
HD  23180    &20.61&1 &14.83&2 &12.84&  &13.28&  &12.18& \\
HD  23478    &20.57&  &14.91&  &12.32&  &13.34&  &12.04& \\
HD  24190    &20.38&  &13.95&  &13.18&  &12.98&  &11.88& \\
HD  24398    &20.67&  &15.26&  &12.45&  &13.32&  &12.51& \\
HD  30122    &20.70&  &14.85&  &12.48&  &13.20&  &12.08& \\
HD  34078    &20.88&  &14.76&  &13.84&  &13.90&  &12.52& \\
HD  36841    &\nodata\tablenotemark{b}& &14.08&  &12.76&  &13.00&  &12.04& \\
HD  37367    &20.61&  &13.85&  &13.51&  &13.08&  &$<$11.53& \\
HD  37903    &20.95&  &13.69&  &13.11&3 &12.96&3 &11.90&3 \\
HD  43818    &\nodata\tablenotemark{b}& &13.87&  &13.18&  &13.04&  &$<$11.79&\\
HD  58510    &20.23&  &13.26&  &13.08&  &12.70&  &$<$11.71& \\
HD  63005    &20.23&  &14.00&  &13.15&  &12.95&  &12.15& \\
HD  91983    &20.23&  &13.79&  &12.52&  &\nodata&&\nodata&\\
HD  93205    &19.83&  &13.15& &$<$12.20&&12.42&  &\nodata&\\
HD  93222    &19.84&  &13.26&  &\nodata&&12.36&  &\nodata&\\
HD  93237    &19.80&  &13.40&  &\nodata&&12.08&4 &\nodata&\\
HD  93840    &19.28&  &13.26&  &12.52&  &12.26&  &\nodata&\\
HD  94454    &20.76&  &14.30&  &\nodata&&12.97&4 &\nodata&\\
HD  96675    &20.86&  &15.28&  &12.69&  &13.36&  &12.77& \\
HD  99872    &20.52&  &14.65&  &13.36&  &13.11&4 &$<$11.79& \\
HD 102065    &20.56&  &13.69&  &13.04&  &12.86&  &$<$12.19& \\
HD 106943    &19.81&  &12.76&  &\nodata&&$<$12.49&4&\nodata&\\
HD 108002    &20.34&  &13.66&  &\nodata&&12.51&4 &\nodata&\\
HD 108610    &19.86&  &\nodata&&\nodata&&$<$12.56&4&\nodata&\\
HD 108639    &20.04&  &13.18&  &\nodata&&$<$12.38&4&\nodata&\\
HD 110434    &19.90&  &12.94&  &\nodata&&$<$12.23&4&\nodata&\\
HD 112999    &20.11& &$<$13.23&&\nodata&&\nodata&&\nodata&\\
HD 114886    &20.34&  &13.61&  &13.28&5 &12.95&5 &\nodata&\\
HD 115071    &20.69&  &14.53&  &\nodata&&12.91&4 &\nodata&\\
HD 115455    &20.58&  &14.08&  &13.23&  &13.23&4 &$<$12.20& \\
HD 116852    &19.83&  &13.30&  &12.54&  &12.23&  &\nodata&\\
HD 122879    &20.36&  &13.11&  &13.08&  &12.38&  &\nodata&\\
HD 124314    &20.52&  &14.20&  &13.18&  &12.94&  &\nodata&\\
HD 137595    &20.62&  &13.89&  &13.26&  &13.08&4 &$<$11.90& \\
HD 140037    &19.34&  &13.61&  &\nodata&&$<$11.85&4&\nodata&\\
HD 144965    &20.79&  &15.28&  &12.88&  &13.15&4 &\nodata&\\
HD 147683    &20.74&  &15.95&  &13.28&  &13.34&4 &\nodata&\\
HD 147888    &20.58&  &15.28&  &12.88&6 &13.34&6 &12.32&6 \\
HD 152590    &20.51&  &13.77&  &13.28&  &13.08&  &\nodata&\\
HD 152723    &20.30&  &13.76&  &13.04&  &13.08&  &\nodata&\\
HD 157857    &20.69&  &14.08&  &13.30&  &12.89&  &$<$12.06& \\
HD 163758    &19.85&  &13.38&  &12.15&  &12.34&  &\nodata&\\
HD 190918    &19.95&  &13.18&  &13.15&  &12.46&  &$<$11.53\\
HD 192035    &20.68&  &15.15&  &12.89&  &13.20&  &12.71& \\
HD 192639    &20.75&7 &13.78&  &13.61&7 &13.45&7 &$<$11.85&8 \\
HD 195965    &20.34&  &14.08&  &\nodata&&\nodata&&\nodata&\\
HD 198781    &20.56&  &15.23&  &12.52&  &13.11&  &12.46& \\
HD 200775    &21.15&  &17.29&  &12.97&9 &13.51&9 &13.08&9 \\
HD 203532    &20.70&  &15.66&  &12.48&  &13.40&  &\nodata&\\
HD 208905    &20.43&  &14.62&  &12.78&6 &12.73&6 &\nodata&\\
HD 209481    &20.54&  &14.60&  &12.72&6 &12.83&6 &\nodata&\\
HD 209975    &20.15&  &14.04&  &13.38&6 &12.93&6 &\nodata&\\
HD 210121    &20.86&  &15.83&10&13.08&  &13.46&  &13.20& \\
HD 210809    &20.00&  &13.36&  &12.88&  &12.74&  &$<$11.81& \\
HD 220057    &20.34&  &14.63&  &12.87&  &13.11&  &12.11& \\
HD 303308    &20.15&  &13.72&  &\nodata&&12.73&  &\nodata&\\
HD 308813    &20.30&  &13.84&  &\nodata&&12.71&  &\nodata&\\
\enddata
\tablenotetext{a}{Supplemented by literature values from listed references.}
\tablenotetext{b}{HD 36841 and HD 43818 are both predicted here to have
log $N$(H$_2$) = 20.4 $\pm$ 0.1, see $\S$ 4.4.}
\tablerefs{
(1) \citealt{Sava77};
(2) \citealt{Shef07};
(3) \citealt{Knau01};
(4) \citealt{Ande02};
(5) \citealt{Gred97};
(6) \citealt{Pan04};
(7) \citealt{Rach02};
(8) \citealt{Thor03};
(9) \citealt{Fede97a};
(10) \citealt{Sonn07}.
}
\end{deluxetable}

\clearpage
\begin{deluxetable}{clrcrrr}
\tabletypesize{\scriptsize}
\tablewidth{0pt}
\tablecaption{Power-Law Fits of Column Density Correlations\tablenotemark{a}}
\tablehead{
\colhead{y} &\colhead{x} &\colhead{n} &\colhead{$r$} &\colhead{CL} &\colhead{$A$} &\colhead{$B$}}
\startdata
log $N$(CO)     &log $N$(H$_2$)            &105 &0.834 &$>$99.99\% &$-$24.4 $\pm$ 3.1 &1.89 $\pm$ 0.15\\
log $N$(CO)     &log $N$(H$_2$) $<$ 20.4   & 50 &0.734 &$>$99.99\% &$-$15.8 $\pm$ 4.5 &1.46 $\pm$ 0.23\\
log $N$(CO)     &log $N$(H$_2$) $\ge$ 20.4 & 55 &0.638 &$>$99.99\% &$-$48.8 $\pm$ 15.0 &3.07 $\pm$ 0.73\\
\\
log $N$(CH)     &log $N$(H$_2$)            & 90 &0.906 &$>$99.99\% &$-$6.80 $\pm$ 1.50&0.97 $\pm$ 0.07\\
log $N$(CH)     &log $N$(H$_2$) $<$ 20.4   & 36 &0.799 &$>$99.99\% &$-$5.87 $\pm$ 3.78&0.92 $\pm$ 0.19\\
log $N$(CH)     &log $N$(H$_2$) $\ge$ 20.4 & 54 &0.740 &$>$99.99\% &$-$9.34 $\pm$ 3.90&1.09 $\pm$ 0.19\\
\\
log $N$(CO)     &log $N$(CH)               & 92 &0.824 &$>$99.99\% &$-$12.3 $\pm$ 2.7 &2.05 $\pm$ 0.21\\
log $N$(CO)     &log $N$(CH) $<$ 13.0      & 42 &0.624 &$>$99.99\% &$-$5.30 $\pm$ 3.8 &1.50 $\pm$ 0.30\\
log $N$(CO)     &log $N$(CH) $\ge$ 13.0    & 50 &0.693 &$>$99.99\% &$-$22.3 $\pm$ 11.3 &2.80 $\pm$ 0.85\\
\\
log $N$(CH$^+$) &log $N$(H$_2$)            & 86 &0.471 &$>$99.99\% &4.41 $\pm$ 2.06 &0.42 $\pm$ 0.10\\
log $N$(CH$^+$) &log $N$(H$_2$) $<$ 20.3   & 26 &0.637 &99.95\%    &$-$2.76 $\pm$ 4.30 &0.78 $\pm$ 0.22\\
log $N$(CH$^+$) &log $N$(H$_2$) $\ge$ 20.3 & 60 &0.089 &50\%       &10.1 $\pm$ 4.3 &0.15 $\pm$ 0.21\\
\\
log $N$(CH$^+$) &log $N$(CO)               & 88 &0.120 &73\%       &12.4 $\pm$ 0.7 &0.04 $\pm$ 0.05\\
log $N$(CH$^+$) &log $N$(CO) $<$ 14.1      & 41 &0.648 &$>$99.99\% &6.83 $\pm$ 1.29&0.46 $\pm$ 0.10\\
log $N$(CH$^+$) &log $N$(CO) $\ge$ 14.1    & 47 &0.268 &93.1\%     &15.2 $\pm$ 1.0 &$-$0.14 $\pm$ 0.07\\
\\
log $N$(CN)     &log $N$(H$_2$)            & 40 &0.669 &$>$99.99\% &$-$18.5 $\pm$ 7.5 &1.49 $\pm$ 0.36\\
log $N$(CN)     &log $N$(H$_2$) $<$ 20.68  & 20 &0.318 &83\%       &$-$9.0 $\pm$ 20.8 &1.02 $\pm$ 1.02\\
log $N$(CN)     &log $N$(H$_2$) $\ge$ 20.68& 20 &0.294 &79\%       &$-$11.5 $\pm$ 24.9 &1.16 $\pm$ 1.19\\
\\
log $N$(CO)     &log $N$(CN)               & 42 &0.836 &$>$99.99\% &$-$2.81 $\pm$ 2.77&1.44 $\pm$ 0.23\\
log $N$(CO)     &log $N$(CN) $<$ 12.31     & 20 &0.453 &95.5\%     &2.69 $\pm$ 5.02 &0.97 $\pm$ 0.42\\
log $N$(CO)     &log $N$(CN) $\ge$ 12.31   & 22 &0.578 &99.5\%     &3.38 $\pm$ 5.70 &0.96 $\pm$ 0.45\\
\enddata
\tablenotetext{a}{BCES(y$\mid$x) results from the \citealt{Akri96} code.}
\end{deluxetable}

\clearpage
\begin{deluxetable}{lccccccc}
\tabletypesize{\scriptsize}
\tablewidth{0pt}
\tablecaption{New Interstellar Excitation Temperatures for H$_2$ and CO}
\tablehead{
\colhead{Star} 
&\colhead{$T_{01}$(H$_2$)} 
&\colhead{$T_{02}$(H$_2$)} 
&\colhead{$T_{03}$(H$_2$)} 
&\colhead{$T_{04}$(H$_2$)} 
&\colhead{$T_{01}$(CO)} 
&\colhead{$T_{02}$(CO)} 
&\colhead{$T_{03}$(CO)}
\\
\colhead{} 
&\colhead{(K)} 
&\colhead{(K)} 
&\colhead{(K)} 
&\colhead{(K)} 
&\colhead{(K)} 
&\colhead{(K)} 
&\colhead{(K)} 
}
\startdata
BD+48 3437   & 83.   &113.   &158.   &246.   & 2.7   &\nodata&\nodata\\
BD+53 2820   & 93.   &120.   &176.   &244.   & 3.3   &\nodata&\nodata\\
CPD$-$69 1743& 79.   &102.   &143.   &213.   & 2.7   &\nodata&\nodata\\
CPD$-$59 2603& 77.   & 95.   &142.   &217.   & 3.0   & 3.3   &\nodata\\
HD  12323    & 82.   &101.   &142.   &217.   & 3.1   & 4.3   &\nodata\\
HD  13268    & 92.   &120.   &167.   &245.   & 3.4   &\nodata&\nodata\\
HD  13745    & 66.   & 93.   &128.   &202.   & 4.0   &\nodata&\nodata\\
HD  14434    & 99.   &129.   &166.   &247.   & 4.4   &\nodata&\nodata\\
HD  15137    &104.   &111.   &153.   &245.   & 3.1   & 4.2   &\nodata\\
HD  23478    & 55.   & 79.   &101.   &171.   & 3.4   & 3.6   & 4.2   \\
HD  24190    & 66.   & 86.   &119.   &193.   & 3.1   & 3.5   &\nodata\\
HD  24398    &\nodata&\nodata&\nodata&\nodata& 3.4   & 3.8   & 4.3   \\
HD  24534    & 54.   & 73.   & 96.   &152.   &\nodata&\nodata&\nodata\\
HD  27778    & 51.   & 78.   &103.   &152.   & 5.3   & 5.5   & 5.6   \\
HD  30122    & 61.   & 86.   &121.   &185.   & 3.8   & 4.0   &\nodata\\
HD  34078    & 75.   & 92.   &128.   &206.   &\nodata&\nodata&\nodata\\
HD  36841    &\nodata&\nodata&\nodata&\nodata& 2.7   & 3.0   &\nodata\\
HD  37367    & 73.   & 82.   &112.   &185.   & 3.2   &\nodata&\nodata\\
HD  37903    & 64.   &121.   &125.   &190.   & 2.7   &\nodata&\nodata\\
HD  43818    &\nodata&\nodata&\nodata&\nodata& 4.1   &\nodata&\nodata\\
HD  58510    & 90.   & 99.   &143.   &212.   & 2.9   &\nodata&\nodata\\
HD  63005    & 78.   & 91.   &129.   &188.   & 3.6   &\nodata&\nodata\\
HD  91983    & 61.   &105.   &144.   &222.   & 2.7   &\nodata&\nodata\\
HD  93205    & 97.   &118.   &167.   &241.   & 2.8   &\nodata&\nodata\\
HD  93222    & 69.   &109.   &162.   &218.   & 3.3   &\nodata&\nodata\\
HD  93237    & 58.   & 85.   &111.   &135.   & 3.1   &\nodata&\nodata\\
HD  93840    & 54.   &112.   &170.   &224.   & 3.1   &\nodata&\nodata\\
HD  94454    & 74.   & 83.   &106.   &167.   & 3.8   &\nodata&\nodata\\
HD  96675    &\nodata&\nodata&\nodata&\nodata& 3.7   & 5.9   &\nodata\\
HD  99872    & 66.   & 94.   &114.   &179.   & 3.7   & 3.8   &\nodata\\
HD 102065    &\nodata&\nodata&\nodata&\nodata& 3.6   &\nodata&\nodata\\
HD 106943    & 96.   &108.   &142.   &214.   & 2.7   &\nodata&\nodata\\
HD 108002    & 77.   & 98.   &133.   &218.   & 3.2   &\nodata&\nodata\\
HD 108610    & 80.   &106.   &138.   &208.   &\nodata&\nodata&\nodata\\
HD 108639    & 88.   &111.   &153.   &219.   & 3.0   &\nodata&\nodata\\
HD 110434    & 87.   &105.   &144.   &216.   & 2.7   &\nodata&\nodata\\
HD 112999    & 96.   &102.   &140.   &231.   & 3.0   &\nodata&\nodata\\
HD 114886    & 92.   &109.   &151.   &214.   & 3.1   &\nodata&\nodata\\
HD 115071    & 71.   & 95.   &133.   &208.   & 3.7   &\nodata&\nodata\\
HD 115455    & 81.   & 96.   &128.   &200.   & 2.9   &\nodata&\nodata\\
HD 116852    & 66.   & 98.   &147.   &200.   & 3.2   &\nodata&\nodata\\
HD 122879    & 90.   &105.   &148.   &200.   & 2.9   &\nodata&\nodata\\
HD 124314    & 74.   & 98.   &138.   &208.   & 3.8   &\nodata&\nodata\\
HD 137595    & 72.   & 94.   &124.   &197.   & 3.9   & 4.4   &\nodata\\
HD 140037    &\nodata&\nodata&\nodata&\nodata& 2.9   &\nodata&\nodata\\
HD 144965    & 70.   & 91.   &125.   &203.   & 4.3   & 5.3   &\nodata\\
HD 147683\tablenotemark{a}    & 58.   & 85.   &116.   &185.   & 5.2   & 6.5   & 6.9\tablenotemark{a}   \\
HD 147888    & 44.   & 90.   &110.   &181.   &13.6   & 9.3   & 9.0   \\
HD 148937    & 69.   & 97.   &132.   &228.   & 3.7   & 4.4   & 5.6   \\
HD 152590    & 64.   & 87.   &125.   &205.   & 4.1   &\nodata&\nodata\\
HD 152723    & 76.   & 96.   &141.   &201.   & 4.0   &\nodata&\nodata\\
HD 154368    & 47.   & 95.   &\nodata&\nodata& 3.0   & 4.3   &\nodata\\
HD 157857    & 86.   & 99.   &133.   &203.   & 4.6   &\nodata&\nodata\\
HD 163758    & 79.   &142.   &204.   &277.   & 4.0   &\nodata&\nodata\\
HD 177989    & 49.   & 85.   &127.   &198.   & 3.3   & 3.5   &\nodata\\
HD 190918    &102.   &156.   &214.   &310.   & 2.7   & 4.0   &\nodata\\
HD 192035    & 68.   & 92.   &126.   &205.   & 3.2   & 3.9   &\nodata\\
HD 195965    & 91.   &103.   &136.   &214.   & 3.0   &\nodata&\nodata\\
HD 198781    & 65.   & 92.   &128.   &191.   & 3.4   & 3.7   &\nodata\\
HD 200775    & 44.   &104.   &104.   &168.   &\nodata&\nodata&\nodata\\
HD 203532    & 47.   & 78.   &102.   &169.   & 5.3   & 4.8   &\nodata\\
HD 208905    & 77.   & 97.   &132.   &214.   & 6.0   &\nodata&\nodata\\
HD 209481    & 78.   & 97.   &137.   &215.   & 2.9   &\nodata&\nodata\\
HD 209975    & 73.   &104.   &149.   &243.   & 2.9   &\nodata&\nodata\\
HD 210121    & 51.   & 83.   &108.   &178.   &\nodata&\nodata&\nodata\\
HD 210809    & 87.   &126.   &187.   &278.   & 3.1   &\nodata&\nodata\\
HD 220057    & 65.   & 87.   &122.   &192.   & 3.0   & 3.8   &\nodata\\
HD 303308    & 91.   &121.   &177.   &300.   & 3.1   &\nodata&\nodata\\
HD 308813    & 73.   & 92.   &129.   &181.   & 3.8   &\nodata&\nodata\\
\enddata
\tablenotetext{a}{HD 147683 presents higher-$J$ lines with $T_{04}$(CO) = 7.7 and
$T_{05}$(CO) = 8.5 K (see Fig. 1).}
\end{deluxetable}

\clearpage
\begin{deluxetable}{llrrrrrccc}
\tablewidth{0pt}
\tabletypesize{\scriptsize}
\tablecaption{Chemical Results from CN Chemistry}
\tablehead{
\colhead{Star}
&\colhead{Cloud\tablenotemark{a}}
&\colhead{$N_{\rm o}$(CH)}
&\colhead{$N_{\rm o}$(C$_2$)}
&\colhead{$N_{\rm p}$(C$_2$)}
&\colhead{$N_{\rm o}$(CN)}
&\colhead{$N_{\rm p}$(CN)}
&\colhead{$T$}
&\colhead{$\tau_{\rm UV}$}
&\colhead{$n_{\rm H}$(CN)} \\
\cline{3-7}
\colhead{}
&\colhead{}
&\multicolumn{5}{c}{($10^{12}$ cm$^{-2}$)}
&\colhead{(K)}
&\colhead{}
&\colhead{(cm$^{-3}$)}}
\startdata
BD+48 3437 & & 4.80 & $\ldots$ & 4.5 & 0.58 & 0.58 & 65 & 2.17 & 475\\
BD+53 2820 & +0.9 & 1.59 & $\ldots$ & $\le$3.9 & $\le$0.93 & $\le$0.93 & 65 & 2.17 & $\le$1600\\
 & +6.9 & 2.49 & $\ldots$ & $\le$5.6 & $\le$0.93 & $\le$0.93 & & & $\le$1425\\
HD 12323 & $-$13.5 & 1.77 & $\ldots$ & 4.1 & 0.89 & 0.72 & 65 & 2.23 & $\sim$1600\\
 & $-$9.7 & 3.36 & $\ldots$ & 6.3 & 1.03 & 1.03 & & & 1200\\
HD 13268 & $-$16.4 & 1.20 & $\ldots$ & $\le$2.1 & $\le$0.42 & $\le$0.42 & 65 & 2.48 & $\le$1450\\
 & $-$10.4 & 2.70 & $\ldots$ & 3.5 & 0.61 & 0.60 & & & 900\\
 & $-$7.4 & 5.42 & $\ldots$ & 7.5 & 1.33 & 1.35 & & & 1000\\
 & $-$1.0 & 1.82 & $\ldots$ & $\le$2.4 & $\le$0.42 & $\le$0.42 & & & $\le$925\\
HD 13745 & $-$43.9 & 5.89 & $\ldots$ & 6.5 & 0.74 & 0.74 & 65 & 1.34 & 850\\
 & $-$18.1 & 3.98 & $\ldots$ & $\le$4.9 & $\le$0.57 & $\le$0.56 & & & $\le$950\\
HD 14434 &  & 9.32 & $\ldots$ & 2.7 & 0.55 & 0.58 & 65 & 3.72 & 125\\
HD 15137 & $-$13.4 & 2.24 & $\ldots$ & $\le$2.0 & $\le$0.42 & $\le$0.41 & 65 & 2.98 & $\le$425\\
 & $-$7.4 & 1.62 & $\ldots$ & $\le$1.8 & $\le$0.42 & $\le$0.42 & & & $\le$900\\
 & $-$0.2 & 2.56 & $\ldots$ & $\le$2.1 & $\le$0.42 & $\le$0.42 & & & $\le$550\\
HD 22951/40 Per &  & 12.0 & 3.6 & 4.4 & 0.64 & 0.42 & 40 & 1.49 & 225\\
HD 23180/$o$ Per & +4.6 & 7.01 & $\le$4.0 & $\le$3.3 & $\le$0.30 & $\le$0.32 & 40 & 1.86 & $\le$200\\
 & +7.3 & 11.96 & 23.0 & 16.0 & 1.65 & 1.91 & & & 625\\
HD 23478 & +4.1 & 13.53 & 7.8 & 8.4 & 1.03 & 0.85 & 50 & 1.67 & 325\\
 & +7.7 & 4.72 & 6.2 & 6.6 & 0.80 & 0.74 & & & 775\\
HD 24190\tablenotemark{b} &  & 9.5 & $\ldots$ & 7.3 & 0.75 & 0.73 & 40 & 1.74 & 375\\
HD 24398/$\zeta$ Per &  & 22.0 & 35.0 & 35.0 & 3.9 & 3.9 & 30 & 2.05 & 700\\
HD 24534/X Per & +5.0 & 6.0 & 3.4 & 3.5 & 0.70 & 0.62 & 20 & 3.84 & 250\\
 & +7.1 & 25.9 & 31.0 & 31.0 & 6.6 & 6.7 & & & 650\\
HD 24912/$\xi$ Per\tablenotemark{b} &  & 12.0 & 7.9 & 4.6 & 0.26 & 0.49 & 70 & 1.44 & $\sim$250\\
HD 27778/62 Tau & +5.0 & 12.8 & 24.0 & 36.0 & 8.9 & 5.8 & 50 & 2.29 & $\sim$1100\\
 & +7.2 & 9.4 & 14.0 & 30.0 & 5.1 & 3.6 & & & $\sim$575\\
HD 30122 &  & 15.72 & $\ldots$ & 10.5 & 1.58 & 1.54 & 65 & 2.48 & 400\\
 &  & 15.72 & $\ldots$ & 6.6 & 1.58 & 1.53 & & 3.72 & 200\\
 &  & 15.72 & $\ldots$ & 13.5 & 1.58 & 1.62 & & 2.11 & 400\\
HD 36841 &  & 9.92 & $\ldots$ & 7.2 & 1.56 & 1.54 & 65 & 3.16 & 475\\
HD 37367 & +3.8 & 3.23 & $\ldots$ & $\le$2.3 & $\le$0.34 & $\le$0.34 & 65 & 2.48 & $\le$425\\
 & +6.2 & 9.76 & $\ldots$ & $\le$2.7 & $\le$0.34 & $\le$0.35 & & & $\le$150\\
HD 43818/11 Gem & $-$7.0 & 3.80 & $\ldots$ & $\le$2.6 & $\le$0.61 & $\le$0.61 & 65 & 3.41 & $\le$425\\
 & $-$3.9 & 2.69 & $\ldots$ & $\le$2.4 & $\le$0.61 & $\le$0.61 & & & $\le$625\\
 & +1.2 & 2.79 & $\ldots$ & $\le$2.4 & $\le$0.61 & $\le$0.61 & & & $\le$600\\
 & +5.2 & 2.02 & $\ldots$ & $\le$2.2 & $\le$0.61 & $\le$0.61 & & & $\le$900\\
HD 58510 &  & 5.11 & $\ldots$ & $\le$4.6 & $\le$0.51 & $\le$0.50 & 65 & 1.92 & $\le$375\\
HD 63005 & +14.3 & 4.64 & $\ldots$ & 10.5 & 1.40 & 1.41 & 65 & 1.74 & 1300\\
 & +21.0 & 4.37 & $\ldots$ & $\le$8.1 & $\le$1.03 & $\le$1.03 & & & $\le$1025\\
HD 96675 &  & 22.76 & $\ldots$ & 49.7 & 6.26 & 6.25 & 50 & 1.58 & 1425\\
HD 99872 &  & 12.6 & $\ldots$ & $\le$6.1 & $\le$0.62 & $\le$0.63 & 65 & 1.90 & $\le$200\\
HD 102065 & +1.0 & 1.1 & $\ldots$ & $\le$1.4 & $\le$0.62 & $\le$0.17 & 65 & 0.90 & $>$1600\\
 & +3.8 & 6.0 & $\ldots$ & $\le$7.9 & $\le$0.94 & $\le$0.93 & & & $\le$1600\\
HD 115455 &  & 17.0 & $\ldots$ & $\le$8.1 & $\le$1.6 & $\le$1.6 & 65 & 3.16 & $\le$275\\
HD 137595 &  & 12.2 & $\ldots$ & $\le$7.7 & $\le$0.80 & $\le$0.82 & 65 & 1.49 & $\le$400\\
HD 148184/$\chi$ Oph &  & 34.0 & 35.0 & 19.0 & 1.3 & 2.7 & 60 & 2.30 & $\sim$300\\
HD 149757/$\zeta$ Oph\tablenotemark{b} &  & 25.0 & 18.0 & 21.0 & 2.6 & 2.2 & 60 & 1.98 & 325\\
HD 154368 & $-$13.1 & 2.1 & $\ldots$ & 0.53 & 0.21 & 0.20 & 50 & 4.77 & 90\\
 & +3.3 & 54.1 & 51.0 & 58.0 & 27.0 & 22.0 & & & 750\\
HD 157857 & +0.0 & 5.02 & $\ldots$ & $\le$6.0 & $\le$1.15 & $\le$1.16 & 65 & 2.67 & $\le$900\\
 & +4.2 & 2.83 & $\ldots$ & $\le$4.9 & $\le$1.15 & $\le$1.10 & & & $\le$1600\\
HD 185418\tablenotemark{b} &  & 13.0 & $\le$10.0 & $\le$1.5 & $\le$0.50 & $\le$0.48 & 65 & 4.46 & $\le$30\\
HD 190918 & +2.1 & 1.73 & $\ldots$ & $\le$1.9 & $\le$0.34 & $\le$0.34 & 65 & 2.54 & $\le$775\\
 & +18.3 & 1.16 & $\ldots$ & $\le$1.8 & $\le$0.34 & $\le$0.34 & & & $\le$1175\\
HD 192035 & +1.4 & 3.31 & $\ldots$ & $\le$4.3 & $\le$0.53 & $\le$0.53 & 65 & 2.05 & $\le$575\\
 & +5.6 & 11.42 & $\ldots$ & 33.4 & 5.07 & 5.11 & & & 1550\\
 & +9.4 & 2.62 & $\ldots$ & $\le$4.1 & $\le$0.53 & $\le$0.53 & & & $\le$725\\
HD 192639\tablenotemark{b} &  & 28.0 & $\le$10.0 & $\le$3.2 & $\le$0.70 & $\le$0.76 & 65 & 3.97 & $\le$40\\
HD 198781 &  & 13.19 & $\ldots$ & 13.1 & 3.29 & 3.25 & 65 & 3.26 & 750\\
HD 208440\tablenotemark{b} &  & 11.7 & $\ldots$ & $\le$8.2 & $\le$0.90 & $\le$0.87 & 65 & 1.80 & $\le$325\\
HD 208905\tablenotemark{b} &  & 5.4 & $\ldots$ & $\le$7.6 & $\le$0.90 & $\le$0.90 & 65 & 1.60 & $\le$850\\
HD 209399\tablenotemark{b} &  & 7.9 & $\ldots$ & $\le$8.1 & $\le$0.90 & $\le$0.91 & 65 & 1.28 & $\le$825\\
HD 209975/19 Cep\tablenotemark{b} &  & 8.5 & $\ldots$ & $\le$8.1 & $\le$0.90 & $\le$0.90 & 65 & 1.52 & $\le$600\\
HD 210121 &  & 286.2 & 65.0 & 55.1 & 17.35 & 19.10 & 50 & 3.35 & 1425\\
HD 210809 & $-$0.6 & 1.61 & $\ldots$ & $\le$4.4 & $\le$0.64 & $\le$0.64 & 65 & 1.98 & $\le$1325\\
 & +3.4 & 3.86 & $\ldots$ & $\le$5.4 & $\le$0.64 & $\le$0.64 & & & $\le$575\\
HD 217035A\tablenotemark{b} &  & 16.8 & $\ldots$ & $\le$9.0 & $\le$0.90 & $\le$0.90 & 65 & 2.47 & $\le$125\\
HD 220057 & $-$1.8 & 9.33 & $\ldots$ & 11.8 & 1.36 & 1.37 & 65 & 1.61 & 750\\
 & +1.9 & 3.86 & $\ldots$ & $\le$5.4 & $\le$0.65 & $\le$0.65 &  & & $\le$850\\
\enddata
\tablecomments{All calculations employ $I_{\rm UV}$ = 1, except for HD 27778/62 Tau and HD 210121,
where $I_{\rm UV}$ = 0.5.} 
\tablenotetext{a}{If more than one cloud containing CN appear along a line of sight, they are designated by
$v_{\rm LSR}$ values from Table 3, having identical values for $T$ and $\tau_{\rm UV}$.}
\tablenotetext{b}{Results are for line of sight because some input 
data are not available for all components.}
\end{deluxetable}

\clearpage
\begin{deluxetable}{lccccc}
\tablecolumns{6}
\tablewidth{0pt}
\tabletypesize{\scriptsize}
\tablecaption{Chemical Results from CH$^+$ Chemistry}
\tablehead{Star\tablenotemark{a} & $N$(CH$^+$) & $N$(CH) & $f$(H$_2$) & $\tau_{\rm UV}$ & $n_{\rm H}$(CH$^+$) \\
 & ($10^{12}$ cm$^{-2}$) & ($10^{12}$ cm$^{-2}$) & & & (cm$^{-3}$) } 
\startdata
BD+53 2820 & 6.40 & 4.08 & 0.11 & 2.17 & 2.1 \\
CPD$-$69 1743 & 15.0 & 2.20 & 0.13 & 1.30 & 1.0 \\
CPD$-$59 2603 & 16.0 & 12.0 & 0.09 & 2.42 & 2.4 \\
HD 15137 & 14.0 & 6.42 & 0.24 & 2.98 & 0.3 \\
HD 30614/$\alpha$ Cam & 20.0 & 6.80 & 0.36 & 1.98 & 0.4 \\
HD 37367 & 32.0 & 13.0 & 0.30 & 2.48 & 0.4 \\
HD 58510 & 12.0 & 5.11 & 0.15 & 1.92 & 1.4 \\
HD 93840 & 3.30 & 1.80 & 0.03 & 0.52 & 31.4 \\
HD 99872 & 23.0 & 12.6 & 0.28 & 1.90 & 1.0 \\
HD 102065 & 11.0 & 7.10 & 0.68 & 0.90 & 1.3 \\
HD 114886 & 19.0 & 9.00 & 0.19 & 2.48 & 0.7 \\
HD 115455 & 17.0 & 17.0 & 0.23 & 3.16 & 0.6 \\
HD 116852 & 3.50 & 1.70 & 0.13 & 1.36 & 3.1 \\
HD 122879 & 12.0 & 3.10 & 0.20 & 2.17 & 0.5 \\
HD 124314($-$19) & 8.60 & 1.80 & 0.33 & 1.37 & 0.5 \\
HD 124314(+2) & 6.20 & 7.10 & 0.22 & 1.37 & 4.2 \\
HD 137595 & 18.0 & 12.0 & 0.60 & 1.49 & 0.8 \\
HD 145502/$\nu$ Sco & 6.30 & 5.90 & 0.10 & 1.17 & 9.4 \\
HD 149038/$\mu$ Nor & 35.0 & 10.0 & 0.36 & 2.36 & 0.2 \\
HD 152590 & 19.0 & 10.0 & 0.22 & 2.67 & 0.5 \\
HD 152723 & 11.0 & 6.40 & 0.13 & 2.60 & 1.1 \\
HD 157857 & 20.0 & 7.85 & 0.33 & 2.67 & 0.3 \\
HD 163758 & 1.40 & 2.20 & 0.08 & 3.07 & 3.4 \\
HD 164353/67 Oph    & 7.40 & 4.50 & 0.26 & 0.74 & 3.5 \\
HD 185418 & 12.0 & 13.0 & 0.40 & 4.46 & 0.1 \\
HD 190918 & 14.0 & 2.89 & 0.07 & 2.54 & 0.8 \\
HD 192639 & 41.0 & 28.0 & 0.34 & 3.97 & 0.1 \\
HD 203532 & 3.00 & 25.0 & 0.34 & 1.92 & 11.5 \\
HD 208440 & 8.70 & 11.7 & 0.21 & 1.80 & 3.5 \\
HD 209339 & 6.60 & 7.90 & 0.17 & 1.28 & 6.5 \\
HD 209975/19 Cep    & 24.0 & 8.50 & 0.17 & 1.52 & 1.5 \\
HD 210809 & 7.50 & 5.47 & 0.10 & 1.98 & 3.2 \\
HD 217035A & 21.0 & 16.8 & 0.38 & 2.47 & 0.6 \\
HD 218376/1 Cas     & 11.0 & 7.60 & 0.24 & 1.36 & 2.4 \\
\enddata
\tablenotetext{a}{If more than one cloud containing CN appear along a 
line of sight, the velocity is given in parentheses.}
\end{deluxetable}

\clearpage
\begin{figure}
\plotone{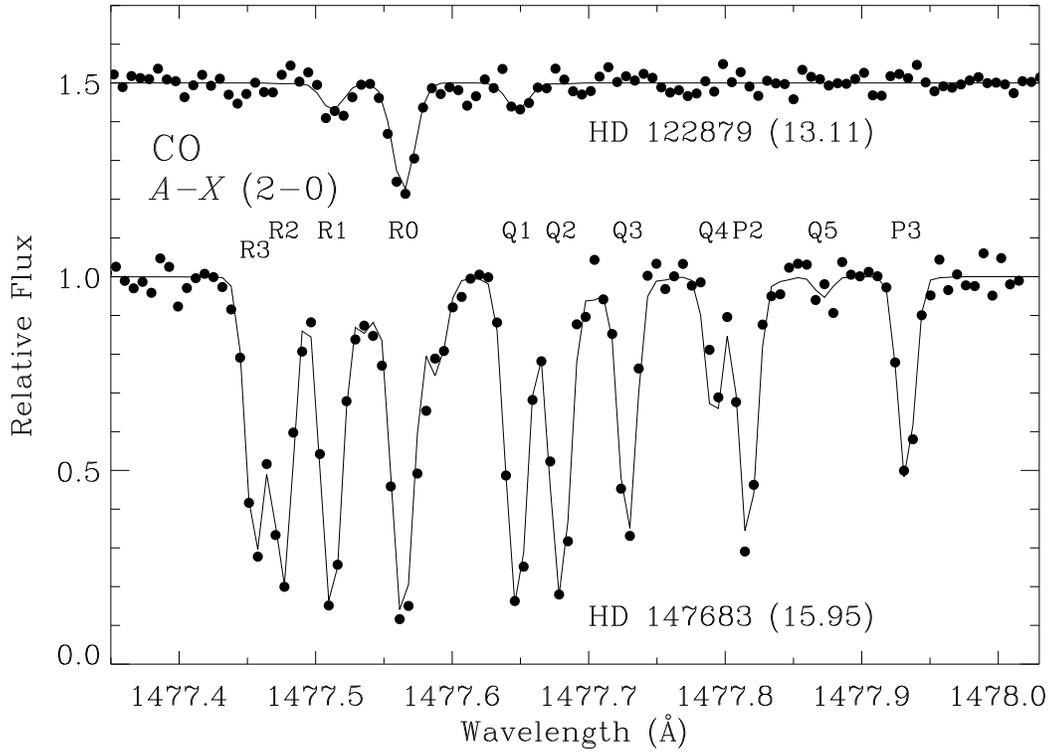}
\caption{Sample CO spectra (dots) from \textit{HST}/STIS and Ismod.f fits (solid lines)
are shown for HD 147683 (log $N$ = 15.95) and HD 122879 (13.11).
The second spectrum has been shifted upward by 0.5 continuum units.} 
\end{figure}

\begin{figure}
\plotone{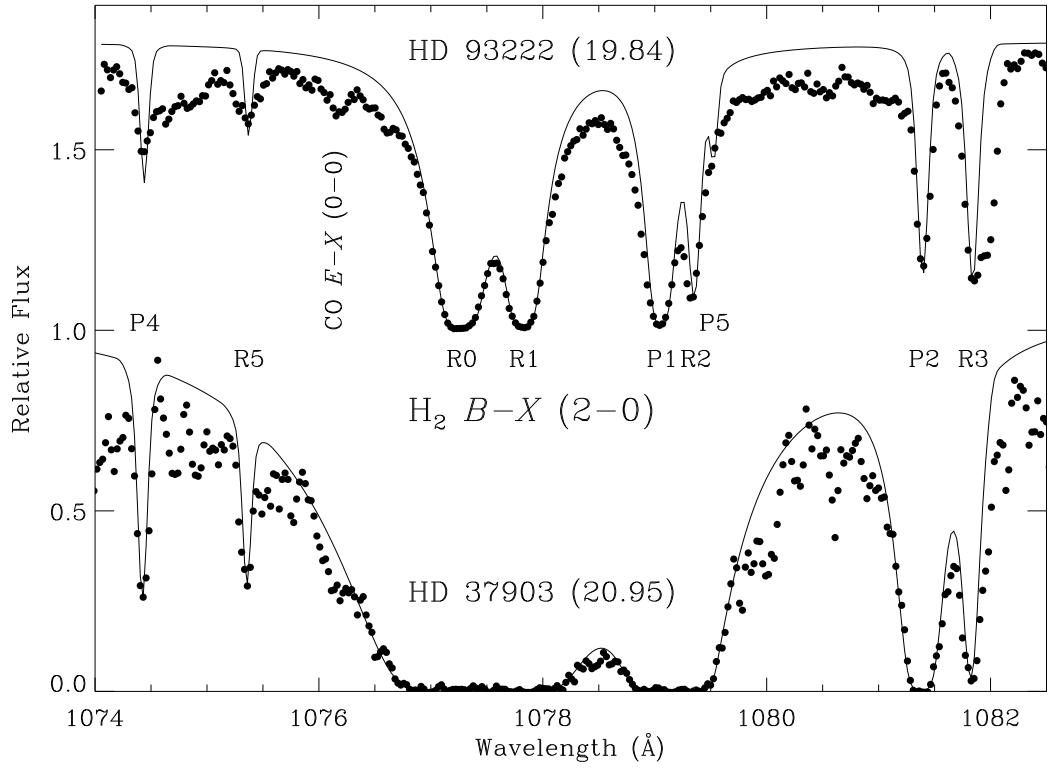}
\caption{Sample H$_2$ spectra (dots) from \textit{FUSE} and Ismod.f fits (solid lines)
are shown for HD 37903 (log $N$ = 20.95) and HD 93222 (19.84).
The second spectrum has been shifted upward by 0.8 continuum units.
Each spectrum synthesis with Ismod.f includes also the (3$-$0) and the (4$-$0) bands.
Also seen is one of the CO Rydberg bands at 1076 \AA.}
\end{figure}

\begin{figure}
\plotone{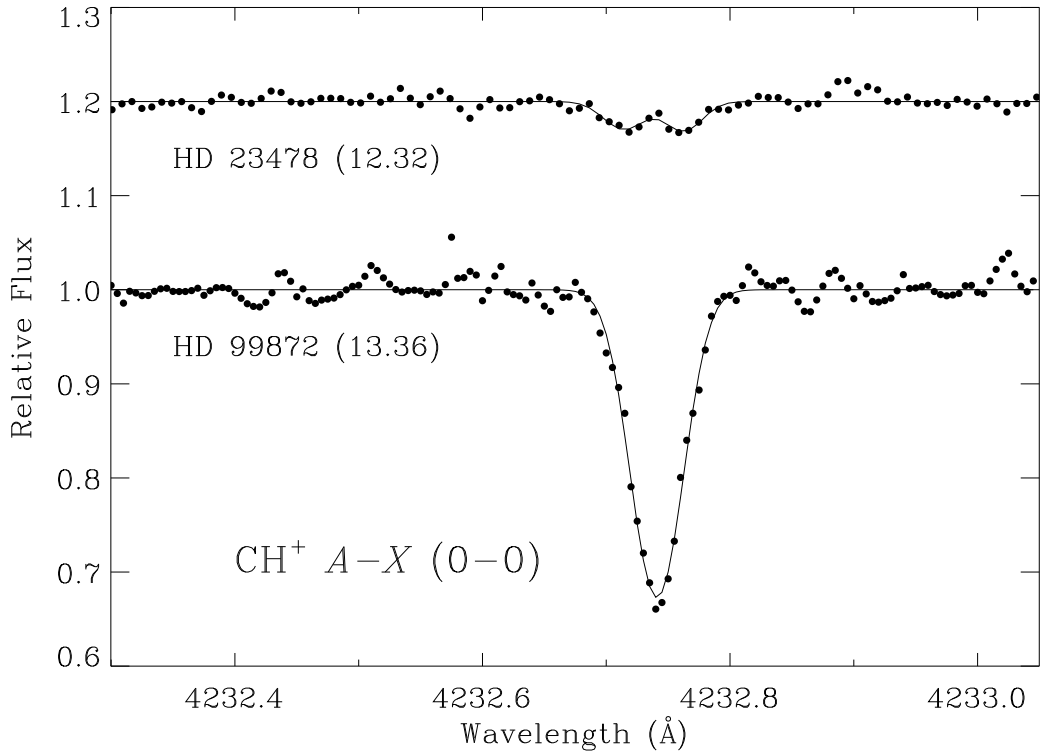}
\caption{Sample optical CH$^+$ spectra (dots) and Ismod.f fits (solid lines)
are shown for HD 99872 (from ESO, log $N$ = 13.36) and HD 23478 (McDonald, 12.32).
The second spectrum has been shifted upward by 0.2 continuum units.
Note the presence of two cloud components along the sight line toward HD 23478
(also seen in CN in Fig. 5).}
\end{figure}

\begin{figure}
\plotone{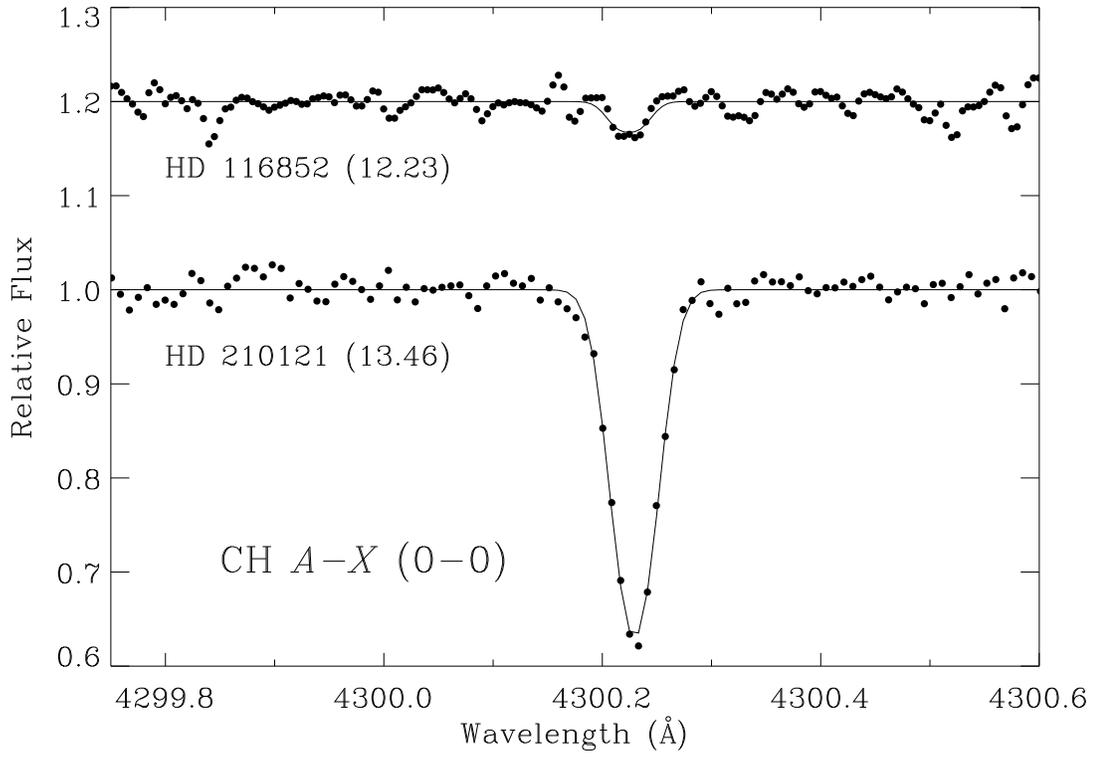}
\caption{Sample optical CH spectra (dots) and Ismod.f fits (solid lines)
are shown for HD 210121 (from McDonald, log $N$ = 13.46) and HD 116852 (ESO, 12.23).
The second spectrum has been shifted upward by 0.2 continuum units.} 
\end{figure}

\begin{figure}
\plotone{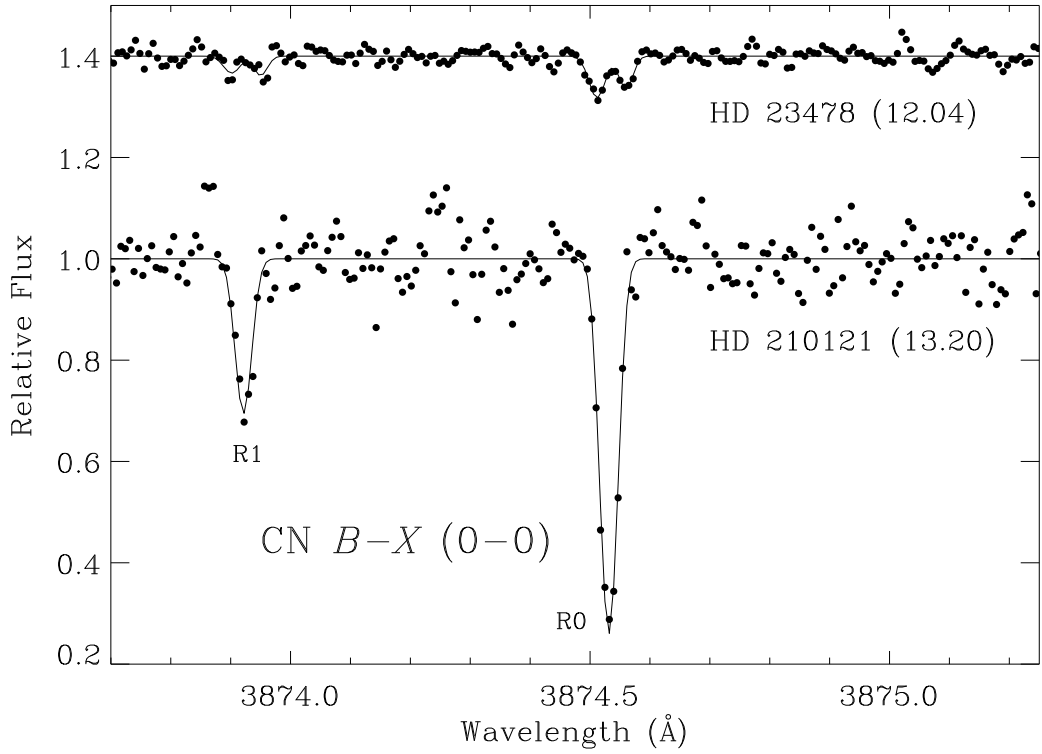}
\caption{Sample optical CN spectra (dots) from McDonald and Ismod.f fits (solid lines)
are shown for HD 210121 (log $N$ = 13.20) and HD 23478 (12.04).
Both R0 and R1 have two cloud
components toward HD 23478, as is the case with CH$^+$ in Fig. 3.} 
\end{figure}

\begin{figure}
\plotone{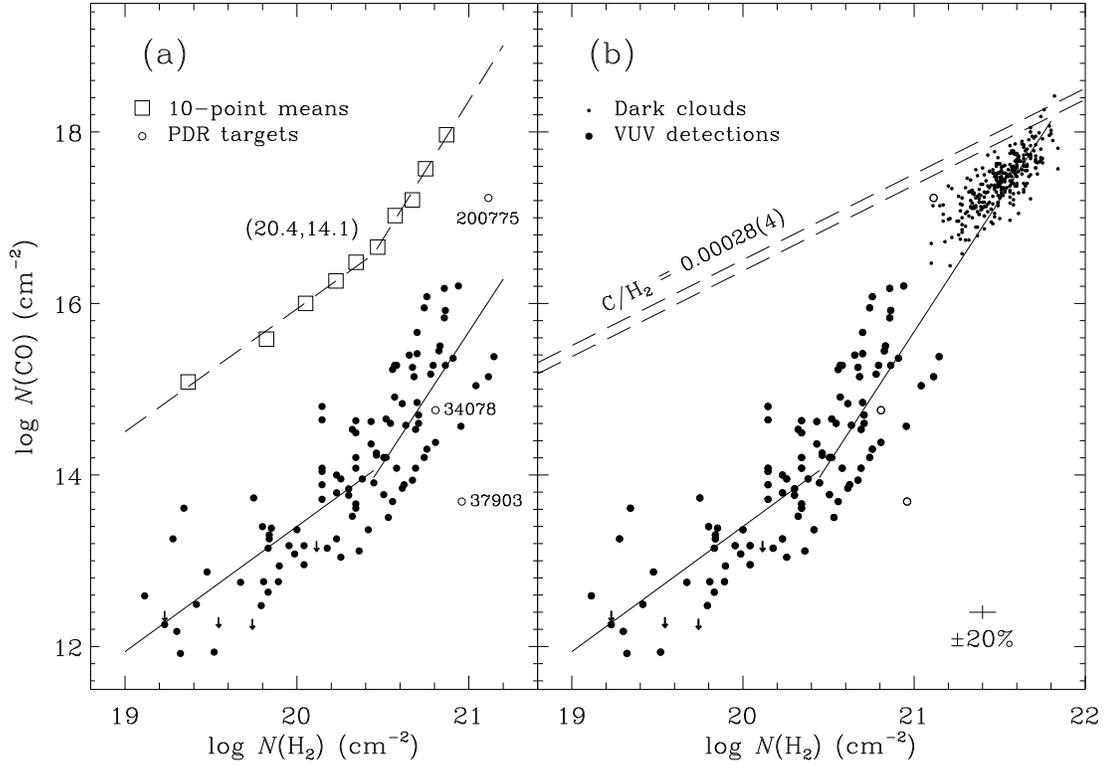}
\caption{CO versus H$_2$ is shown for our sample of diffuse clouds.
Open circles here and in subsequent figures
represent sight lines probing prominent PDRs.
In panel (a) the sample is fit with two power laws, as revealed by the 10-point means,
with a break at log $N$ = (20.4, 14.1).
Panel (b) expands the view to include CO derived for dark clouds (smaller dots, \citealt{Fede90}).}
\end{figure}

\begin{figure}
\plotone{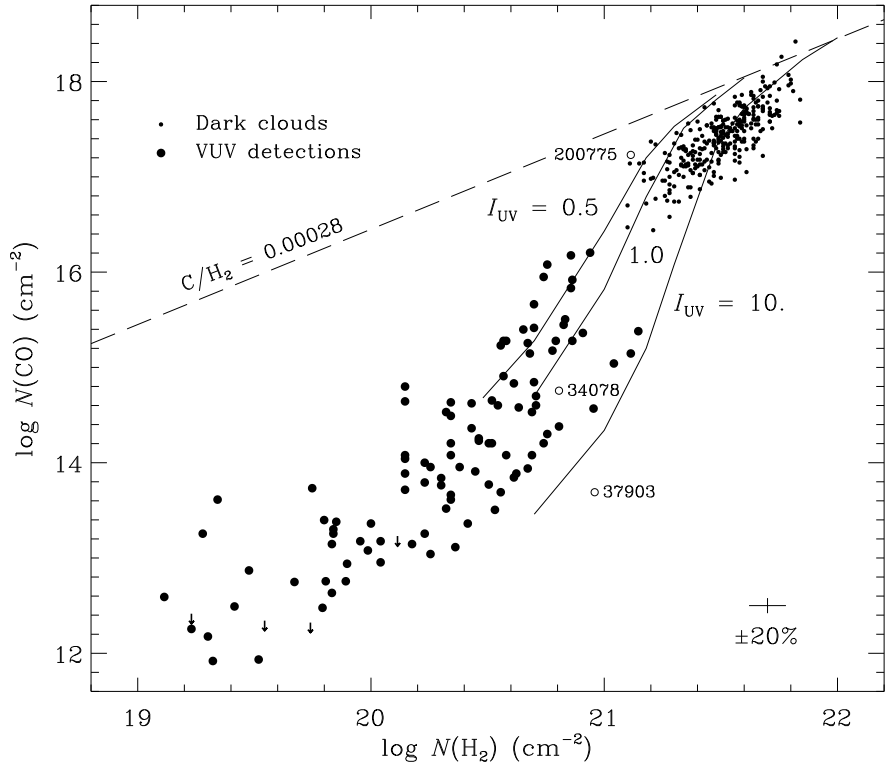}
\caption{CO versus H$_2$ distribution for diffuse and dark clouds is compared
with the H ($I_{\rm UV}$ = 0.5), T (1.0), and I (10) theoretical models for translucent
clouds from \citet{vanD88}.
Note the overall agreement between the shape of model curves
and the transition region from diffuse to dark cloud regimes, as well as
with the observed slopes of each type of clouds.}
\end{figure}

\begin{figure}
\plotone{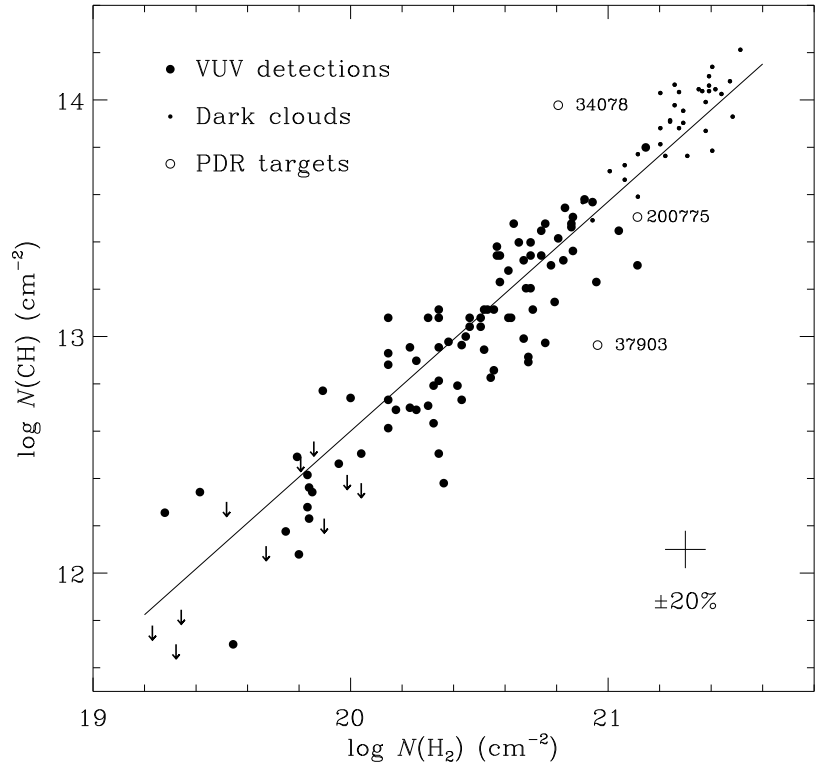}
\caption{CH column density is shown to have a linear correlation with $N$(H$_2$),
characterized by a single slope of 0.97 $\pm$ 0.07. The optical-data fit is seen
to match the \citet{Matt86} dark cloud extension of the CH versus H$_2$ relationship.}
\end{figure}

\clearpage
\begin{figure}
\plotone{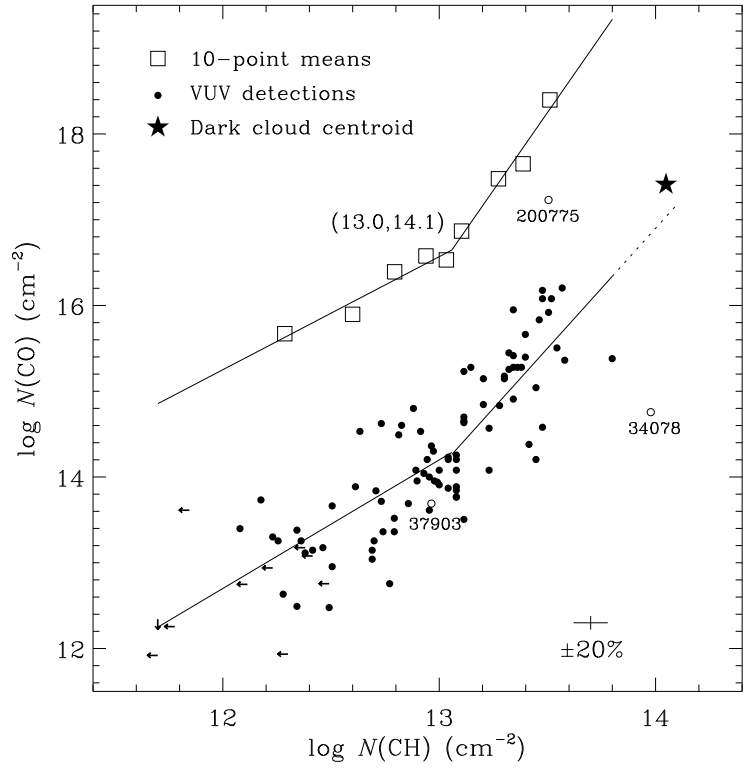}
\caption{
CO has a dual-slope relationship with CH as it has
with H$_2$ (cf. Fig. 6).
The break in slopes is found at
log $N$(CH) = 13.0 and log $N$(CO) = 14.1, in excellent agreement with the CO versus H$_2$
break found in Fig. 6 and with log $<$CH/H$_2$$>$ = $-$7.5.}
\end{figure}

\begin{figure}
\plotone{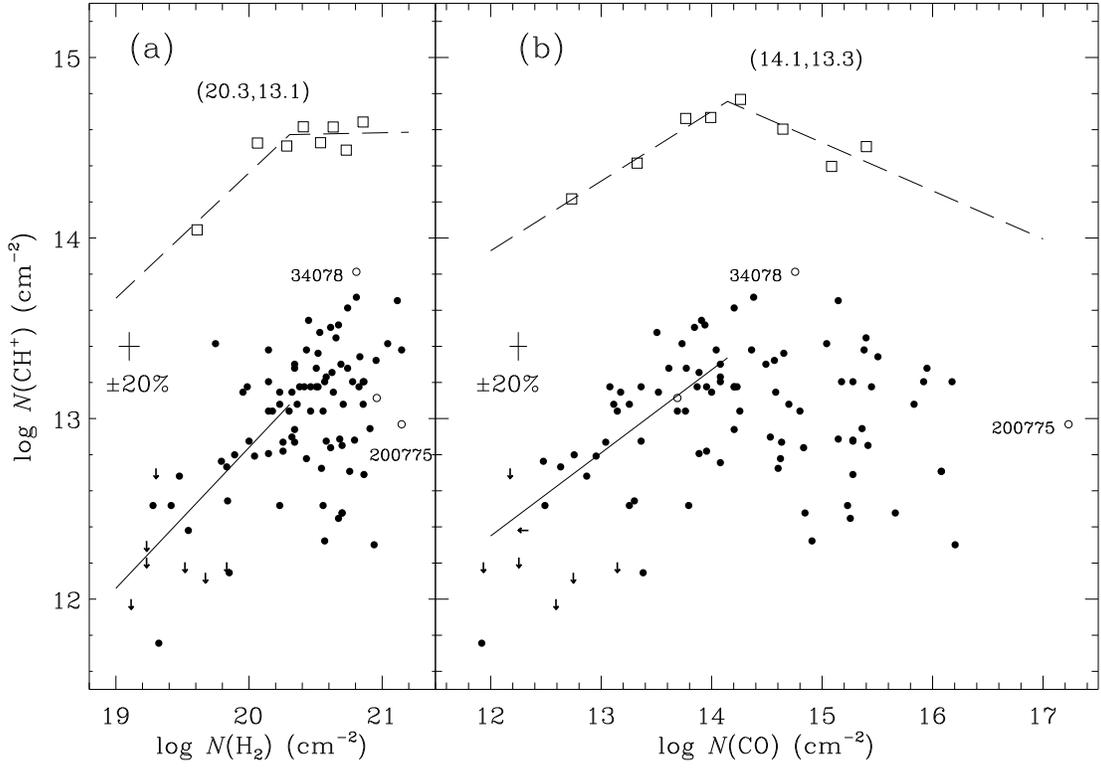}
\caption{
In panel (a), log $N$(CH$^+$) is seen to be correlated with log $N$(H$_2$) $\leq$ 20.3.
Similarly in (b), CH$^+$ is seen to be well correlated with CO below log $N$ = 14.1.
These two power-law breaks are in excellent agreement with the break of CO versus H$_2$
(Fig. 6), as well as in excellent mutual agreement that the CH$^+$ abundance stops increasing
at $<$log $N$$>$ = 13.2 $\pm$ 0.1.}
\end{figure}

\begin{figure}
\plotone{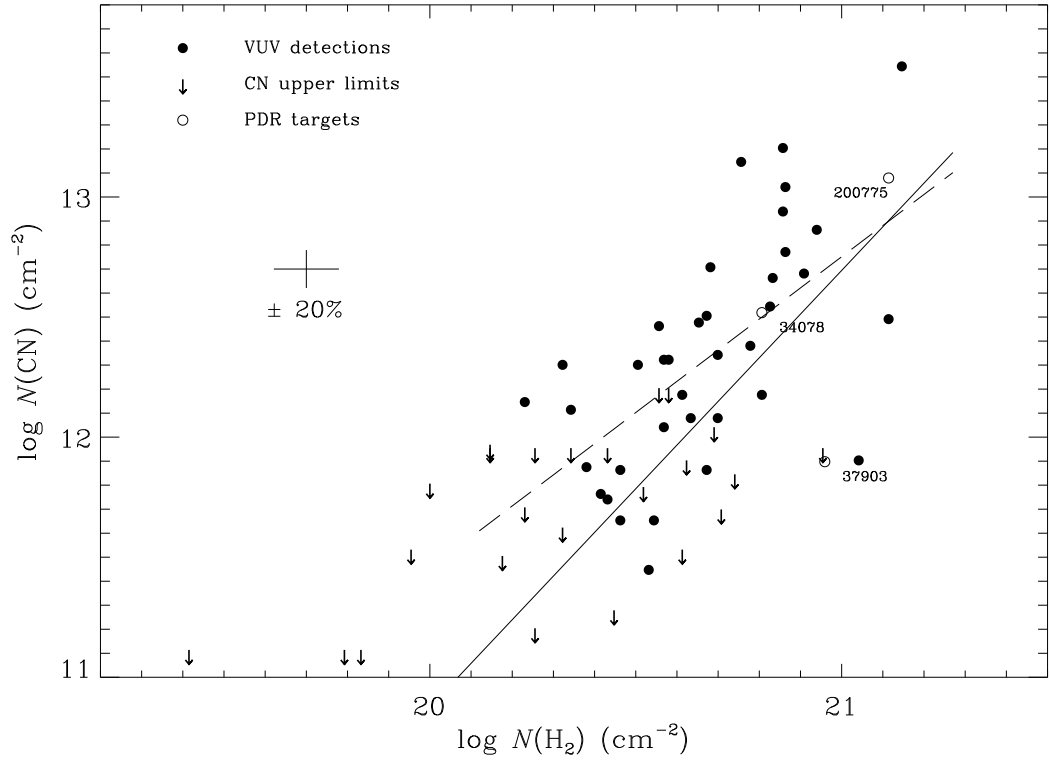}
\caption{Fitting CN detections only versus H$_2$ returns a slope of $B$ = 1.5 $\pm$ 0.4. 
(dashed line). A slope of $B$ = 1.8 $\pm$ 0.3 results from using the Buckley-James
regression method for censored data.}
\end{figure}

\begin{figure}
\plotone{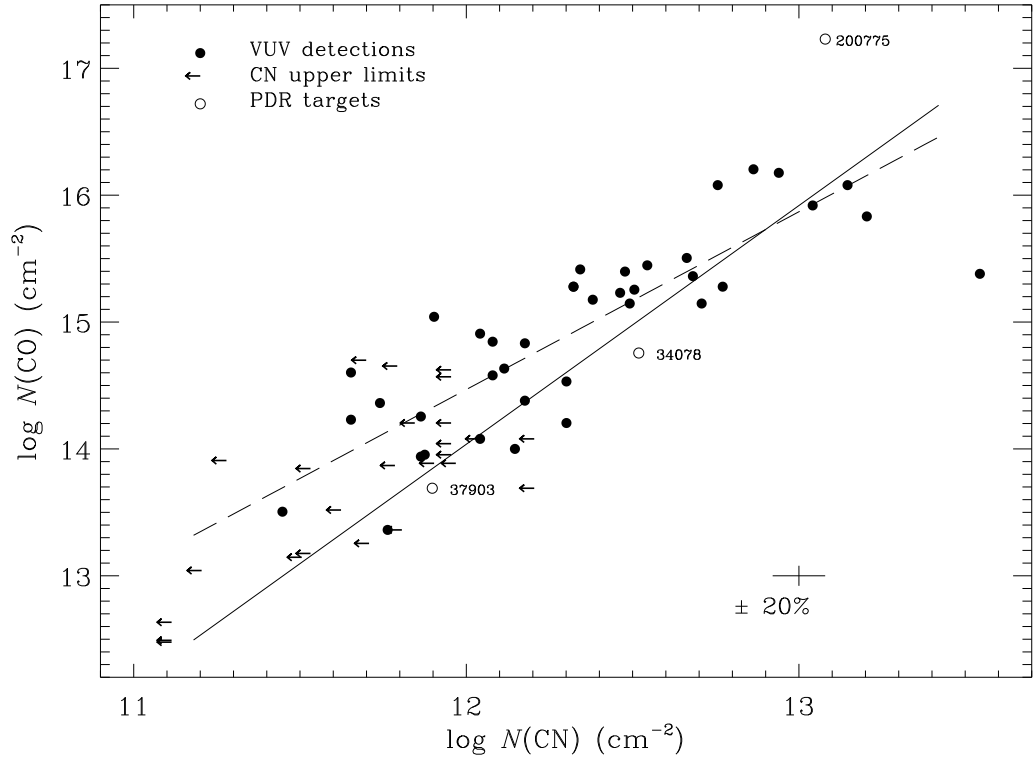}
\caption{Two heavier diatomic molecules, CO and CN, are shown to be well
correlated. This confirms earlier clues that CO and CN are found together in the
same colder and denser clumps of gas. The dashed line shows a fit of CN
detections only with $B$ = 1.4 $\pm$ 0.2, and the solid line is a Buckley-James regression
that includes all CN upper limits (censored data, $B$ = 1.9 $\pm$ 0.2).}
\end{figure}

\begin{figure}
\plotone{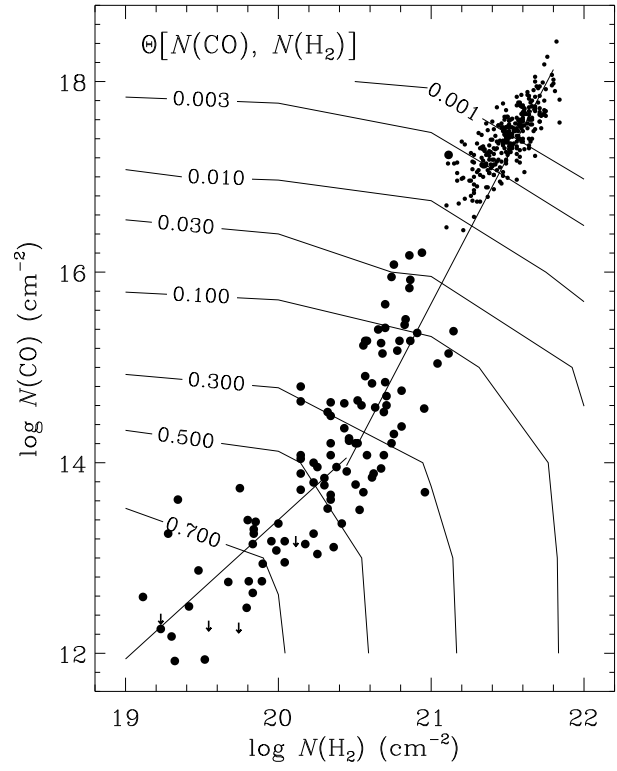}
\caption{Interpolated values of the shielding function $\Theta$ of \citet{vanD88} are plotted as
contours over the observed distribution of $N$(CO) versus $N$(H$_2$).
UV shielding plays a role in the steepening slope beyond log $N$(CO) $\approx$ 15.}
\end{figure}

\begin{figure}
\plotone{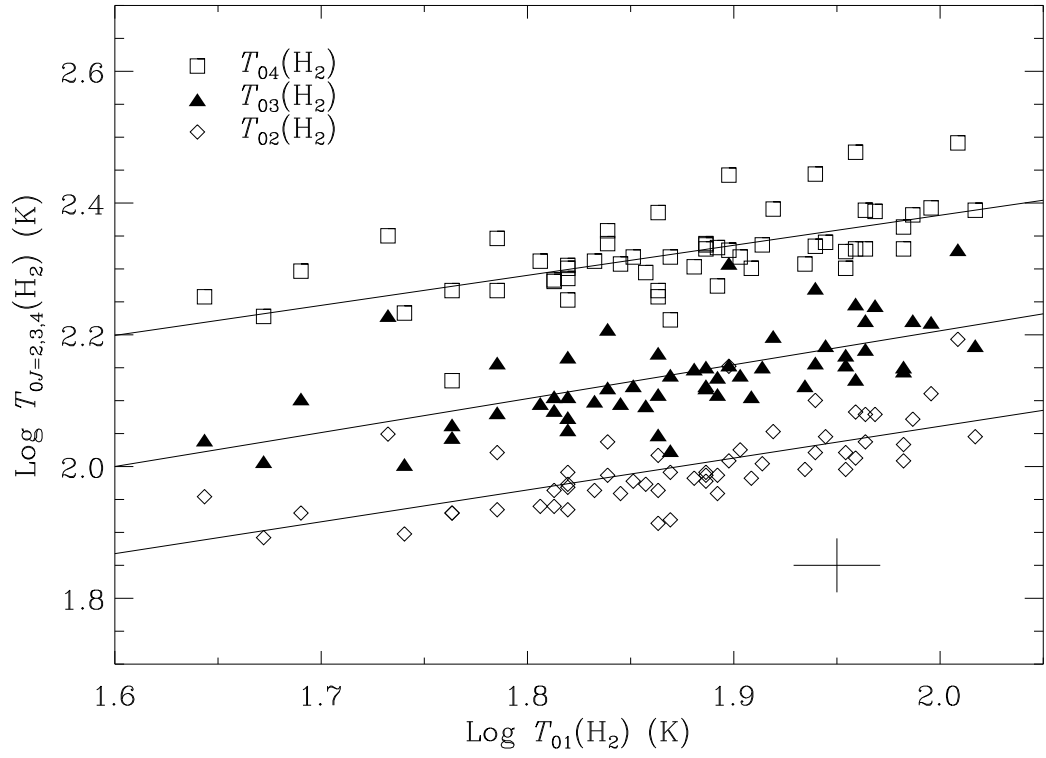}
\caption{Three higher-$J$ excitation temperatures of H$_2$ are plotted versus
$T_{\rm ex}$ of the $J$ = 1 level, showing three indistinguishable positive correlations
with $<B>$ = 0.48 $\pm$ 0.08.
The adopted global uncertainties are $\pm$5\% and $\pm$10\% in $T_{01}$ and
$T_{0J}$ ($J >$ 1), respectively.}
\end{figure}

\begin{figure}
\plotone{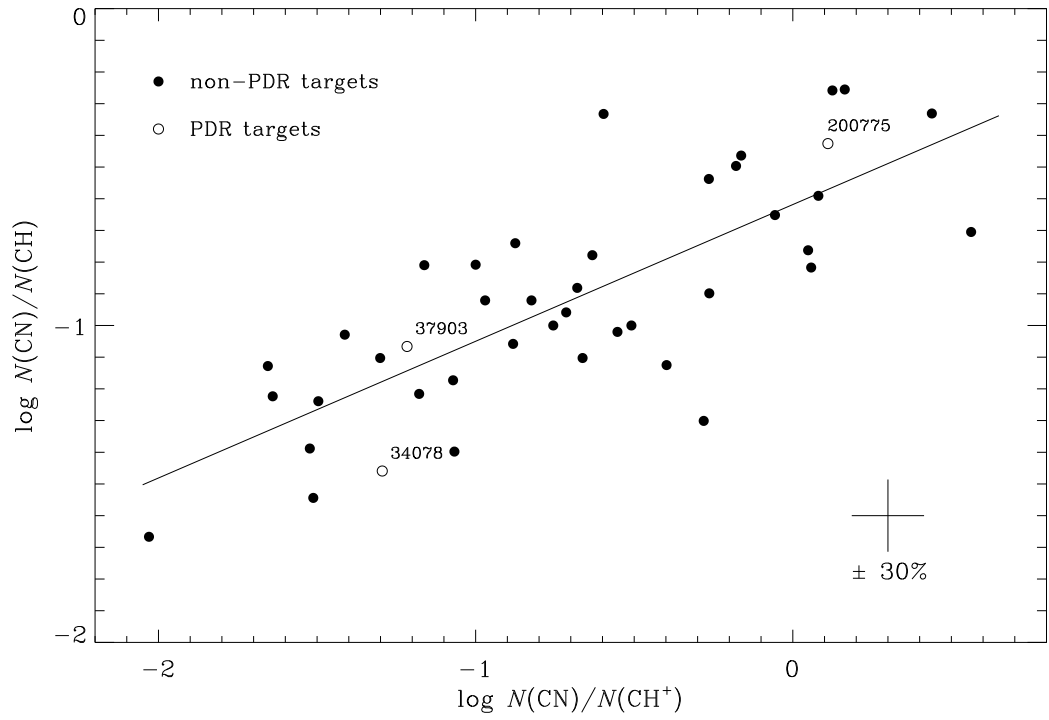}
\caption{This plot shows the empirical density indicator, CN/CH, to be well correlated with
CN/CH$^+$, so that gas density is increasing with either quantity. The relationship has a slope
of 0.43 $\pm$ 0.06 and $r$ = 0.77.}
\end{figure}

\begin{figure}
\plotone{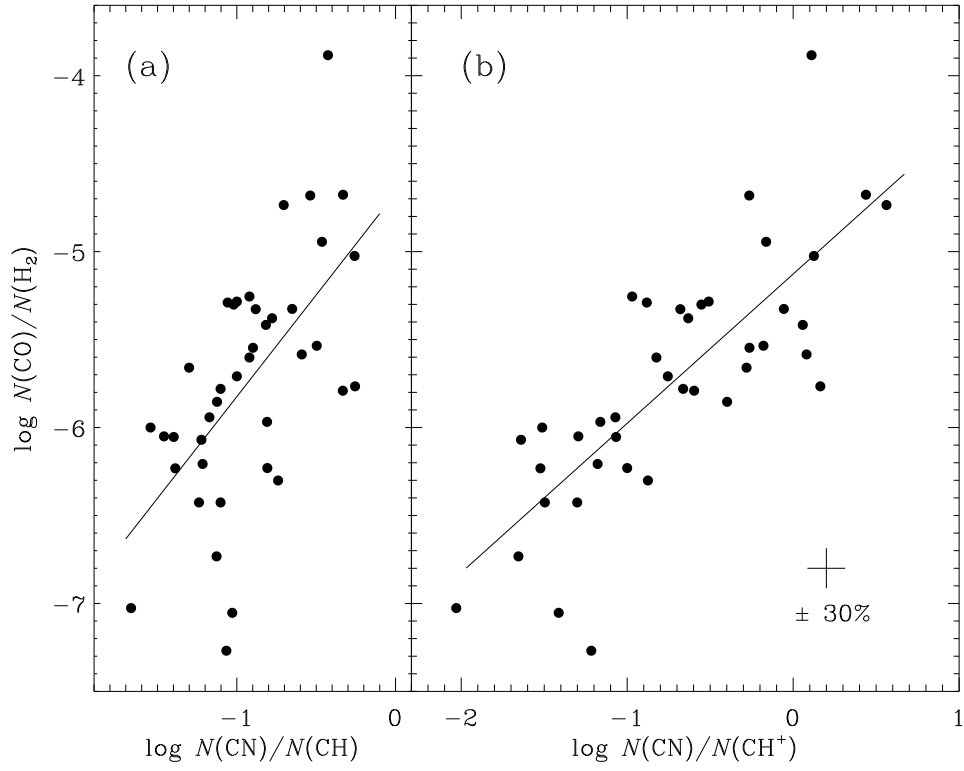}
\caption{Abundance of CO relative to H$_2$ is plotted versus two empirical
indicators of gas density: CN/CH in (a) and CN/CH$^+$ in (b). The sample of
CO/H$_2$ data points shows a tighter correlation with CN/CH$^+$ ($r$ = 0.784) than with
CN/CH ($r$ = 0.604).}
\end{figure}

\clearpage
\begin{figure}
\plotone{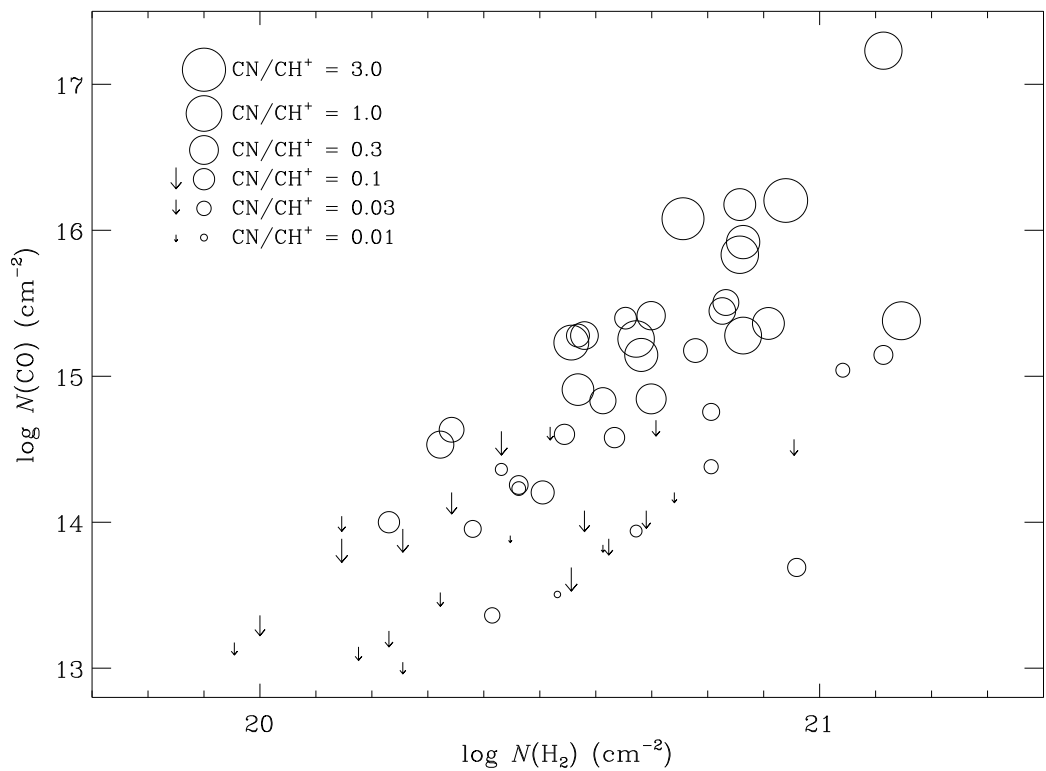}
\caption{
Open circles denote values of the density indicator $N$(CN)/$N$(CH$^+$),
showing that higher gas density is associated with higher $N$(CO)
along the upper envelope of the distribution, as well as
with higher $N$(H$_2$) (from left to right along the diagonal).}
\end{figure}

\begin{figure}
\plotone{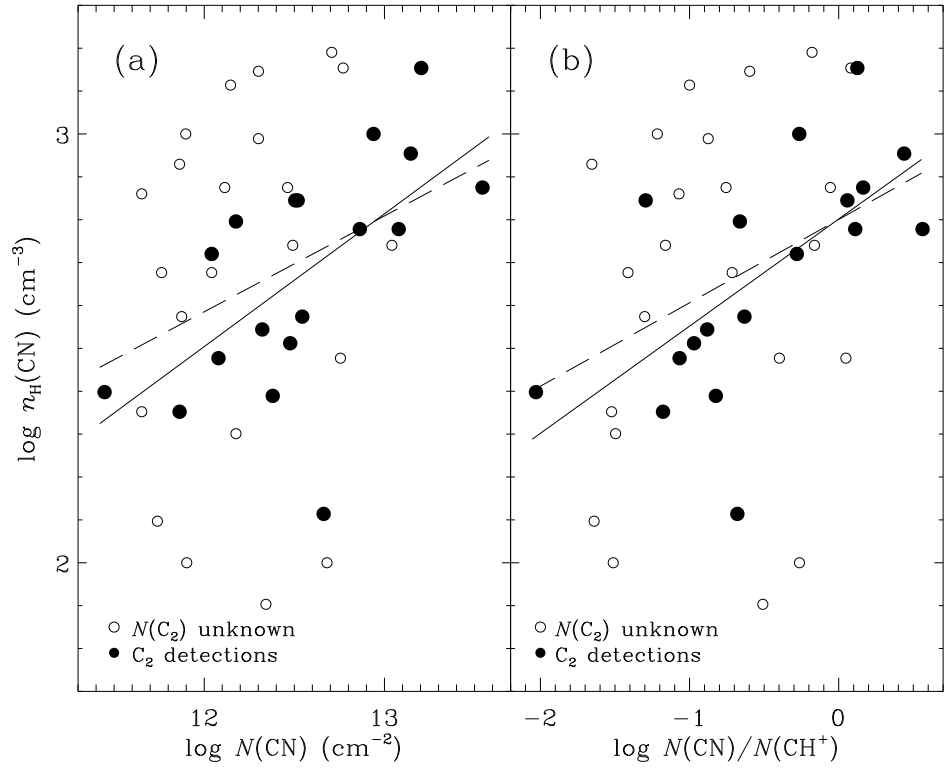}
\caption{Panel (a) shows density from the CN chemical analysis versus $N$(CN).
Sight lines with known $N$(C$_2$) (filled circles) have $r$ = 0.61 (solid line) instead of 0.32
(CL increases from 95 to 99\%).
Panel (b) shows $n_{\rm H}$ versus $N$(CN)/$N$(CH$^+$),
where $r$ increases from 0.36 to 0.64, and CL from 98 to 99.5\% owing to restricting
the sample to those sight lines with known $N$(C$_2$).}
\end{figure}

\begin{figure}
\plotone{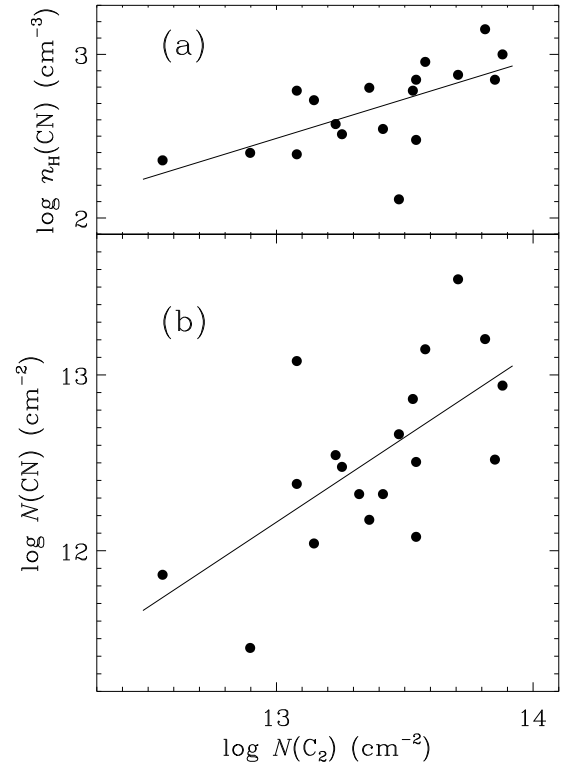}
\caption{Values of $n_{\rm H}$(CN) from the chemical analysis are seen in (a) to be well
correlated with $N$(C$_2$) values. The slope is $B$ = 0.48 $\pm$ 0.15 and $r$ = 0.63.
In (b) CN is shown to be linearly related to C$_2$, since the slope of
their abundance relationship is 0.97 $\pm$ 0.28 ($r$ = 0.64).}
\end{figure}

\begin{figure}
\plotone{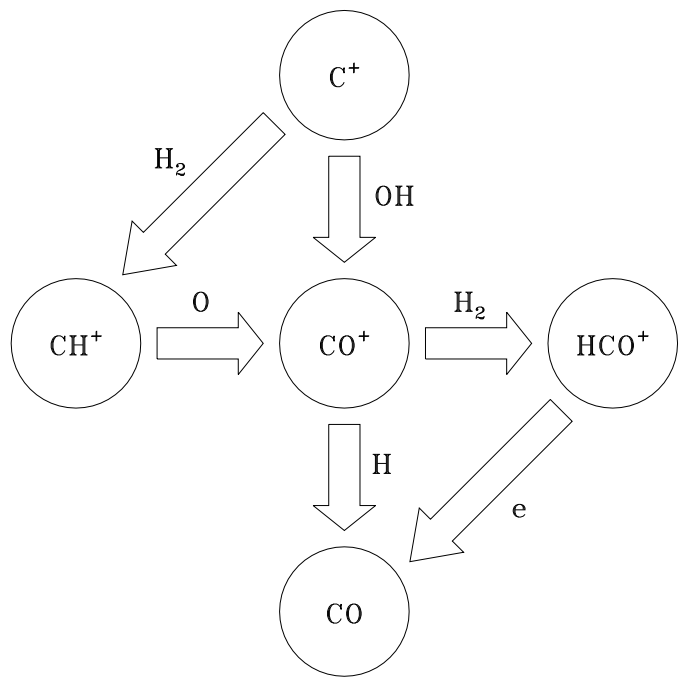}
\caption{Schematic formation routes from C$^+$ to CO involving the most common
intermediate gas-phase chemical reactants and products.}
\end{figure}

\begin{figure}
\plotone{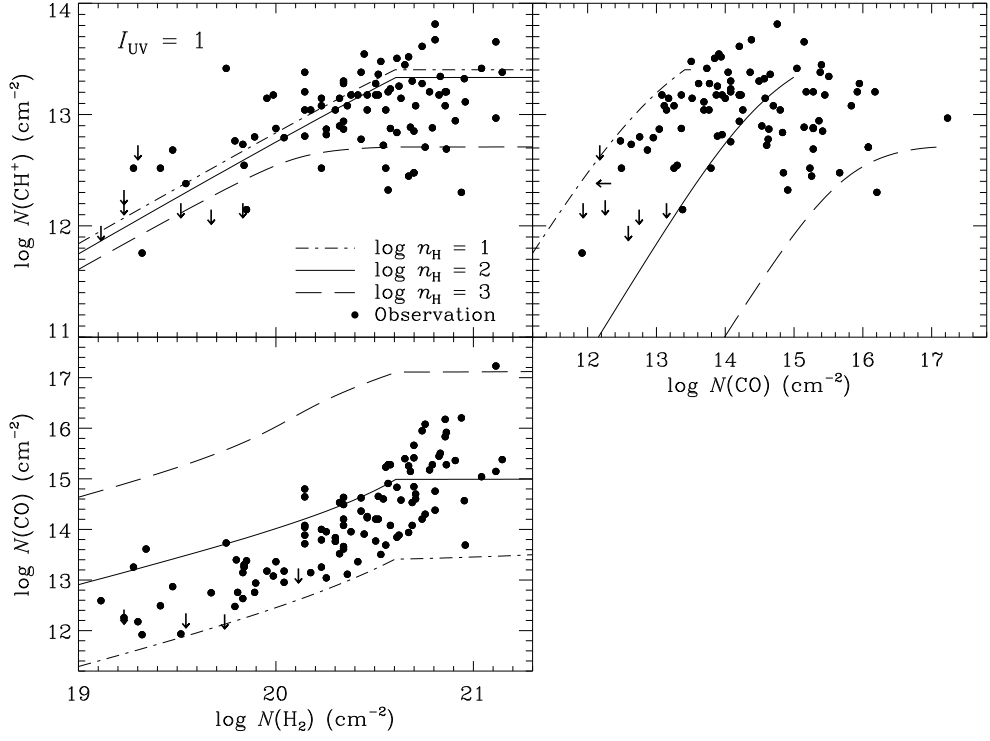}
\caption{CH$^+$ versus H$_2$, CH$^+$ versus CO, and CO versus H$_2$ as a
function of
$n_{\rm H}$, for the average value of the far-UV interstellar radiation 
field ($I_{\rm UV}$ = 1) and $\Delta v_{\rm turb}$ = 3.3 km s$^{-1}$.}
\end{figure}

\begin{figure}
\plotone{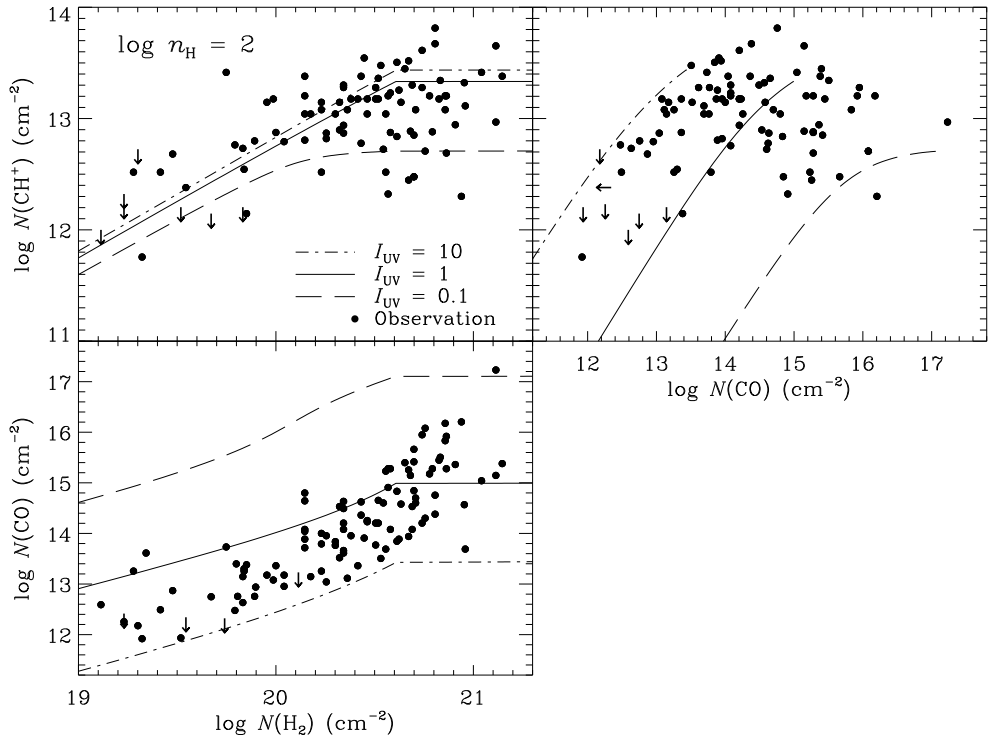}
\caption{CH$^+$ versus H$_2$, CH$^+$ versus CO, and CO versus H$_2$ as a
function of
$I_{\rm UV}$, for $n_{\rm H}$ = 100 cm$^{-3}$ and $\Delta v_{\rm turb}$ = 3.3 km
s$^{-1}$.}
\end{figure}

\begin{figure}
\plotone{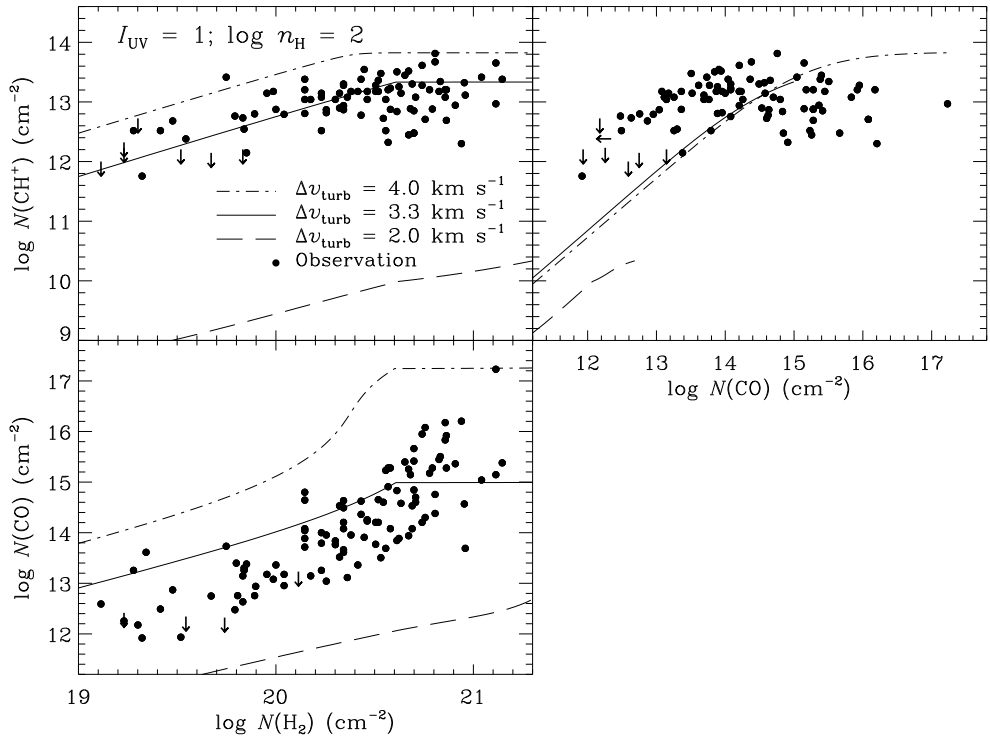}
\caption{CH$^+$ versus H$_2$, CH$^+$ versus CO, and CO versus H$_2$ as a
function of
$\Delta v_{\rm turb}$, for $I_{\rm UV}$ = 1 and $n_{\rm H}$ = 100 cm$^{-3}$.}
\end{figure}

\end{document}